\documentclass[twocolumn,times,astrosymb,tighten,twocolappendix]{aastex631}

\usepackage{graphicx,subfigure,amsmath, amsfonts, amssymb,aas_macros,times,soul}
\hypersetup{linkcolor=ProcessBlue,
			citecolor=blue,
			filecolor=RoyalPurple,
			urlcolor=Emerald}
\newcommand{\magphys}{{\sc magphys}}

\begin{document}
\title{\Large Characterizing the Molecular Gas in Infrared Bright Galaxies with CARMA}

\correspondingauthor{Katherine Alatalo}
\email{kalatalo@stsci.edu}

\author[0000-0002-4261-2326]{Katherine Alatalo}
\affiliation{Space Telescope Science Institute, 3700 San Martin Dr., Baltimore, MD 21218, USA}
\affiliation{William H. Miller III Department of Physics and Astronomy, Johns Hopkins University, Baltimore, MD 21218, USA}

\author[0000-0003-4030-3455]{Andreea O. Petric}
\affiliation{Space Telescope Science Institute, 3700 San Martin Dr., Baltimore, MD 21218, USA}
\affiliation{William H. Miller III Department of Physics and Astronomy, Johns Hopkins University, Baltimore, MD 21218, USA}

\author[0000-0002-3249-8224]{Lauranne Lanz}
\affiliation{The College of New Jersey, 2000 Pennington Road, Ewing, NJ 08628, USA}

\author[0000-0001-7883-8434]{Kate Rowlands}
\affiliation{AURA for ESA, Space Telescope Science Institute, 3700 San Martin Drive, Baltimore, MD, USA}
\affiliation{William H. Miller III Department of Physics and Astronomy, Johns Hopkins University, Baltimore, MD 21218, USA}

\author[0000-0002-1912-0024]{Vivian U}
\affiliation{Department of Physics and Astronomy, 900 University Avenue, University of California, Riverside CA 92521, USA}
\altaffiliation{University of California Chancellor's Postdoctoral Fellow}
\affiliation{Department of Physics and Astronomy, 4129 Frederick Reines Hall, University of California, Irvine, CA 92697, USA}

\author[0000-0003-3917-6460]{Kirsten L. Larson}
\affiliation{AURA for ESA, Space Telescope Science Institute, 3700 San Martin Drive, Baltimore, MD, USA}

\author[0000-0003-3498-2973]{Lee Armus}
\affiliation{Infrared Processing and Analysis Center, California Institute of Technology, 1200 E. California Blvd, Pasadena, CA, 91125, USA}

\author[0000-0003-0057-8892]{Loreto Barcos-Mu\~noz}
\affiliation{National Radio Astronomy Observatory, 520 Edgemont Road, Charlottesville, VA 22903, USA}
\affiliation{Department of Astronomy, University of Virginia, 530 McCormick Road, Charlottesville, VA 22903, USA}

\author[0000-0003-2638-1334]{Aaron~S. Evans}
\affiliation{Department of Astronomy, University of Virginia, 530 McCormick Road, Charlottesville, VA 22903, USA}
\affiliation{National Radio Astronomy Observatory, 520 Edgemont Road, Charlottesville, VA 22903, USA}

\author[0000-0002-8762-7863]{Jin Koda}
\affiliation{Dept. of Physics and Astronomy, Stony Brook University, Stony Brook, NY 11794-3800}

\author[0000-0002-0696-6952]{Yuanze Luo}
\affiliation{William H. Miller III Department of Physics and Astronomy, Johns Hopkins University, Baltimore, MD 21218, USA}

\author[0000-0001-7421-2944]{Anne~M. Medling}
\affiliation{Ritter Astrophysical Research Center and Department of Physics \& Astronomy, University of Toledo, Toledo, OH 43606, USA}

\author[0000-0003-1991-370X]{Kristina~E. Nyland}
\affiliation{U.S. Naval Research Laboratory, 4555 Overlook Ave SW, Washington, DC 20375, USA}

\author[0000-0003-3191-9039]{Justin A. Otter}
\affiliation{William H. Miller III Department of Physics and Astronomy, Johns Hopkins University, Baltimore, MD 21218, USA}

\author[0000-0002-9471-8499]{Pallavi Patil}
\affiliation{William H. Miller III Department of Physics and Astronomy, Johns Hopkins University, Baltimore, MD 21218, USA}

\author[0000-0003-1839-5859]{Fernando Pe\~naloza}
\affiliation{Instituto de Astrof\'isica, Facultad de F\'isica, Pontificia Universidad Cat\'olica de Chile, Campus San Joaqu\'in, Av. Vicu\~na Mackenna
3 4860, Macul Santiago, Chile, 7820436}

\author[0000-0002-2858-3506]{Diane Salim}
\affiliation{Physics \& Astronomy, Rutgers, The State University of New Jersey, 136 Frelinghuysen Road, Piscataway, NJ 08854-8019}

\author[0000-0002-1233-9998]{David~B. Sanders}
\affiliation{Institute for Astronomy, 2680 Woodlawn Drive, Honolulu, HI 96822, USA}

\author[0000-0001-6245-5121]{Elizaveta Sazonova}
\affiliation{Department of Physics and Astronomy, University of Waterloo, 200 University Avenue West, Waterloo, Ontario, Canada N2L 3G1}

\author[0009-0004-0844-0657]{Maya Skarbinski}
\affiliation{William H. Miller III Department of Physics and Astronomy, Johns Hopkins University, Baltimore, MD 21218, USA}

\author[0000-0002-3139-3041]{Yiqing Song}
\affiliation{European Southern Observatory, Alonso de C\'ordova, 3107, Vitacura, Santiago, 763-0355, Chile}
\affiliation{Joint ALMA Observatory, Alonso de C\'ordova, 3107, Vitacura, Santiago, 763-0355, Chile}
\altaffiliation{ESA-ALMA Fellow}

\author[0000-0001-7568-6412]{Ezequiel Treister}
\affiliation{Instituto de Astrof\'isica, Facultad de F\'isica, Pontificia Universidad Cat\'olica de Chile, Campus San Joaqu\'in, Av. Vicu\~na Mackenna
3 4860, Macul Santiago, Chile, 7820436}

\author[0000-0002-0745-9792]{C. Meg Urry}
\affiliation{Yale Center for Astronomy \& Astrophysics, Physics Department, P.O. Box 208120, New Haven, CT 06520, USA}
\affiliation{Department of Physics, Yale University, P.O. Box 208121, New Haven, CT 06520, USA}

\begin{abstract}
We present the CO(1--0) maps of 28 infrared-bright galaxies from the Great Observatories All-Sky Luminous Infrared Galaxy Survey (GOALS) taken with the Combined Array for Research in Millimeter Astronomy (CARMA). 
We detect 100\,GHz continuum in 16 of 28 galaxies, which trace both active galactic nuclei (AGNs) and compact star-forming cores.
The GOALS galaxies show a variety of molecular gas morphologies, though in the majority of cases, the average velocity fields show a gradient consistent with rotation.
We fit the full continuum SEDs of each of the source using either \magphys\ or SED3FIT (if there are signs of an AGN) to derive the total stellar mass, dust mass, and star formation rates of each object.
We adopt a value determined from luminous and ultraluminous infrared galaxies (LIRGs and ULIRGs) of $\alpha_{\rm CO}$\,=\,$1.5^{+1.3}_{-0.8}$\,M$_\odot$ (K~km~s$^{-1}$~pc$^2$)$^{-1}$, which leads to more physical values for $f_{\rm mol}$ and the gas-to-dust ratio. Mergers tend to have the highest gas-to-dust ratios.
We assume the cospatiality of the molecular gas and star formation, and plot the sample on the Schmidt-Kennicutt relation, we find that they preferentially lie above the line set by normal star-forming galaxies. This hyper-efficiency is likely due to the increased turbulence in these systems, which decreases the freefall time compared to star-forming galaxies, leading to ``enhanced'' star formation efficiency.
Line wings are present in a non-negligible subsample (11/28) of the CARMA GOALS sources and are likely due to outflows driven by AGNs or star formation, gas inflows, or additional decoupled gas components.
\end{abstract}


\section{Introduction}
\label{sec:intro}
There is a bimodality seen between blue, star-forming spiral galaxies to red, quiescent early-type (lenticular and elliptical) galaxies \citep{hubble26,hubble36,baade58}, with a dearth of galaxies seen with transitional (green) colors \citep{tinsley78,strateva+01,baldry+04}. This suggests that galaxies must transform rapidly from blue spirals and red early-types \citep{faber+07}.

Galaxy interactions play a crucial role in galaxy evolution, having been identified as an effective way to transform star-forming disk galaxies into quiescent ellipticals \citep{holmberg41,toomre72}, through the loss of gas angular momentum and the growth of a bulge. These interactions are also efficient at rapidly ceasing the star formation in a galaxy through a combination of gas expulsion from the system \citep{dimatteo+05,hopkins+08} and driving gas into the circumnuclear region, where it undergoes a starburst \citep{mihos+96,bryant+99,he+23}.

Interactions also tend to be the locations of the most prolific star formation in the local universe and have long been studied in an attempt to understand the evolution of galaxies from star-forming spirals into passive, bulge-dominated systems. In that quest, the molecular gas in these systems has been considered a linchpin to the understanding of how star formation first rises rapidly, and then ultimately ceases.

Simulations of galaxy interactions seem to show a common narrative as the merger progresses (e.g., \citealt{hopkins+08,teyssier+10,renaud+14,fensch+21,he+23}). Initially, (1) the galaxies form a small group or pair, (2) there is an interaction and a first pass until ultimately (3) the galaxies coalesce and go through the star-bursting infrared-bright (ULIRG) phase. Then (4) the supermassive black hole in the center ignites and blows all of the remaining circumnuclear gas out of the system \citep{ricci+17}, ending in a (5) quasar phase (in which the AGN is visible), which then (6) decays into a post-starburst ``E+A'' galaxy \citep{zabludoff+96,quintero+04} and ultimately a (7) quiescent galaxy. The timescale for this transformation is $\sim$\,1\,Gyr \citep{hopkins+08,lanz+14}.

Recent observations of molecular gas in post-merger systems have called into question whether the molecular gas is truly dissipated in this way. AGN-driven molecular gas outflows have now been seen in a substantial number of infrared-bright galaxies \citep{feruglio+10,combes+13,garcia-burillo+14,garcia-burillo+15,sakamoto+13,sakamoto+14,cicone+14,luo+22}, challenging whether these outflows can simultaneously be ubiquitous and dramatically transform the molecular content of these galaxies on very short ($\sim$10s of Myr) timescales. Observations of Mrk\,231 by \citet{alatalo15} suggest that the importance of the molecular outflows observed in these galaxies may have much less impact than previously thought, namely, that the majority of the molecular gas taking part in the outflow will not escape and deplete from the system, due to being below the escape velocity. This lengthened the depletion time to better match the age of stellar populations in the system \citep{canalizo+00} and is able to explain the ubiquity of these outflows.

%%%%%%%%%%%%%%% Table 1 %%%%%%%%%%%%%%%
\begin{table*}[t]
\centering
\caption{CARMA GOALS Properties and Observing Parameters} \vspace{-1mm}
\begin{tabular}{l r r r c c l r r r r}
\hline \hline
{\bf Object} & {\bf RA}~~~~~\, & {\bf Dec}~~~~ & \textbf{\em D}~~~ & {\bf Morph.} & {\bf log(}$L_{\rm IR}{\bf)}$ & {\bf Semester} & {\bf Total} & {\bf Gain Cal} & {\bf Beam} & {\bf KperJy} \\
{\bf Name} & J2000~~~ & J2000~~~ & Mpc~ & {\bf Class} & $L_\odot$ & & {\bf Hours} & & \arcsec$\times$\arcsec & \\
~~~~~~~(1) & (2)~~~~~ & (3)~~~~~ & (4)~~ & (5) &(6) & (7) & (8) & (9) & (10) & (11)\\
\hline
MCG+12-02-001    & 00:54:03.6 & +73:05:12 & 68.1  & m & 11.4 & 2010a     & 5.0  & 0217+738 & 5.05$\times$4.01 & 4.55   \\
CGCG\,436-030    & 01:20:02.7 & +14:21:43 & 137.0 & m & 11.6 & 2010b     & 4.9  & 0108+015 & 1.88$\times$1.67 & 29.24  \\
III\,Zw\,35      & 01:44:30.5 & +17:06:05 & 119.8 & em & 11.6 & 2008b     & 4.5  & 0238+166 & 4.08$\times$3.37 & 6.71   \\
NGC\,695         & 01:51:14.2 & +22:34:57 & 142.5 & nm & 11.6 & 2010a     & 2.4  & 0205+322 & 4.67$\times$3.37 & 5.85   \\
NGC\,958         & 02:30:42.8 & -02:56:20 & 83.2  & nm & 11.2 & 2010b     & 12.5 & 0239-025 & 3.11$\times$2.96 & 10.00  \\
UGC\,02369       & 02:54:01.8 & +14:58:25 & 136.8 & em & 11.6 & 2008b     & 2.0  & 0238+166 & 4.20$\times$3.19 & 6.87   \\
UGC\,02608       & 03:15:01.4 & +42:02:09 & 101.8 & nm & 11.4 & 2010b     & 3.7  & 3C84     & 2.86$\times$2.46 & 13.06  \\
IRAS\,03582+6012 & 04:02:32.5 & +60:20:40 & 131.5 & m & 11.4 & 2010b     & 8.7  & 0359+509 & 2.97$\times$2.38 & 13.03  \\
NGC\,1614        & 04:33:59.8 & -08:34:44 & 69.1  & m & 11.6 & 2012b     & 8.9  & 0423-013 & 3.21$\times$2.93 & 9.81   \\
CGCG\,468-002    & 05:08:20.4 & +17:21:59 & 79.0  & em & 11.1 & 2011a     & 10.3 & 0530+135 & 3.10$\times$2.74 & 10.86  \\
NGC\,2146        & 06:18:37.7 & +78:21:25 & 12.8  & m & 11.1 & 2010b     & 5.6  & 0841+708 & 3.26$\times$2.94 & 9.61   \\
NGC\,2623        & 08:38:24.1 & +25:45:17 & 80.4  & m & 11.5 & 2012a     & 5.9  & 0854+201 & 2.94$\times$2.39 & 13.12  \\
Arp\,55	         & 09:15:55.1 & +44:19:55 & 173.4 & m & 11.7 & 2012a     & 7.5  & 0920+446 & 4.37$\times$3.43 & 6.15   \\
UGC\,05101       & 09:35:51.6 & +61:21:11 & 173.7 & m & 12.0 & 2012b     & 4.9  & 0958+655 & 1.76$\times$1.44 & 36.26  \\
Arp\,148         & 11:03:53.6 & +40:50:57 & 151.7 & m & 11.6 & 2008b     & 3.2  & 0927+390 & 3.74$\times$3.34 & 7.89   \\
Arp\,299         & 11:28:32.3 & +58:33:45 & 44.5  & m & 11.9 & 2013a,14a & 7.4  & 1153+495 & 2.33$\times$2.25 & 17.58  \\
NGC\,4418        & 12:26:54.6 & -00:52:39 & 31.3  & nm & 11.1 & 2011a     & 6.5  & 3C273    & 0.83$\times$0.70 & 159.00 \\
NGC\,4922        & 13:01:24.9 & +29:18:46 & 102.9 & m & 11.3 & 2010b     & 4.8  & 1310+323 & 2.00$\times$1.31 & 35.16  \\
IC\,860	         & 13:15:03.5 & +24:37:08 & 48.2  & nm & 11.2 & 2012a,14a & 5.0  & 1310+323 & 2.29$\times$1.29 & 32.00  \\
VV\,250	         & 13:15:32.8 & +62:07:37 & 135.2 & m & 11.7 & 2010a     & 13.9 & 0958+655 & 4.46$\times$3.63 & 5.69   \\
NGC\,5256        & 13:38:17.5 & +48:16:37 & 121.9 & m & 11.5 & 2010a     & 4.1  & 1310+323 & 3.86$\times$3.33 & 7.16   \\
CGCG\,142-034    & 18:16:40.6 & +22:06:46 & 81.2  & m & 11.1 & 2011a     & 12.1 & 1751+096 & 2.44$\times$1.86 & 20.32  \\
NGC\,6670        & 18:33:35.4 & +59:53:20 & 126.3 & m & 11.6 & 2008b     & 5.5  & 1642+689 & 1.50$\times$0.95 & 65.05  \\
NGC\,6786        & 19:10:53.9 & +73:24:37 & 109.2 & m & 11.4 & 2008b     & 10.1 & 1927+739 & 3.83$\times$3.33 & 7.22   \\
NGC\,6926        & 20:33:06.1 & -02:01:39 & 85.3  & m & 11.3 & 2010b     & 5.5  & 2134-018 & 2.89$\times$2.62 & 12.15  \\
II\,Zw\,96       & 20:57:23.9 & +17:07:39 & 158.9 & m & 11.9 & 2008b     & 5.5  & 2148+069 & 3.22$\times$2.55 & 11.22  \\
IC\,5298         & 23:16:00.7 & +25:33:24 & 119.9 & nm & 11.5 & 2010a     & 6.9  & 2236+284 & 3.02$\times$2.92 & 10.42  \\
NGC\,7674        & 23:27:56.7 & +08:46:45 & 126.6 & em & 11.5 & 2010b     & 7.3  & 0010+109 & 2.54$\times$2.42 & 15.02  \\ \\
\hline \hline
\end{tabular} \\
\label{tab:params}
\raggedright{\footnotesize Column (1): Object name. 
Columns (2-3): RA/declination of central CO pointing. 
Column (4): Luminosity distance to object. 
Column (5): Morphological class using the classification scheme of \citet{petric+18}, of ``m'' for merger, ``em'' for early merger, and ``nm'' for non-merger.
Column (6): Total infrared luminosity from \citet{sanders+03}. 
Column (7): CARMA observing semester(s).
Column (8): Total CARMA on-source observing hours.
Column (9): The gain calibrator for observations.
Column (10): The synthesized beam (in arcseconds).
Column (11): The total Kelvin per Jansky factor of the observation.\\
}
\end{table*}

Additionally, there is further evidence that the molecular gas in these post-merger systems is able to remain beyond star formation quenching. Transitioning galaxies have now been observed to contain non-negligible reservoirs of molecular gas \citep{french+15,rowlands+15,a16_spogco}, suggesting that star formation quenches post-merger, before the entirety of the molecular reservoir has been removed. It is possible that an injection of kinetic energy is able to suppress star formation sufficiently that the depletion timescale of the molecular gas is even further extended \citep{a15_sfsupp,a15_hcgco,guillard+15,lanz+16,salim+20}. In merging systems, turbulence may have the opposite effect \citep{mihos+96,federrath13,renaud+14,salim+15,sparre+16,bustamante+18,thorp+22}, enhancing the star formation efficiency of the nuclear molecular gas, suggesting that the role of turbulence in these transitioning galaxies is multifaceted and complex.

Shedding light on the fate of molecular gas in merging systems requires an in-depth analysis of the processes taking place in molecular gas during all phases of a galaxy's transformation, from the time of first-pass in a merger, through coalescence, to the post-starburst and quiescent phases. An ideal sample to investigate the physics of molecular gas during the ULIRG/coalescence phase of interacting galaxies is the Great Observatory All-sky LIRG Survey (GOALS; \citealt{goals})\footnote{\href{http://goals.ipac.caltech.edu/}{http://goals.ipac.caltech.edu/}}. GOALS is drawn from the IRAS Revised Bright Galaxy sample \citep{sanders+03}. The galaxies in the GOALS sample represent galaxies with infrared luminosities above 10$^{11}$\,$L_\odot$. These infrared luminosities infer the presence of large quantities of dust and thus molecular gas.

Recent studies have been undertaken on merging systems and galaxy pairs, investigating the impact that the interactions have on star formation \citep{pan+18} and finding that with decreased separation, $f_{\rm mol}$, $\Sigma_{\rm mol}$ and star formation rate all increase, though star formation efficiency only increases at the closest separations. \citet{thorp+22} was able to expand upon this work using resolved mapping of star formation and molecular gas in galaxies pairs identified by the Mapping Nearby Galaxies at APO (MaNGA) survey \citep{bundy+15}, concluding that the enhanced star formation was due to a combination of enhanced gas fraction and higher star formation efficiency, implying that interactions have an impact on star formation along many different axes. While previous studies have focused on the molecular gas properties of samples of U/LIRGs \citep{downes+98,gao+solomon04,ueda+14,yamashita+17}, a large scale interferometic investigation of the GOALS galaxies, especially toward the lower end of the L$_{\rm IR}$ range has not been completed to-date. This study brings together CO imaging data available for 28 of the GOALS sample, including 15 objects with $L_{\rm IR} \lesssim 10^{11.5}$\,$L_\odot$, providing new insights into the behavior of star formation in a sample distinguished by its infrared luminosity, where much of the previous work has focused on the objects at the highest end of the $L_{\rm IR}$ spectrum, independent of merger configuration (or lack thereof).

The Combined Array for Research in Millimeter Astronomy (CARMA)\footnote{\href{http://www.mmarray.org}{http://www.mmarray.org}} is an interferometric array of 15 radio dishes (6$\times$10.4m and 9$\times$6.1m) located in the Eastern Sierras in California \citep{carma} that is well-suited for morphological studies of molecular gas in interacting galaxies. Given that these galaxies tend to contain large reservoirs of molecular gas, the combination of the high fidelity imaging of CARMA and the strong detections allows for us to trace the morphologies of the molecular gas in detail and compare these distributions with other properties of the interacting galaxies, including the star formation rate (SFR), the degree of morphological disruptions, and the stage along the interaction lifecycle.

We present and discuss molecular gas observations of GOALS galaxies taken with CARMA. In \S\ref{sec:sample}, we describe the selection to build the CARMA GOALS sample. In \S\ref{sec:obs}, we describe how the data were reduced, and how data products were created for each object. In \S\ref{sec:disc}, we derive the physical properties of the molecular gas, and discuss the implications of these observations. In \S\ref{sec:conclusions}, we present our conclusions. The cosmological parameters $H_0$\,=\,70\,km~s$^{-1}$, $\Omega_{\rm m}$\,=\,0.3 and $\Omega_\Lambda$\,=\,0.7 \citep{wmap} are used throughout.

%%%%%%%%%%%%%%% Table 2 %%%%%%%%%%%%%%%
\begin{table*}[t]
\centering
\caption{CARMA GOALS Measured Values}
\begin{tabular}{l c c c c c c c c c c}
\hline \hline
{\bf Name} & $v_{\rm sys}$ & {\bf Chan. width} & {\bf Vel. range} & $S_{\rm 100}$ & $\sigma_{\rm RMS}$ & $S_{\rm CO}$ & \multicolumn{2}{c}{\bf Area} & $L_{\rm CO}$ \\
& km~s$^{-1}$ & km~s$^{-1}$ & km~s$^{-1}$ & mJy & mJy & Jy~km~s$^{-1}$ & sq.arcsec & kpc$^2$& 10$^4$\,$L_\odot$ \\
(1) & (2) & (3) & (4) & (5) & (6) & (7) & (8) &(9) & (10)\\
\hline
MCG+12-02-001 & 4780 & 20.6 & 4559--4972 & $45.16\pm0.48$ & 6.61 & $501.8\pm0.8$ & 470.8 & 49.5 & $28.23\pm0.05$\\
CGCG\,436-030 & 9370 & 21.3 & 9222--9563 & $9.54\pm0.37$ & 4.60 & $91.6\pm0.4$ & 33.9 & 13.2 & $19.99\pm0.09$\\
III\,Zw\,35 & 8230 & 84.5 & 7993--8500 & $<13.35$ & 3.65 & $155.0\pm1.0$ & 145.6 & 44.2 & $26.05\pm0.17$\\
NGC\,695 & 9720 & 21.3 & 9564--9862 & $<3.18$ & 5.23 & $225.0\pm0.6$ & 499.8 & 209.5 & $52.94\pm0.14$\\
NGC\,958 & 5770 & 20.8 & 5451--6075 & $<0.89$ & 5.66 & $610.7\pm0.7$ & 1289.5 & 196.3 & $50.20\pm0.06$\\
UGC\,2369N & 9430 & 81.3 & 9071--9559 & $<16.24$ & 5.28 & $108.9\pm1.6$ & 132.6 & 52.4 & $24.08\pm0.35$\\
UGC\,2369S & 9450 & 81.3 & 9071--9559 & $-$ & -- & $206.2\pm1.6$ & 119.3 & 47.3 & $45.81\pm0.35$\\
UGC\,02608 & 7040 & 20.9 & 6725--7207 & $6.02\pm0.18$ & 4.53 & $253.5\pm0.6$ & 357.7 & 80.3 & $31.11\pm0.07$\\
IRAS\,03582+6012 & 8900 & 21.2 & 8634--9100 & $0.86\pm0.18$ & 3.61 & $99.8\pm0.4$ & 133.0 & 47.1 & $19.66\pm0.08$\\
NGC\,1614 & 4730 & 20.6 & 4456--4972 & $25.76\pm0.34$ & 5.52 & $603.1\pm0.6$ & 275.8 & 28.5 & $33.29\pm0.03$\\
CGCG\,468-002W & 5030 & 20.7 & 4766--5510 & $2.77\pm0.20$ & 2.88 & $60.2\pm0.4$ & 65.6 & 7.6 & $3.76\pm0.02$\\
CGCG\,468-002E & 5140 & 20.7 & 4869--5614 & $3.58\pm0.20$ & -- & $102.6\pm0.4$ & 68.0 & 8.3 & $6.69\pm0.02$\\
NGC\,2146 & 874 & 10.1 & 653--1136 & $<12.43$ & 35.1 & $4913.\pm3.1$ & 776.0 & 2.8 & $9.19\pm0.01$\\
NGC\,2623 & 5510 & 20.8 & 5275--5773 & $43.97\pm0.35$ & 4.41 & $295.3\pm0.5$ & 143.3 & 20.0 & $22.18\pm0.04$\\
Arp\,55 & 11960 & 32.4 & 11544--12160 & $5.47\pm0.25$ & 4.64 & $338.8\pm0.7$ & 422.5 & 263.4 & $121.19\pm0.26$\\
UGC\,05101 & 11780 & 21.6 & 11412--12147 & $24.35\pm0.41$ & 4.49 & $174.5\pm0.6$ & 49.9 & 30.2 & $60.52\pm0.21$\\
Arp\,148 & 10350 & 84.1 & 10014--10939 & $<185.1$ & 4.14 & $211.9\pm1.3$ & 123.9 & 58.6 & $56.56\pm0.34$\\
Arp\,299W & 3080 & 20.4 & 2827--3337 & $48.62\pm0.53$ & 7.56 & $587.2\pm0.9$ & 203.6 & 9.0 & $13.68\pm0.02$\\
Arp\,299E & 3170 & 20.4 & 2908--3419 & $17.99\pm0.53$ & -- & $346.8\pm0.9$ & 239.5 & 11.2 & $8.55\pm0.02$\\
NGC\,4418 & 2100 & 20.3 & 1878--2446 & $0.84\pm0.16$ & 3.25 & $141.1\pm0.4$ & 7.5 & 0.1 & $1.53\pm0.01$\\
NGC\,4922 & 7080 & 21.0 & 6797--7342 & $<1.42$ & 6.58 & $108.4\pm0.7$ & 38.2 & 8.7 & $13.48\pm0.09$\\
IC\,860 & 3880 & 20.5 & 3684--4136 & $30.86\pm0.57$ & 7.11 & $145.7\pm0.7$ & 35.3 & 2.5 & $5.39\pm0.03$\\
VV250W & 9420 & 21.3 & 9129--9725 & $3.90\pm0.31$ & 4.76 & $58.7\pm0.6$ & 118.4 & 46.7 & $12.96\pm0.13$\\
VV250E & 9400 & 21.3 & 9108--9704 & $6.63\pm0.31$ & -- & $93.4\pm0.6$ & 105.6 & 41.5 & $20.55\pm0.14$\\
NGC\,5256 & 8460 & 21.1 & 7916--8762 & $15.74\pm0.41$ & 5.74 & $230.9\pm0.8$ & 167.1 & 53.6 & $41.05\pm0.15$\\
CGCG\,142-034W & 5540 & 20.8 & 4881--5793 & $<1.03$ & 3.70 & $228.9\pm0.6$ & 113.6 & 16.0 & $17.37\pm0.04$\\
CGCG\,142-034E & 5470 & 20.7 & 4757--5709 & $-$ & -- & $93.3\pm0.6$ & 172.7 & 23.7 & $6.90\pm0.04$\\
NGC\,6670W & 8540 & 81.3 & 8033--8846 & $<21.07$ & 4.27 & $984.0\pm1.2$ & 105.1 & 34.3 & $178.37\pm0.22$\\
NGC\,6670E & 8720 & 81.3 & 8196--9008 & $-$ & -- & $248.6\pm1.3$ & 32.1 & 10.9 & $46.97\pm0.24$\\
NGC\,6786 & 7500 & 84.1 & 7244--7748 & $<6.28$ & 2.86 & $245.1\pm0.7$ & 422.8 & 107.3 & $34.16\pm0.10$\\
NGC\,6926 & 6040 & 20.8 & 5695--6361 & $<0.94$ & 5.42 & $364.8\pm0.7$ & 870.0 & 145.0 & $32.91\pm0.06$\\
II\,Zw\,96 & 10920 & 85.9 & 10662--11264 & $<23.10$ & 3.41 & $222.7\pm1.1$ & 153.0 & 80.2 & $66.26\pm0.33$\\
IC\,5298 & 8200 & 10.6 & 7980--8423 & $3.57\pm0.30$ & 3.64 & $95.9\pm0.3$ & 96.7 & 29.2 & $16.03\pm0.05$\\
NGC\,7674 & 8630 & 10.0 & 8507--8778 & $11.64\pm0.47$ & 5.96 & $252.3\pm0.4$ & 496.7 & 165.6 & $46.71\pm0.07$\\
 \\
\hline \hline
\end{tabular} \\
\label{tab:properties}
\raggedright{\footnotesize Column (1): Object name. 
Column (2): Systemic velocity of the CO emission in km~s$^{-1}$, defined as the center of the CO emission. 
Column (3): Channel width in km~s$^{-1}$. 
Column (4): Velocity range of detected CO emission in km~s$^{-1}$. 
Column (5): 100\,GHz continuum emission in mJy.
Column (6): Root mean square noise per channel in mJy~beam$^{-1}$.
Column (7): CO(1--0) flux in Jy~km~s$^{-1}$
Columns (8-9): The total area subtended by the CO, in celestial (sq.arcsec) and physical (kpc$^2$) units.
Column (10): The total CO luminosity in 10$^4$\,$L_\odot$.\\
}
\end{table*}

%%%%%%%%%%%%%%% Table 4 %%%%%%%%%%%%%%%
\begin{table*}[t]
\centering
\caption{CARMA GOALS \magphys\ Properties} \vspace{-1mm}
\begin{tabular}{l l c c c c c}
\hline \hline
Name & $\chi^2$ & log$_{10}$(${M}_*$) & log$_{10}$($M_{\mathrm{dust}}$) & log$_{10}$($L_{\mathrm{dust}}$) & log$_{10}$(SFR) & flag\_agn \\
(1) & (2) & (3) & (4) & (5) & (6) & (7) \\
\hline \hline
MCG+12-02-001 & 32.79 & $10.48^{+0.25}_{-0.23}$ & $7.8^{+0.0}_{-0.03}$ & $11.49^{+0.0}_{-0.02}$ & $1.47^{+0.04}_{-0.04}$ & 0 \\
CGCG\,436-030 & 61.08 & $10.4^{+0.0}_{-0.0}$ & $7.67^{+0.0}_{-0.0}$ & $11.66^{+0.0}_{-0.0}$ & $1.76^{+0.0}_{-0.04}$ & 1 \\
III\,Zw\,35 & 70.54 & $10.61^{+0.02}_{-0.0}$ & $7.79^{+0.0}_{-0.0}$ & $11.72^{+0.01}_{-0.0}$ & $1.82^{+0.01}_{-0.0}$ & 0 \\
NGC\,695 & 20.39 & $10.92^{+0.12}_{-0.07}$ & $8.26^{+0.03}_{-0.04}$ & $11.68^{+0.01}_{-0.02}$ & $1.59^{+0.03}_{-0.05}$ & 0 \\
NGC\,958 & 8.95 & $11.28^{+0.08}_{-0.1}$ & $8.3^{+0.03}_{-0.05}$ & $11.25^{+0.01}_{-0.01}$ & $1.06^{+0.04}_{-0.05}$ & 0 \\
UGC\,02369 & 28.55 & $11.11^{+0.1}_{-0.11}$ & $7.99^{+0.01}_{-0.04}$ & $11.57^{+0.02}_{-0.02}$ & $1.48^{+0.09}_{-0.06}$ & 0 \\
UGC\,02608 & 11.58 & $10.59^{+0.12}_{-0.06}$ & $7.87^{+0.02}_{-0.01}$ & $11.38^{+0.01}_{-0.04}$ & $1.28^{+0.03}_{-0.02}$ & 1 \\
IRAS\,03582+6012 & 36.5 & $9.85^{+0.09}_{-0.17}$ & $7.61^{+0.02}_{-0.04}$ & $11.32^{+0.08}_{-0.09}$ & $1.41^{+0.07}_{-0.04}$ & 1 \\
NGC\,1614 & 66.38 & $10.24^{+0.52}_{-0.0}$ & $7.62^{+0.03}_{-0.0}$ & $11.7^{+0.0}_{-0.05}$ & $1.77^{+0.02}_{-0.0}$ & 0 \\
CGCG\,468-002 & 35.84 & $10.57^{+0.0}_{-0.0}$ & $7.46^{+0.0}_{-0.0}$ & $11.15^{+0.0}_{-0.0}$ & $1.11^{+0.02}_{-0.02}$ & 1 \\
NGC\,2146 & 28.86 & $10.28^{+0.12}_{-0.14}$ & $7.27^{+0.02}_{-0.02}$ & $10.91^{+0.01}_{-0.01}$ & $0.81^{+0.06}_{-0.06}$ & 0 \\
NGC\,2623 & 90.42 & $10.52^{+0.0}_{-0.0}$ & $7.75^{+0.0}_{-0.0}$ & $11.57^{+0.0}_{-0.0}$ & $1.7^{+0.0}_{-0.21}$ & 1 \\
Arp\,55 & 47.62 & $10.83^{+0.07}_{-0.04}$ & $8.3^{+0.0}_{-0.0}$ & $11.64^{+0.0}_{-0.01}$ & $1.51^{+0.01}_{-0.01}$ & 0 \\
UGC\,05101 & 25.87 & $10.92^{+0.0}_{-0.02}$ & $8.51^{+0.0}_{-0.02}$ & $11.95^{+0.0}_{-0.02}$ & $1.89^{+0.04}_{-0.01}$ & 1 \\
Arp\,148 & 52.75 & $10.42^{+0.04}_{-0.06}$ & $8.03^{+0.0}_{-0.01}$ & $11.56^{+0.0}_{-0.0}$ & $1.47^{+0.05}_{-0.01}$ & 0 \\
Arp\,299 & 51.94 & $10.77^{+0.0}_{-0.0}$ & $7.75^{+0.0}_{-0.0}$ & $11.83^{+0.01}_{-0.0}$ & $1.89^{+0.0}_{-0.0}$ & 1 \\
NGC\,4418 & 26.7 & $9.68^{+0.26}_{-0.24}$ & $6.91^{+0.1}_{-0.02}$ & $11.0^{+0.01}_{-0.07}$ & $1.07^{+0.05}_{-0.06}$ & 0 \\
NGC\,4922N & 48.74 & $10.41^{+0.0}_{-0.0}$ & $7.62^{+0.0}_{-0.0}$ & $11.28^{+0.0}_{-0.0}$ & $1.21^{+0.08}_{-0.01}$ & 1 \\
IC\,860 & 104.98 & $10.47^{+0.0}_{-0.0}$ & $7.21^{+0.0}_{-0.0}$ & $11.03^{+0.0}_{-0.01}$ & $1.07^{+0.0}_{-0.01}$ & 0 \\
VV\,250 & 40.91 & $10.33^{+0.26}_{-0.06}$ & $7.82^{+0.0}_{-0.04}$ & $11.66^{+0.03}_{-0.01}$ & $1.67^{+0.08}_{-0.03}$ & 0 \\
NGC\,5256 & 22.52 & $10.87^{+0.09}_{-0.07}$ & $7.84^{+0.01}_{-0.0}$ & $11.46^{+0.0}_{-0.01}$ & $1.37^{+0.02}_{-0.02}$ & 0 \\
CGCG\,142-034 & 9.96 & $10.79^{+0.08}_{-0.12}$ & $7.71^{+0.05}_{-0.01}$ & $11.06^{+0.02}_{-0.01}$ & $0.81^{+0.07}_{-0.07}$ & 1 \\
NGC\,6670 & 11.87 & $10.88^{+0.12}_{-0.13}$ & $8.12^{+0.09}_{-0.05}$ & $11.59^{+0.01}_{-0.03}$ & $1.47^{+0.04}_{-0.04}$ & 0 \\
NGC\,6786 & 6.36 & $10.6^{+0.13}_{-0.12}$ & $7.76^{+0.02}_{-0.0}$ & $11.17^{+0.01}_{-0.01}$ & $1.15^{+0.03}_{-0.04}$ & 0 \\
NGC\,6926 & 11.29 & $11.02^{+0.05}_{-0.12}$ & $8.05^{+0.04}_{-0.03}$ & $11.25^{+0.01}_{-0.01}$ & $1.17^{+0.06}_{-0.05}$ & 1 \\
II\,Zw\,96 & 37.4 & $10.22^{+0.21}_{-0.3}$ & $7.88^{+0.01}_{-0.09}$ & $11.87^{+0.0}_{-0.03}$ & $1.94^{+0.03}_{-0.02}$ & 0 \\
IC\,5298 & 83.5 & $11.15^{+0.0}_{-0.0}$ & $7.84^{+0.0}_{-0.0}$ & $11.44^{+0.0}_{-0.0}$ & $1.44^{+0.01}_{-0.0}$ & 1 \\
NGC\,7674 & 17.82 & $11.01^{+0.04}_{-0.0}$ & $8.04^{+0.0}_{-0.04}$ & $11.44^{+0.0}_{-0.02}$ & $1.48^{+0.03}_{-0.02}$ & 1 \\ \\
\hline \hline
\end{tabular} \\
\label{tab:magphys}
\raggedright{\footnotesize Column (1): Object name. 
Column (2): Best fit $\chi^2$. 
Column (3): Stellar mass.
Column (4): Dust mass.
Column (5): Dust luminosity.
Column (6): SFR averaged over the last $10^7$ years.
Uncertainties are indicated by the median 84th--16th percentile range from each individual parameter PDF. 
Column (7): AGN flag.
\\
}
\end{table*}

%%%%%%%%%%  Figure 1  %%%%%%%%%%
\begin{figure*}[t!]
\centering
\includegraphics[width=\textwidth]{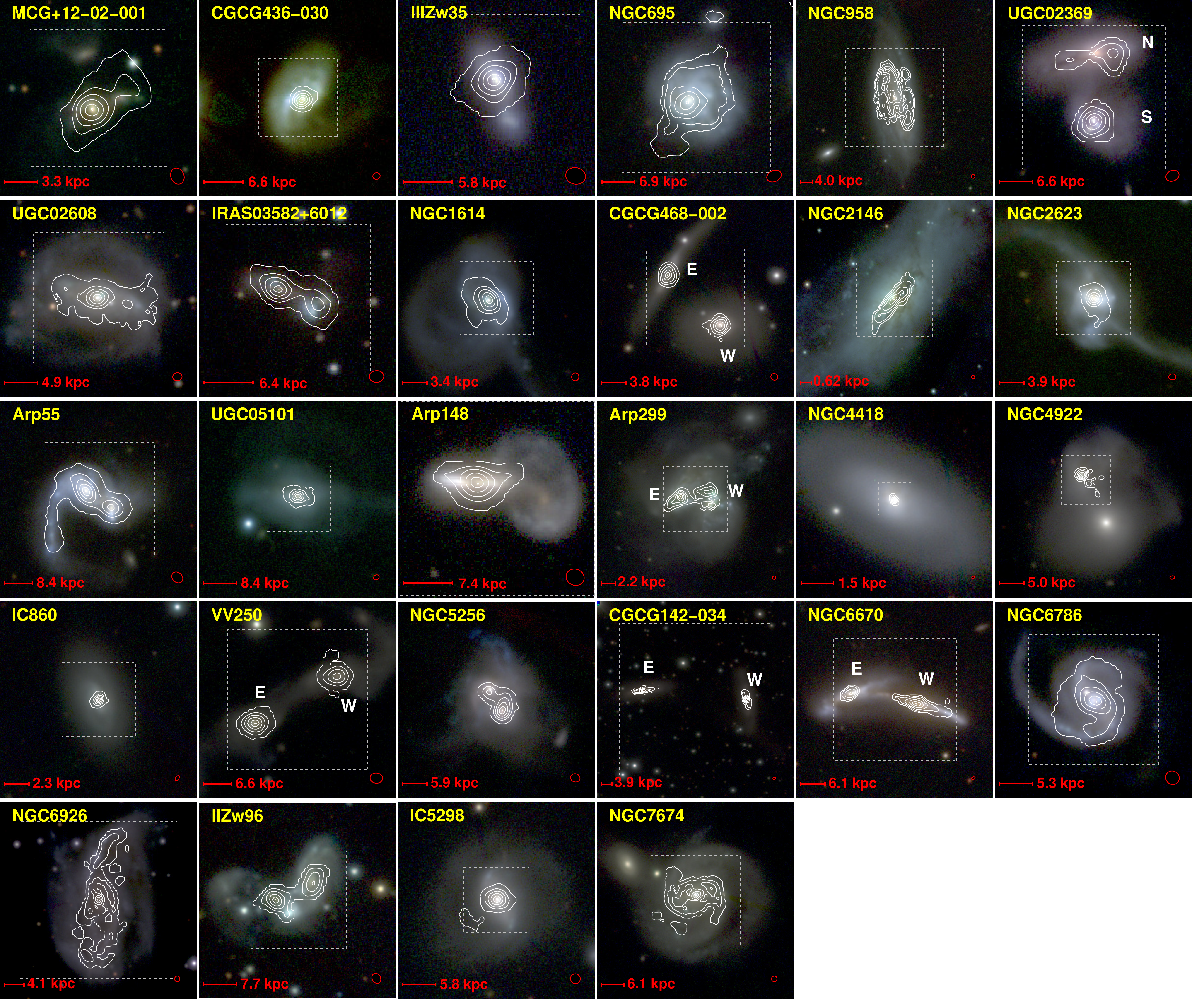} \vskip -1.75mm
\caption{The CO(1--0) integrated intensity (moment0) maps from CARMA (white contours) are overlaid on the 3-color {\em g\,r\,i} images from PanSTARRS \citep{panstarrs} for the 28 CARMA GOALS galaxies, ordered based on their RA. For each map, north is up and east is to the left. The red bar in the lower lefthand corner demarcates 10$''$ in each field (with the corresponding physical scale listed). The beam is shown in the lower right corner in red. The gray dotted borders represent the size of the CO-only moments from Figure~\ref{fig:mom0} and \ref{fig:mom1}.}
\label{fig:co+opt}
\end{figure*}

%%%%%%%%%%  Figure 2  %%%%%%%%%%
\begin{figure*}[t!]
\centering
\includegraphics[width=\textwidth]{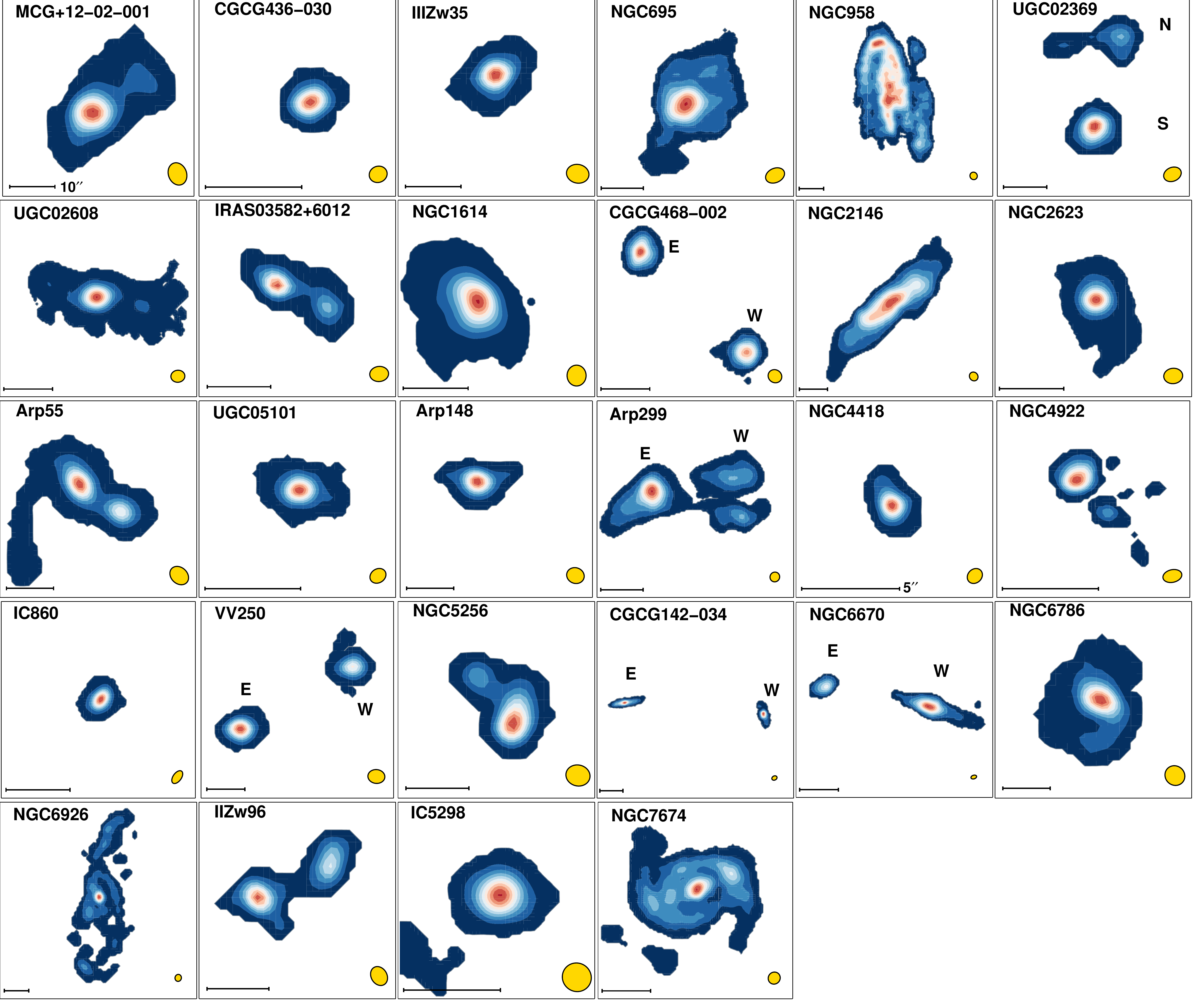} \vskip -1.75mm
\caption{The CO(1--0) integrated intensity (moment0) maps of the 28 CARMA GOALS galaxies. The black bar in the lower lefthand corner demarcates 10$''$ in each field. The beam is shown in the lower right corner in yellow, confirming that all galaxies are well-sampled.}
\label{fig:mom0}
\end{figure*}

%%%%%%%%%%  Figure 3  %%%%%%%%%%
\begin{figure*}[t!]
\centering
\includegraphics[width=\textwidth]{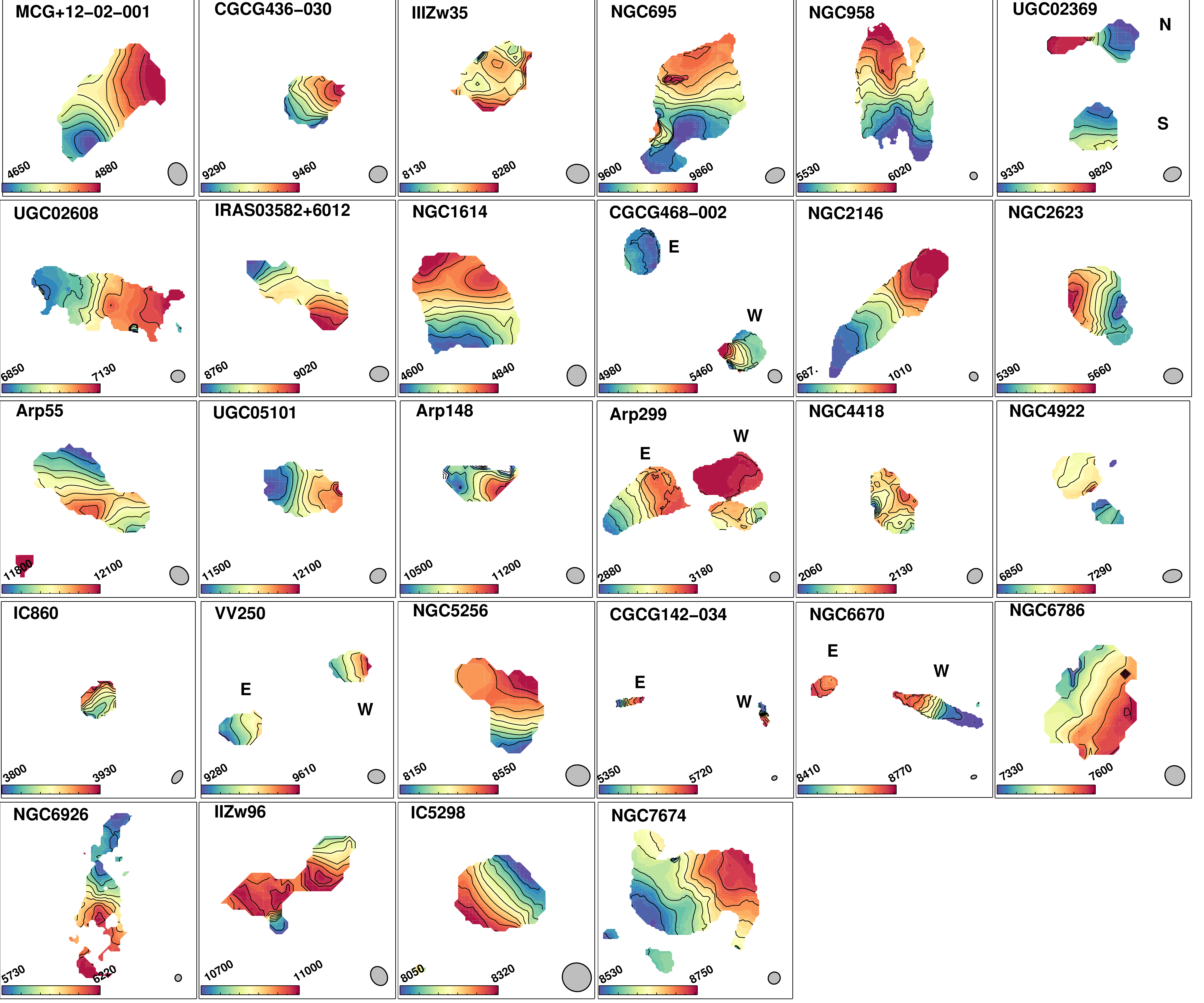} \vskip -1.75mm
\caption{The mean velocity (moment1) maps of CO(1--0) of the 28 CARMA GOALS galaxies. The size scales are identical to those used in Figure~\ref{fig:mom0}.}
\label{fig:mom1}
\end{figure*}

%%%%%%%%%%  Figure 4  %%%%%%%%%%
\begin{figure*}[t!]
\includegraphics[width=\textwidth]{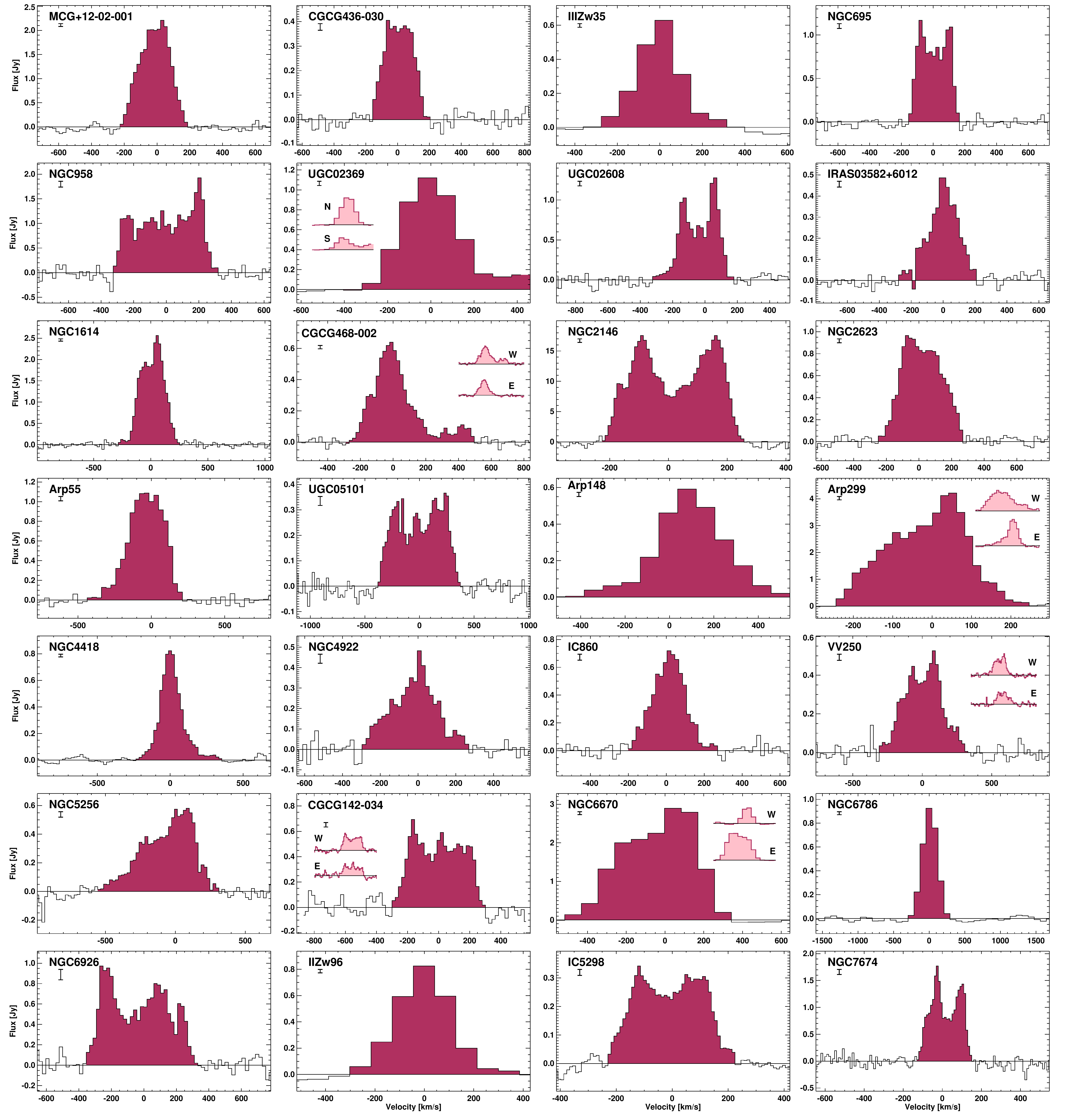} \vskip -1.75mm
\caption{The integrated CO(1--0) spectra for the 28 CARMA GOALS galaxies. Each spectrum was created by using the moment0 map (shown in Fig.\,\ref{fig:mom0}) as a clip mask and totaling all flux within the mask in each channel. The areas shaded in fuschia denote the channels used to calculate the total flux. The RMS noise per channel is shown in the upper left corner below the galaxy name. In all cases, the CO(1--0) is detected to be very high signal to noise. The six CARMA objects (UGC\,02369, CGCG\,468-002, Arp\,299, VV\,250, CGCG\,142-034, and NGC\,6670), where two different galaxies were differentiable in our data have had their spectra separated and plotted in the corner of the panel.}
\label{fig:specs}
\end{figure*}

%%%%%%%%%%  Figure 5  %%%%%%%%%%
\begin{figure}[t!]
\includegraphics[width=0.49\textwidth]{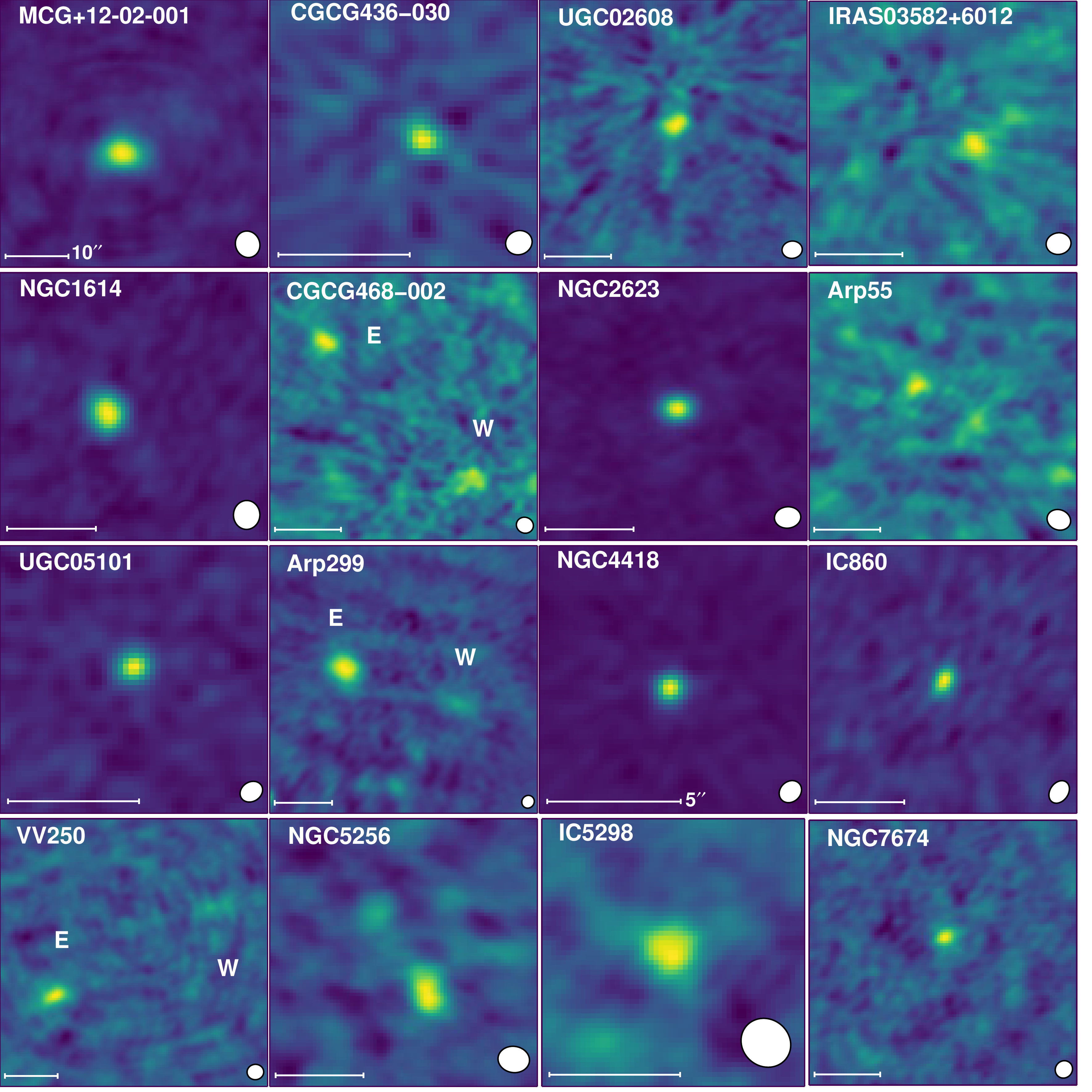} \vskip -1.75mm
\caption{The 100\,GHz continuum images of the 16 continuum-detected CARMA GOALS objects. The field of view is equivalent to what was used in Figures~\ref{fig:mom0} and \ref{fig:mom1}. In most cases, the continuum is unresolved by CARMA.}
\label{fig:cont}
\end{figure}

%%%%%%%%%%  Figure 6  %%%%%%%%%%
\begin{figure*}[t!]
\centering
\subfigure{\includegraphics[width=0.49\textwidth]{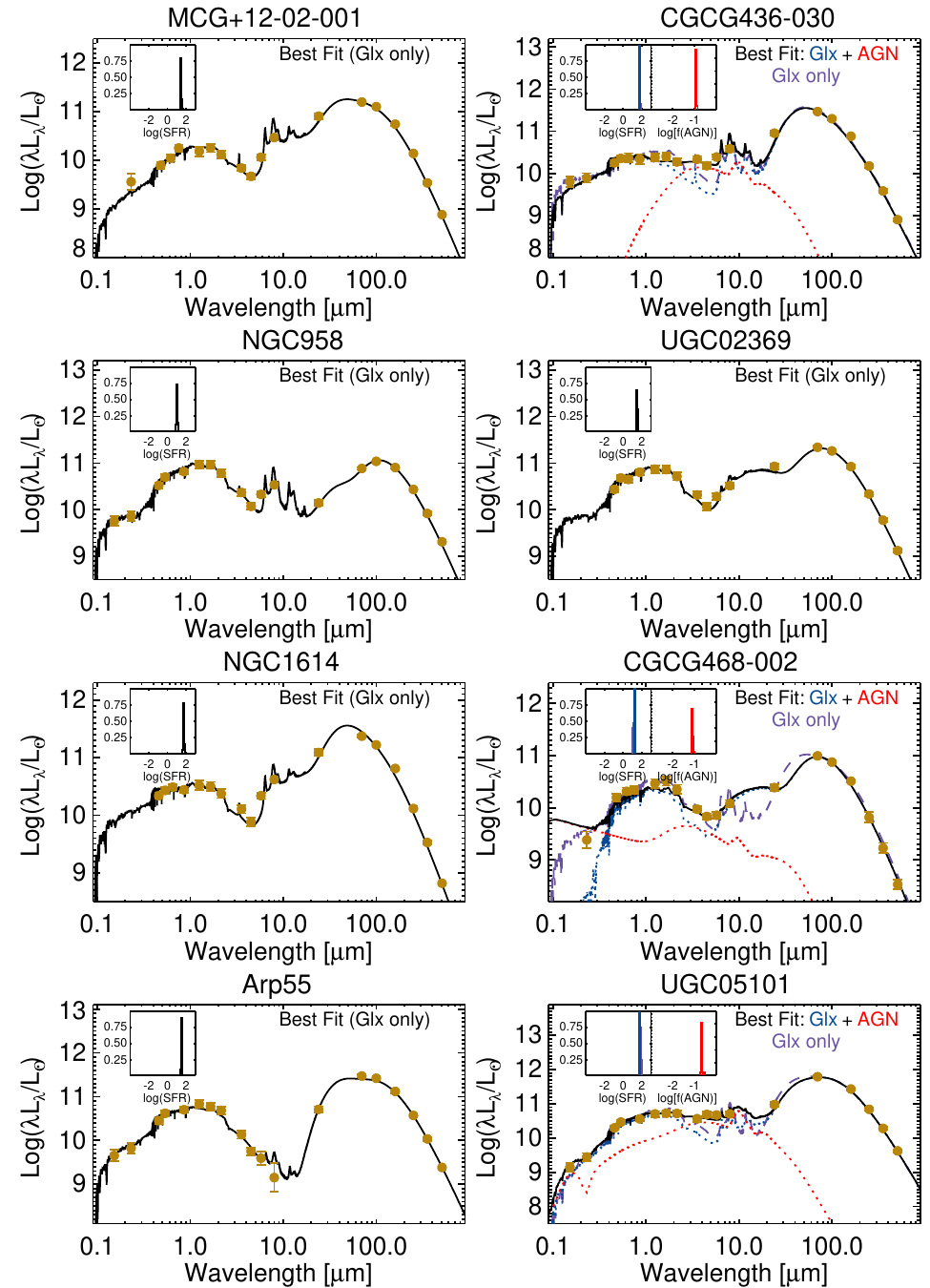}}
\subfigure{\includegraphics[width=0.49\textwidth]{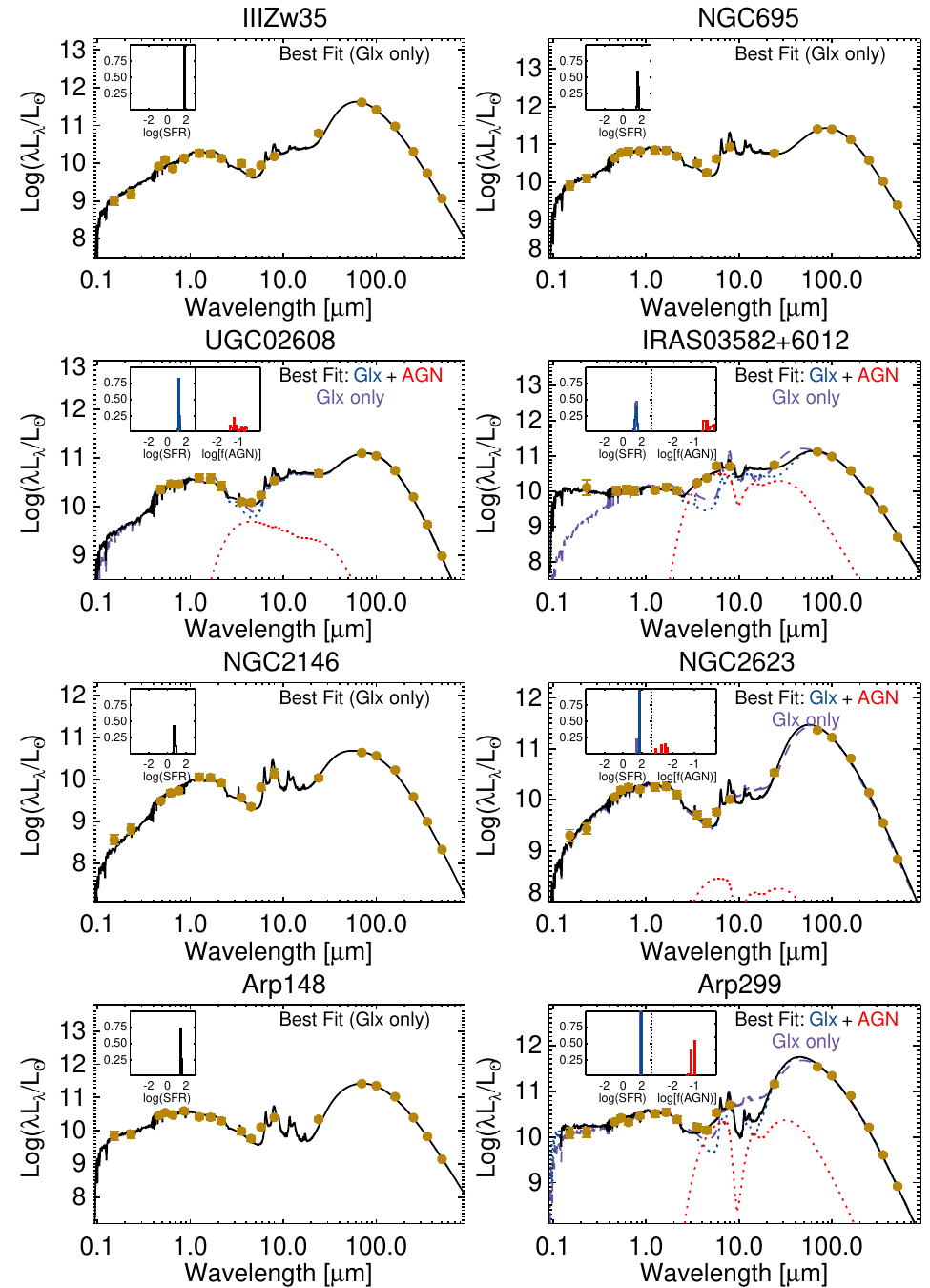}} \vskip -4mm
\subfigure{\includegraphics[width=0.49\textwidth,clip,trim=0cm 5.6cm 0cm 0cm]{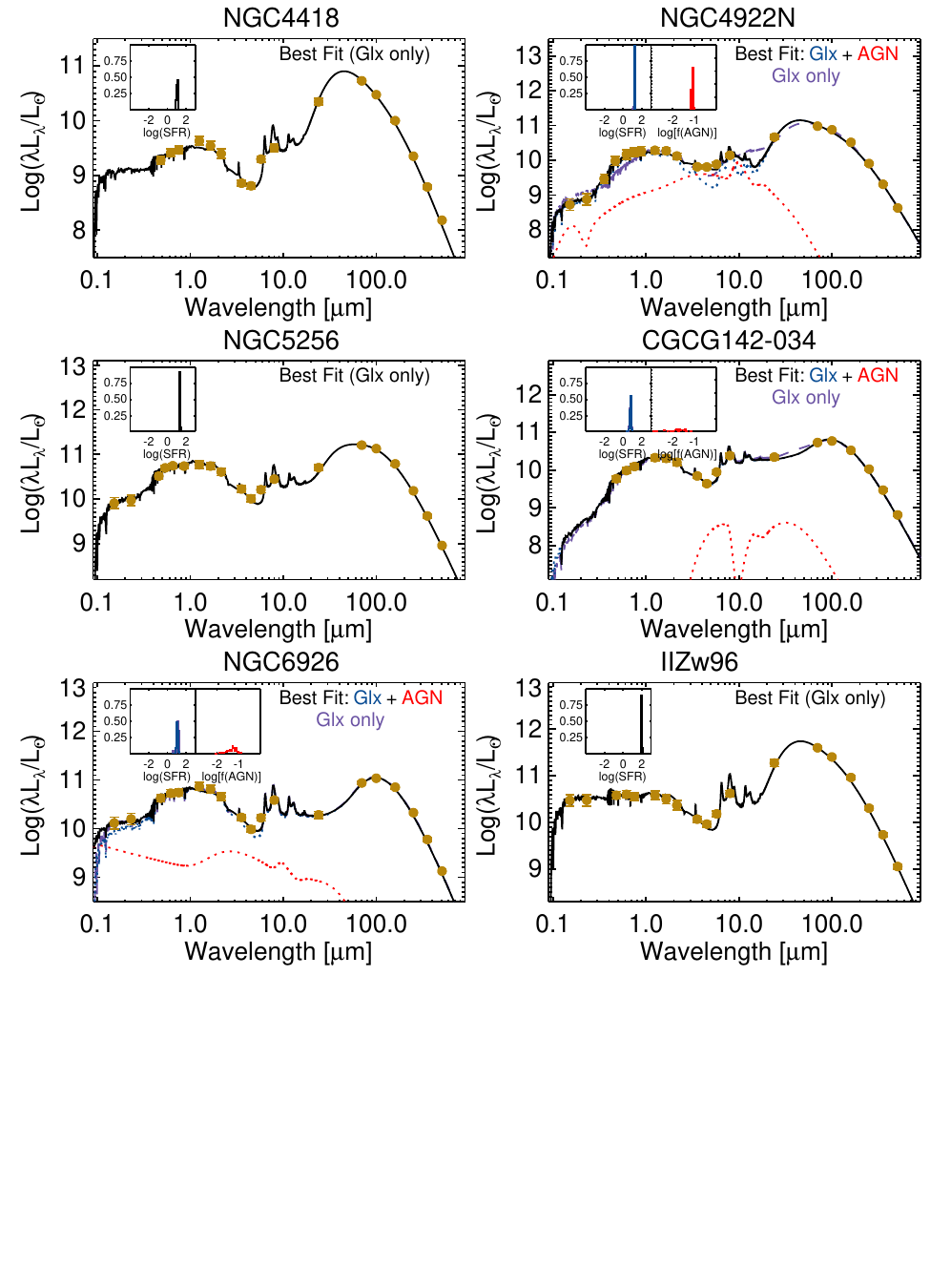}}
\subfigure{\includegraphics[width=0.49\textwidth,clip,trim=0cm 5.6cm 0cm 0cm]{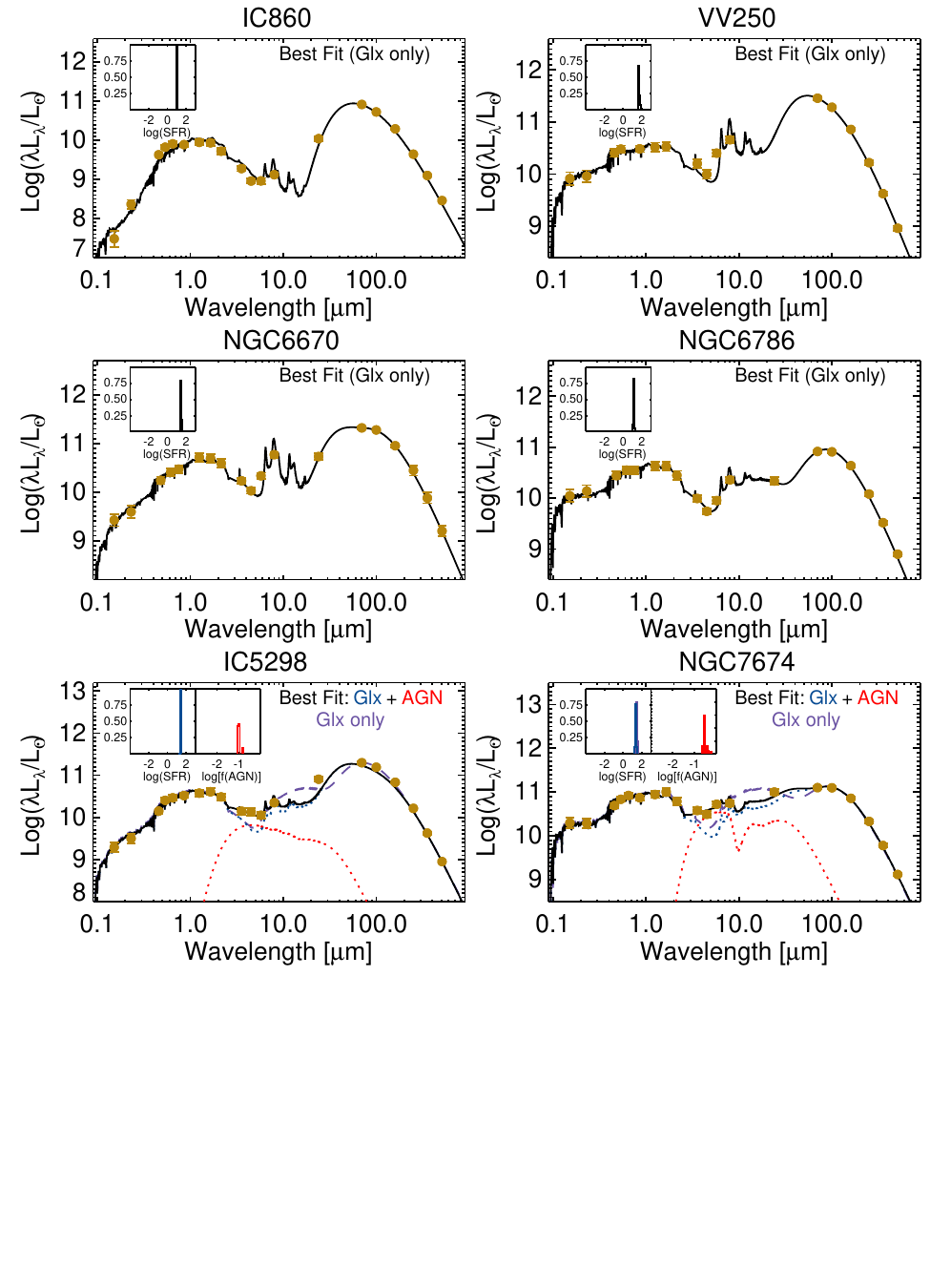}} \vskip -1.75mm
\caption{Best-fitting models to the SEDs. The black line shows the best overall model. In some cases, the fit was improved by including a torus component from the AGN (shown in red) in addition to the galactic emission (shown in blue). The fit achieved with just galactic emission in those cases is shown in purple. See Section \ref{sec:sed} for discussion on the fitting process. Probability density functions (PDFs) for SFR in the last 10 Myr and, as relevant, fraction of infrared luminosity arising from the AGN component are shown as insets. When a galaxy is modeled both with and without an AGN component, both of the resulting SFR PDFs are shown. These generally do not show much difference, indicating that the inclusion of an AGN component does not affect our derived SFR significantly.}
\label{fig:seds}
\end{figure*}

\section{The Sample}
\label{sec:sample}
The sample presented in this work was drawn by cross-referencing the GOALS \citep{goals} sample with objects available within the CARMA archive.\footnote{\href{http://carma-server.ncsa.uiuc.edu:8181/}{http://carma-server.ncsa.uiuc.edu:8181/}} To derive the sample presented in this paper, we ran a $\approx$1\arcmin\ proximity match for the 202 objects in GOALS \citep{goals}, searching for all datasets that contained observations for the CO(1--0) line and requiring complete velocity coverage of the line.
%To derive the sample presented in this paper, we ran a $\approx$1\arcmin\ proximity match for the sample of 202 objects that make up the GOALS sample \citep{goals}, searching for all datasets that contained observations for the CO(1--0) line on GOALS objects, requiring complete velocity coverage of the line. 
This overall cross-matching resulted in 31 objects. We report on 28 objects, excluding Arp\,220 and NGC\,6240, which are described in detail in \citet{manohar+17}, and Markarian\,231, which is described in detail in \citet{alatalo15}. Table~\ref{tab:params} lists the names and characteristics of the cross-matched GOALS galaxies, as well as the CARMA observing parameters, detailed below. The objects in the CARMA GOALS sample span the range of IR luminosities defined for LIRGs and ULIRGs, from 10$^{11.1-12.5}$~$L_\odot$. Figure~\ref{fig:co+opt} shows the PanSTARRS {\em g\,r\,i} data \citep{panstarrs} of the sources represent a large swath of morphologies, from clear interactions (like II\,Zw\,96) to semi-settled and quiescent objects (like NGC\,4418).

\section{Observations and Analysis}
\label{sec:obs}
\subsection{CARMA Observations}
CARMA observations for GOALS sources with CO(1--0) data were downloaded from the archive. Observations were taken between July 2008 and June 2014, in a large set of array configurations and correlator configurations. The field of view of these observations corresponds to the primary beam of the 6m antennas at 3mm of $\approx$\,100\arcsec. Data reduction was done in identical fashion to \citet{alatalo+13}, utilizing the Multichannel Image Reconstruction, Interactive Analysis and Display software ({\sc miriad}; \citealt{miriad}). Table~\ref{tab:params} presents the semester of observations, synthesized beam, and gain calibrators used for each CARMA GOALS object.

The available correlator configurations varied significantly from semester to semester over the 6 years that these observations were taken. Observations taken in early semesters (e.g., III\,Zw\,35, Arp\,148) often required use of a broadband low-resolution configuration to capture the entire footprint of the CO(1--0) line. In these cases, the channel width is often 80\,km~s$^{-1}$. Despite this, all early observed galaxies have CO lines that have been resolved, with at least 7 channels across them. In later semesters, the observations were taken using the upgraded correlator, which simultaneously provided large bandwidth and high resolution and provided a marked improvement in resolution across the line.

\subsubsection{Calibration and Imaging}
\label{sec:calibration}
The resulting moment maps were constructed in an identical fashion to \citet{alatalo+13}\footnote{Full description available in \S 3.2}.  Figures~\ref{fig:co+opt} and \ref{fig:mom0} showcase the CARMA integrated intensity (moment0) data for each individual GOALS source, overlaid on PanSTARRS {\em g\,r\,i} data \citep{panstarrs}, and on its own, respectively. Figure~\ref{fig:mom1} presents the mean velocity (moment1) maps of the GOALS galaxies.

The integrated spectra are constructed by using the moment0 map to create a clip-mask and integrating the flux within the moment0-defined (unmasked) aperture.  This was done separately for each object. Clearly differentiable doubles, defined as those objects that could be unambiguously separated into two distinct forms in their moment0 maps, combined with confirmation of two obvious sources in the optical data, were also identified. These doubles include UGC\,02369, CGCG\,468-002, Arp\,299, VV\,250, CGCG\,142-034, and NGC\,6670. In these cases, the moment0 map clipmask was separated to distinguish between the gas distributions of each object. Figure~\ref{fig:specs} shows the CO(1--0) spectra integrated in this way, with the channels used to calculate the integrated line flux for each galaxy shaded in violet. The six objects deemed to be doubles we also show spectra of each of the individual objects as a separate spectrum in the panel. 

The root mean square (RMS) per channel is calculated by (1) taking the standard deviation of all pixels within the cube that were outside the moment0 aperture, (2) applying an additional noise up-correction of 30\% to account for the oversampling of the maps (see: \citealt{a15_co13} for details), and (3) multiplying by the square root of the total number of beams represented in the moment0 aperture.

To calculate the integrated line flux for each galaxy, we summed the shaded channels in Fig.~\ref{fig:specs} and multiplied by the velocity width of each channel.  The line flux RMS was calculated by multiplying the RMS per channel by the channel velocity width and the square root of the total number of channels containing line emission. Table~\ref{tab:properties} presents the line flux, total CO(1--0) area subtended, and other derived molecular gas properties of the CARMA GOALS sample. It is important to note that the errors quoted also do not include the 20\% flux calibration uncertainties associated with CARMA observations.\footnote{https://www.mmarray.org/memos/carma\_memo59.pdf}

\subsubsection{Continuum imaging}
We isolated the line-free channels using the {\sc miriad} task {\tt uvlin}\footnote{\href{http://www.atnf.csiro.au/computing/software/miriad/doc/uvlin.html}{http://www.atnf.csiro.au/computing/software/miriad/doc/uvlin.html}}. We then inverted the line-free visibilities using the multifrequency synthesis options with {\sc miriad}, creating an integrated image to search for 100\,GHz continuum emission present in each of the CARMA GOALS galaxies. Of the 28 galaxies that are part of this sample, 16 contained detectable 100\,GHz continuum emission. Figure~\ref{fig:cont} shows the continuum maps of these objects.

To calculate the total 100\,GHz flux for each of the detected objects, we pinpointed the location of the peak in the image using the IDL routine {\tt find\_galaxy}.\footnote{\href{http://www-astro.physics.ox.ac.uk/~mxc/software/}{http://www-astro.physics.ox.ac.uk/$\sim$mxc/software/}} We then created a box centered on the peak emission with twice the major axis of the beam on a side and summed all emission inside this box. The detected 100\,GHz continuum emission in these sources is all are centrally concentrated, and therefore it is unlikely that significant extended emission is being missed. To calculate the RMS noise, we used this box to mask emission and calculated the standard deviation of the unmasked pixels with IDL routine {\tt robust\_sigma}.\footnote{\href{https://idlastro.gsfc.nasa.gov/ftp/pro/robust/robust_sigma.pro}{https://idlastro.gsfc.nasa.gov/ftp/pro/robust/robust\_sigma.pro}} In the cases of 100\,GHz non-detected sources, we calculated the 3$\sigma$ upper limits in an equivalent area (twice the size of the beam major axis) to the detected sources. Table~\ref{tab:properties} lists the 100\,GHz fluxes, upper limits, and RMS noise.

%\subsection{Determination of Star Formation Rate}
%\label{sec:SFRdet}

\subsection{Multi-wavelength Photometry}
All of the galaxies in our sample were observed with the {\em Herschel} Space Observatory \citet{herschel}. These photometric measurements were published in  \citet{chu+17}. Over half of our sample was included in the SED analysis of \citet{U+12}. For these galaxies, we used the {\em Galaxy Evolution Explorer} ({\em GALEX}; \citealt{galex}), optical (UBVRI), the 2 Micron All Sky Survey (2MASS; \citealt{2mass}), and {\em Spitzer} \citep{spitzer} photometry from \citet{U+12}. We applied foreground extinction corrections to the UV and optical bands, as described in the next two sections, using the Galactic extinction values from \citet{schlafly+11} provided in the NASA Extragalactic Database (NED). For 11 the remaining CARMA GOALS sources detailed below, we determined the apertures for photometry based on a visual inspection of the PanSTARRs optical imaging, then applied extinction corrections. We elected to use SDSS for NGC\,4922N due to anomalies in the PanSTARRs data. Tables~\ref{tab:photData}, \ref{tab:mirData} and \ref{tab:Ecorr} present these data.

In the UV, optical, near- and mid-infrared we convolve the catalogue error in quadrature with a calibration error of 20, 10, 15 and 10\% of the flux respectively, to allow for differences in the methods used to measure total photometry and errors in the spectral synthesis models used to fit the underlying stellar populations, following \citet{rowlands+12}. Due to the additional uncertainties in the aperture photometry of the highly blended source NGC4922N we convolve the catalogue error with a calibration error of 20\% in the optical.

\subsubsection{GALEX Photometry}
\label{sec:GALEX}

Of the total 23 galaxies in our sample observed with {\em GALEX}, eight were not included in the set examined by \citet{howell+10, U+12}. The observations for these galaxies were retrieved from the Mikulski Achive for Space Telescopes using GalexView version 1.4.11. To calculate fluxes from the background-subtracted count rate, we used the conversions provided by \citet{goddard+04}\footnote{\href{https://asd.gsfc.nasa.gov/archive/galex/FAQ/counts_background.html}{https://asd.gsfc.nasa.gov/archive/galex/FAQ/counts\_background.html}}: 
\begin{eqnarray*}
    f_{NUV} &=& 2.06 \times 10^{-16} \times ct_{\rm NUV} {\rm ~~~and} \\
    f_{FUV} &=& 1.40 \times 10^{-15} \times ct_{\rm FUV}, 
\end{eqnarray*}
where $ct_{\rm NUV}$ and $ct_{\rm FUV}$ are background corrected count rates and the fluxes are given in erg\,s$^{-1}$\,cm$^{-2}$\,\AA$^{-1}$. The background rate is estimated  in a source-free region near each galaxy. We also masked out foreground and background objects within the source aperture. The observed UV fluxes are corrected for foreground Galactic extinction using the relations given by \citet{wyder+05}, where E(B--V) is calculated from the \citet{schlafly+11} A$_{V}$ assuming an R$_{V}=3.1$ reddening law \citep{fitzpatrick99}:
\begin{eqnarray*}
    A_{NUV} &=& 8.376 \times {\rm E(B-V)} {\rm ~and~} \\
    A_{FUV} &=& 8.741 \times {\rm E(B-V)}.
\end{eqnarray*}
Photometric uncertainties consist of the 10\% calibration uncertainty \citep{goddard+04} combined in quadrature with Poisson uncertainty.

\subsubsection{PanSTARRS photometry}
Calibrated PanSTARRS cutout images in the $g$, $r$, and $i$ bands were downloaded from the archive\footnote{\href{http://ps1images.stsci.edu/cgi-bin/ps1cutouts}{http://ps1images.stsci.edu/cgi-bin/ps1cutouts}}.  Per PanSTARRS documentation\footnote{\href{https://outerspace.stsci.edu/display/PANSTARRS/}{https://outerspace.stsci.edu/display/PANSTARRS/}}, the CD matrices in the FITS headers of these files were first updated with the IDL routine \verb|FITS_CD_FIX|. Given the photometric zeropoint of 25 + 2.5 $\log(t_{\rm exp})$ (D. Jones, private communication) and standard $f_0 = 3631$ Jy \citep{oke+gunn83}, we obtained the fluxes in Jy via the following: 

\begin{eqnarray*}
    m_{g,r,i} &=& -2.5 \log(ct_{g,r,i}) + 2.5 \log(t_{\rm exp}) + 25, \\ % {\rm and} \\
    f_{g,r,i} &=& 3631 \times 10^{\frac{m_{g,r,i}}{-2.5}},
\end{eqnarray*}
where $t_{\rm exp}$ is the exposure time in seconds, while $m_{g,r,i}$, $ct_{g,r,i}$, and $f_{g,r,i}$ are the magnitude, count, and flux in Jansky for the $g$, $r$, and $i$ filters, respectively. A foreground extinction was subsequently applied in the same manner as was done for the {\em GALEX} photometry but directly using the extinction magnitudes from \citet{schlafly+11} provided by NED. Photometric errors in these images dominated the uncertainty and were conservatively estimated at the 10\% level.  

Since the optical images from PanSTARRS have sufficient spatial resolution to resolve the large-scale tidal features of each merger system, we employed mask photometry to properly measure the total flux within these optical bands. Specifically, masks have been generated based on isophotes of 24.5 mag in the PanSTARRS $i$-band images for consistent comparisons to the global optical fluxes from previous GOALS studies~\citep{howell+10,vavilkin11,U+12}. Within the wavelength regime covered by the PanSTARRS filters, the masks are much larger than the molecular gas footprint and hence aperture corrections are not needed in these bands. The mask apertures generally cover similar regions to those used for the other bands, yielding photometry consistent within the errors with that derived using the elliptical aperture from the other bands.  

\subsubsection{2MASS Photometry}
All of our galaxies were observed as part of 2MASS. We retrieved mosaics for the 11 galaxies not included in \citet{U+12} from the NASA/IPAC Infrared Science Archive (IRSA). Using the magnitude zeropoints given in the header of each image, we converted the background subtracted counts, measured in the same aperture as the {\em GALEX} fluxes, to magnitudes and then to fluxes using the conversions of \citet{cohen+03}. Background estimates were measured using source-free regions near each galaxy, and photometric uncertainties are the sum in quadrature of Poisson uncertainty, uncertainty due to uncertainty in the flux conversion factor, and a 3\% calibration uncertainty \citep{cutri+06}.

\subsubsection{Spitzer Photometry}

In the MIR, we used photometry from the Infrared Array Camera (IRAC; \citealt{spitzerIRAC}) and the 24$\mu$m band of the Multiband Imaging Photometer (MIPS; \citealt{spitzerMIPS}) on {\em Spitzer}. Mosaics generated via the automatic pipeline were retrieved from the {\em Spitzer} Heritage Archive. Using the same aperture as the {\em GALEX} and 2MASS photometry, we calculated background-subtracted fluxes. We used the extended source aperture corrections given in the instrument handbooks\footnote{\href{http://irsa.ipac.caltech.edu/data/SPITZER/docs/irac/iracinstrumenthandbook/29/}{http://irsa.ipac.caltech.edu/data/SPITZER/docs/irac/ \\ iracinstrumenthandbook/29/} and \href{https://irsa.ipac.caltech.edu/data/SPITZER/docs/mips/mipsinstrumenthandbook/50/}{https://irsa.ipac.caltech.edu/data/SPITZER/ \\ docs/mips/mipsinstrumenthandbook/50/}}, using the effective radius of the elliptical aperture given by the geometrical mean of its axes. Photometric uncertainties consist of the sum in quadrature of the calibration uncertainty (3\% for IRAC, \citealt{cohen+03}; 4\% for MIPS, \citealt{engelbracht+07}) and the error measured on the uncertainty mosaics.

\subsection{SED fitting}
\label{sec:sed}
Because at least 11/28 of the CARMA GOALS sources contain AGNs, we used two, related SED fitting codes to extract galactic properties from their broadband photometry: \magphys\ \citep{magphys}, which is an SED fitting code that balances energy, and {\sc SED3FIT} \citep{berta+13}, a modified version of \magphys\ that also includes AGN templates. %We fit our SEDs with both to test the necessity of including an AGN component to fit the observed emission. 
The sections below summarize the two fitting programs.

The decision whether to include an AGN component in the SED fitting was based on the AGN diagnostics in \citet{petric+11}, which determined the AGN contribution using mid-IR emission. If this fraction was above 0, we identified the system as an AGN composite. We also double-checked these determinations in the $\chi^2$ from fitting with \magphys\ only vs. SED3FIT. Additionally, an AGN component was included for NGC\,4922 which a Seyfert 2 galaxy \citep{saade+22} and because the SED fit is significantly improved with an AGN component.

%To determine which fit is best, we first look at the $\chi^2$ statistic of both. When the fit with the AGN has a smaller statistic, we check that the AGN torus model is applicable as part of the overall model\footnote{We did not accept two slightly statistically better fits with unobscured Type I AGN models for CGCG468-002E and VV250.}. If the fits are statistically similar, we checked the literature for other evidence of AGN activity, such as detection of the 3mm continuum (IC\,5298) or the presence of a megamaser (III\,Zw\,35). In the absence of such evidence or improvement with the inclusion of an AGN component, we use the \magphys\ fit.

\subsubsection{MAGPHYS}
\label{sec:magphys}
%{\color{red}Describe MAGPHYS and SED3FIT in brief. Note that the fit was done with both for each. Comment on selection of one fit vs. the other. Maybe we want to comment on the SFR excess plot?}
\magphys\footnote{The \citet*{magphys} models are publicly available as a
user-friendly model package {\sc magphys} at www.iap.fr/magphys/.}(DCE08) uses a physically motivated approach, balancing the UV-optical radiation from stars absorbed by dust and its re-radiation in the FIR. These two wavelength regimes are covered by separate model libraries combined via this energy balance. The stellar UV - NIR library contains 50,000 spectral models generated via the population synthesis code of \citet{bruzual+charlot03} assuming an exponentially declining star formation history (SFH) onto which random bursts are superimposed (creating stochastic SFHs). These 50,000 spectra cover a wide range of parameters including number and timing of SF bursts, dust mass (and attenuation), metallicity, and stellar mass. For each SFH, the average SFR over the last 10 Myr and 100 Myr are calculated. Attenuation by dust on these stellar models is implemented using the two-component model of \citet{charlot+fall00}, which uses a larger attenuation for stars in birth clouds relative to attenuation by the ambient ISM for older stars. 

The dust emission is also modeled with a library of 50,000 infrared spectra, composed out of four different components. Denser ISM associated with stellar birth clouds consists of polycyclic aromatic hydrocarbons (PAHs), stochastically-heated hot dust (temperatures of $130-250$\,K), and warm dust ($30-60$\,K) in thermal equilibrium. PAH emission is modeled with a fixed template derived from observations of nearby star-forming galaxies, while the two dust components are modeled as gray-body emission. For the diffuse ambient ISM, the relative fractions of these three components are fixed and an additional cold dust ($15-25$\,K) component is included. \magphys\ assumes a dust mass absorption coefficient $\kappa_{\lambda} \propto \lambda^{-\beta}$ which has a normalisation of $\kappa_{850}=0.077\,\rm{m}^2 \,\rm{kg}^{-1}$. A dust emissivity index of $\beta=1.5$ is assumed for warm dust, and $\beta=2.0$ for cold dust.

\magphys\ uses a Bayesian approach to determine the likelihood that models paired by the energy balance requirement could generate the observed photometry. Model photometry is generated using the response functions for the observed filters. Probability density functions (PDFs) are generated for parameters of interest by marginalizing over the other parameters. We plot these for the average SFR in the last 10 Myr as insets in Figure \ref{fig:seds}.

\subsubsection{SED3FIT}
\label{sec:sed3fit}
{\sc SED3FIT} \citep{berta+13} is a variant on \magphys\ in which an additional set of AGN torus templates based on the updated \citet{fritz+06} models described in \citet{feltre+12}. These models cover a range of lines of sight ($0-90^{\circ}$) through a variety of equatorial optical depths ($0.1-6$). Torus parameters, including outer and inner radii and aperture angle are held fixed at values of $R_\mathrm{out}/R\mathrm{in}=30$ and $40^{\circ}$, respectively. The density distribution along the horizontal and vertical axes and the slope of the power law in the NIR-MIR domain are all held fixed at zero.
Due to the additional parameter space added by these models, using the full set of models can quickly become computationally expensive.  We therefore reach a compromise between run time and sampling of the prior space and run the code with 10000 optical, 1000 infrared and 10 torus model libraries randomly drawn from the full sample. Statistical constraints are obtained in the same way as \magphys, but with the additional torus parameters. In this work, we limit our analysis to the fraction of infrared luminosity, $L(8-1000\mu$m), from the torus component. For galaxies in which the AGN component significantly improves the fit, Figure \ref{fig:seds} shows the fit both with and without the AGN component, and we include the PDF of the fraction of the infrared luminosity contributed by the torus.

\section{Molecular gas properties of CARMA GOALS galaxies}
To derive molecular gas properties, we convert the CO luminosity to a molecular gas mass (including helium) assuming a Milky Way-like $L_{\rm CO}/M{\rm(H_2)}$ conversion \citep{bolatto+13}. It was argued in \citet{downes+98} that the standard conversion often over-estimates the total molecular gas mass in interacting galaxies, due in part to the disrupted nature of the gas. This leads to a different distribution of molecular gas than in the Milky Way and similar galaxies. Namely, that there is a substantial fraction of molecular gas in a diffuse component within interacting galaxies, rather than the conglomeration of finite clumps and molecular complexes like in the Milky Way (and by extension, normal star-forming galaxies). Given this fact, converting from $L_{\rm CO}$ to M(H$_2$) is not clear cut, and the conversion factor for each individual galaxy may be different (rather than being able to make assumptions over the whole of the population.)

\section{Discussion}
\label{sec:disc}

\subsection{The nature of the 100\,GHz continuum}
Mergers and interactions have a high incidence of AGN activity \citep{barnes+92,kormendy+13}. Despite this, those AGN can be difficult to detect due to the extreme (and often Compton-thick) column densities that often obscure the AGN. For this reason, traditional searches including optical emission line ratios and even X-rays can sometimes fall short. In the era of {\em Spitzer}, identifying AGN in these circumstances became simpler, due to the fact that the hot dust radiating from the accretion disk and torus region influence the mid-IR colors \citep{stern+05,stern+12,lacy+07,lacy+13}. Mid-IR spectroscopy is an even more powerful diagnostic, with mid-IR lines that can unambiguously identify AGN and being far less affected by extinction \citep{petric+11}. While the mid-IR selection is powerful, if an AGN is present but not radiatively dominant, especially when buried under dense gas, these methods can also fail to detect an AGN's presence.

\citet{kawamuro+22} investigated a large sample of AGN using ALMA, focusing on the $\lambda$\,$\approx$\,1mm regime. They found that there is significant emission, and that the emission cannot be explained as originating from star formation, and must originate from the ultracompact coronae of the AGN, and that there was a clear correlation with the AGN luminosity. This leads to the possibility that compact millimeter continuum might be an independent tracer of AGN activity, including in sources with star formation dominating the infrared emission, and Compton thick molecular gas columns obscuring the hard X-rays. Our result is consistent with the findings of \citet{ricci+23} of a definitive connection between bright compact millimeter emission and AGN activity. All of our sources except for one (NGC\,958) are detected in the VLA Sky Survey (VLASS; \citealt{vlass}).  However, it is non-trivial to unambiguously distinguish between AGN jet and star formation origins for the radio continuum emission using only the VLASS data.

\begin{figure}[t!]
\includegraphics[width=0.49\textwidth]{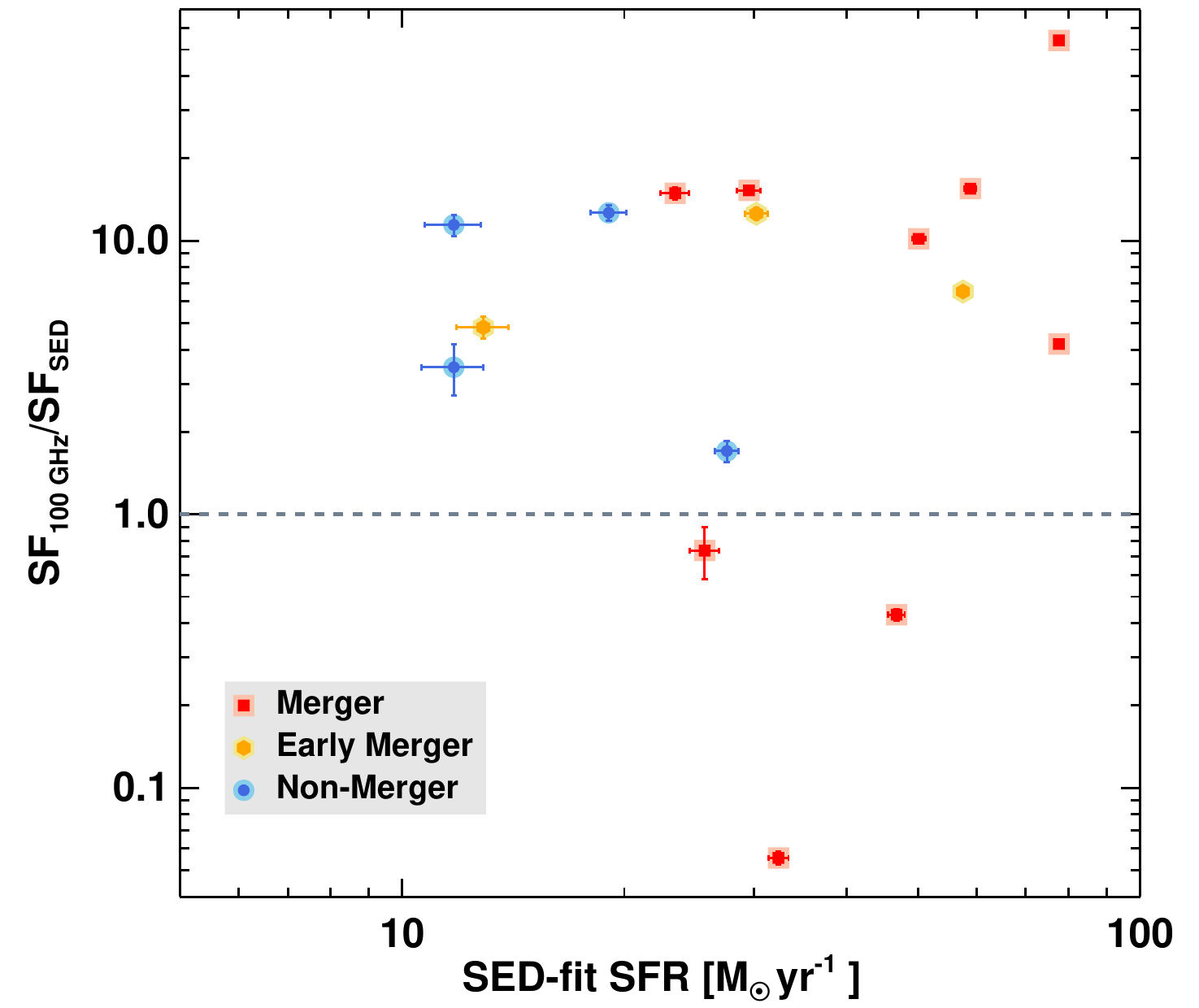} \vskip -1.75mm
\caption{The ratio between the SF predicted by the 100\,GHz continuum and SED fitting, showing that for the majority of the these sources, the 100 GHz emission cannot come from a star formation origin.}
\label{fig:sf100}
\end{figure}

In many of our CARMA GOALS sources, the correlator setup available for our observations allowed us to measure radio continuum in conjunction with the molecular line emission, which we were able to detect and image in 16 of our sources. Figure~\ref{fig:cont} presents the imaging of the 100\,GHz continuum for these sources. We note that the majority of the 3mm continuum sources are unresolved, consistent with an AGN origin. But it is also possible that in many cases, we are tracing ultra-compact starbursting regions. In order to test this hypothesis, we convert the 100\,GHz emission to a star formation rate using equation~10 from \citet{murphy+12} (assuming that 100\,GHz emission originates from star formation.) Figure~\ref{fig:sf100} presents the comparison between the 100\,GHz SFRs and the SED fit SFRs. We see that in the case for all but three sources (IRAS\,03582+6012, Arp\,55 and VV\,250), the 100\,GHz derived SFRs exceed the \magphys-derived SFRs by factors of at least 2, and up to 50. This suggests that in many cases, the 100\,GHz emission cannot come from a star formation origin, and must instead be from an alternative.

\citet{petric+11} measured AGN contributions to the mid-IR for the GOALS sample, including all 16 of our 100\,GHz detected sources. They found AGN contributions in all except for MCG+12-02-001, NGC\,1614, NGC\,4418 and VV\,250. This points to an AGN origin for much of the 100\,GHz emission that we measure, though given that these sources are all LIRGs and ULIRGs, we also cannot rule out that the Rayleigh-Jeans tail of dust emission is also contributing to these detections. While we find a ratio of SFR$_{100}$/SFR$_{\rm SED}$ of 15 for NGC\,1614, \citet{konig+16} used ALMA to observe NGC\,1614, and found the 3mm continuum region to be centrally peaked, but with components that traced the dust lanes and CO emission as well. This points to there being multiple origins of the 100\,GHz emission in CARMA GOALS sources, and suggests that in this sample, a clean separation of emission from the AGN and star formation is difficult due to their substantial molecular reservoirs, significant star formation activity, and opaque column densities even into the X-rays.

Higher resolution 100\,GHz continuum measurement will be able to spatially distinguish between emission originating from star formation and the AGN, as was done in \citet{konig+16}, but that is beyond the scope of this paper.

%%%%%%%%%%  Figure 7  %%%%%%%%%%

%%%%%%%%%%  Figure 5  %%%%%%%%%%
\begin{figure*}[t!]
\includegraphics[width=\textwidth]{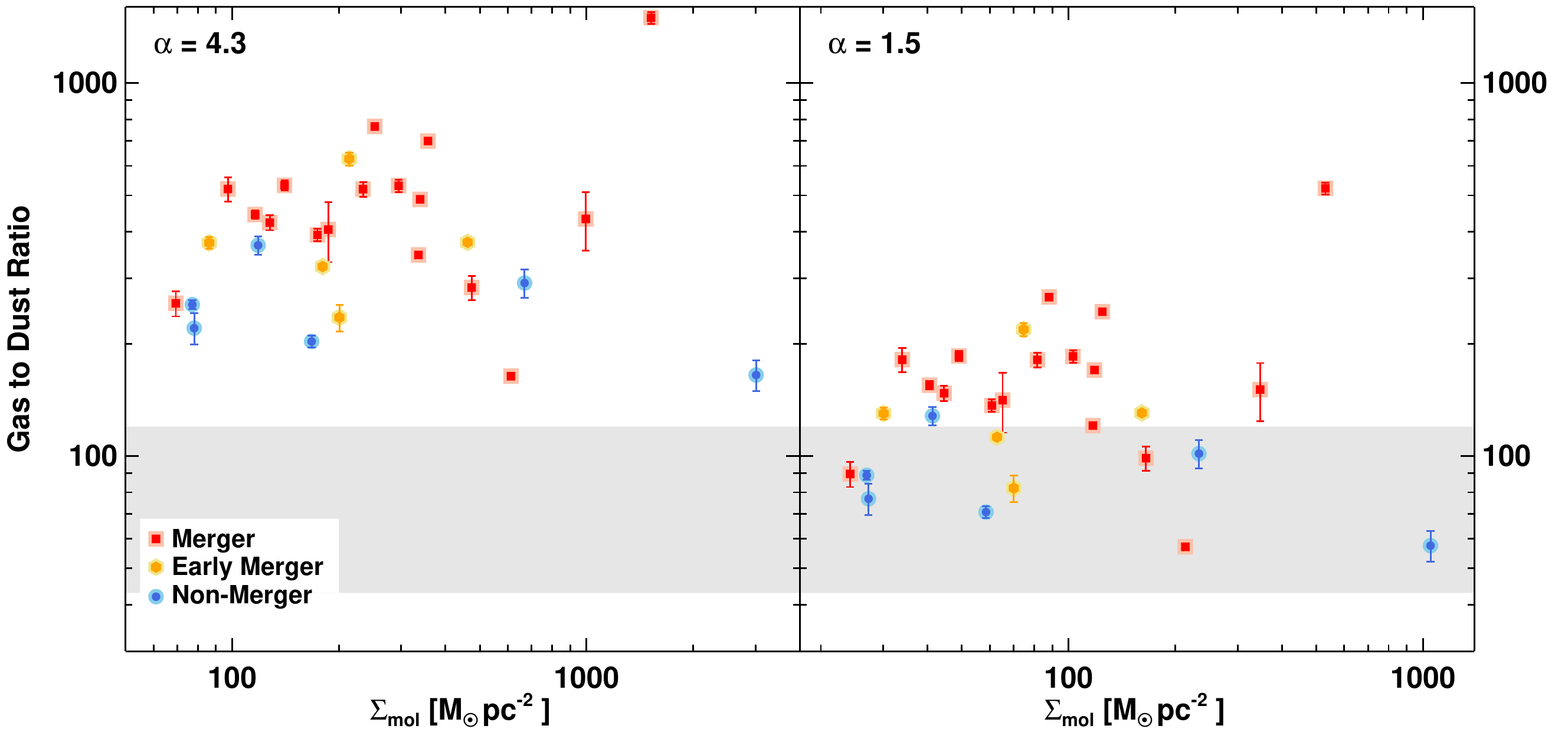} \vskip -1.75mm
\caption{The gas-to-dust ratio for all CARMA GOALS sources labeled based on their merger classification from \citet{petric+18}: red squares for "m"/mergers, orange hexagons for "em"/early-mergers and blue circles for "nm"/non-mergers. With the left plot using the standard Milky Way conversion factor ($\alpha$\,=\,4). The range of gas-to-dust ratios for normal star-forming galaxies from \citet{sandstrom+13} is shaded in gray. If we assume that $L_{\rm CO}$-to-M(H$_2$) conversion is not consistent with a Milky Way conversion and use the conversion from \citet{herrero-illana+19}, we find that the measured gas-to-dust ratios in our sources are much more consistent with the range found in normal galaxies. The plot on the right uses the $\alpha_{\rm CO}$ determined for GOALS sources in \citet{herrero-illana+19}. }
\label{fig:gastodust}
\end{figure*}

\subsection{Morphologies of the Molecular Gas}
If we focus on the spectra (Fig.~\ref{fig:specs}), we observe a large variety of shapes. In many objects, we see the familiar double-horned profile consistent with the molecular gas tracing the rotation out to the flat part of the rotation curve. Other spectra show gaussian-dominated profiles, although in many cases, these are the objects with 80\,km~s$^{-1}$ channel widths, which likely resolve out some of the sub-components of the velocity. Originally, we may have assumed that these gaussian profiles were observed in objects with significant amounts of disordered, turbulence-dominated molecular gas, but the average velocity maps (Fig.~\ref{fig:mom1}) reveal different properties. The vast majority of CARMA GOALS sources show clear velocity gradients, likely representing that the molecular gas is overall rotation-dominated. Even in galaxies that are at end-stages of mergers, there is ordered motion in the molecular gas. This result is consistent with \citet{ueda+14}, which also found that the majority of merger remnants exhibited ordered motion in their molecular gas velocity profiles. It is possible that this signifies that the molecular gas rapidly resettles into a disk configuration, even while other signs of morphological disruption remain at other wavelengths. This means that the presence of velocity gradients in high redshift galaxies and submillimeter galaxies \citep{genzel+20,stanley+23} does not imply that they are dynamically settled systems as one might have assumed from the rotation domination observed in the gas.

The moment maps and distributions of the CARMA data shown in Figures~\ref{fig:co+opt}--\ref{fig:mom1} display the various gas morphologies present in our sample. Although the majority of sources show signs of order and rotation, 11/28 sources are disordered (III\,Zw\,35, Arp\,299W, NGC\,4418) or have abnormal gas components (NGC\,695, UGC\,02369, NGC\,2623, NGC\,4922, NGC\,5256), and some show molecular bridges between two components (IRAS\,03582+6012, Arp\,55, II\,Zw\,96). While many of these disrupted gas morphologies appear in galaxies classified as mergers \citep{larson+16,stierwalt+14,petric+18}, there are at least a few that are found in either early mergers or non-merging galaxies (II\,Zw\,35, IC\,860, NGC\,4418), suggesting that the disruption in the molecular gas does not necessarily perfectly match the optical or infrared morphological characteristics of the system.

In the cases where there is no obvious merger, the disrupted molecular gas may be pointing to a previous merger event or a minor merger \citep{a14_stelpop}, where the dense molecular gas has been driven into the center and is igniting star formation and/or an AGN, which in turn is creating the substantial infrared flux that is observed. Indeed, both IC\,860 and NGC\,4418 are known to be compact obscured nuclei (CONs; \citealt{falstad+21}), it is probable that the AGN in these obscured sources is not only providing additional radiative power that we observe in the infrared, but also stirring up the molecular gas.

\subsection{The gas to dust ratio of CARMA GOALS sources}
\label{sec:gastodust}
The ratio of dust mass to gas mass in galaxies is thought to fit into a small range of values, as has been seen in star-forming galaxies \citep{sandstrom+13}, ranging from 43--120, for galaxies with solar metallicities. Previous studies of ULIRGs have used this range of values to calibrate the conversion between $L_{\rm CO}$ and M(H$_2$) \citep{herrero-illana+19}. This method takes into account the fact that ULIRGs and LIRGs have a different distribution of molecular gas than is seen in either the Milky Way, the most commonly used $\alpha_{\rm CO}$ calibrator, or normal star-forming galaxies \citep{downes+98}.

Figure~\ref{fig:gastodust} shows the gas-to-dust ratios for the CARMA GOALS sources. Dust masses were calculated using \magphys\ (or in the cases of galaxies with significant AGN contributions, SED3fit) discussed in Section\,\ref{sec:sed}. Gas masses were calculated using different assumptions for $\alpha_{\rm CO}$, including a Milky Way conversion, as well as the $\alpha_{\rm CO}$ found by \citet{herrero-illana+19} to be appropriate for GOALS sources. Assuming a Milky Way conversion of 4.3\,M$_\odot$\,(K\,km\,s$^{-1}$\,pc$^2$)$^{-1}$, we find that the gas-to-dust ratio of the CARMA GOALS sources ranges between 164-1500, significantly higher than the range of ratios seen in star-forming galaxies. Using the \citet{herrero-illana+19} value of 1.5\,M$_\odot$\,(K\,km\,s$^{-1}$\,pc$^2$)$^{-1}$, the gas-to-dust ratio ranges between 57-522, which while still higher than what is seen in the star-forming sample, span a much more reasonable set of values.

Although this adopted value does not completely reconcile the CARMA GOALS sources and star-forming galaxies, it is possible that these sources have an intrinsically higher gas-to-dust ratio than star-forming galaxies. Most of the CARMA GOALS sources have undergone an interaction, allowing shocks to pass through the dense gas and dust, destroying a non-negligible amount of the dust in their wake. Figure~\ref{fig:gastodust} also codifies the merger classification from \citet{petric+18}, including mergers, early mergers and non-mergers. What we see is no matter the $\alpha_{\rm CO}$ chosen, mergers tend to have the highest gas-to-dust ratios, with a median gas-to-dust of 444, followed by early mergers (368), followed by non-mergers (254). This supports the idea that the molecular gas is more stirred up during a collision, possibly producing both effects that are seen: less dust and more diffuse gas, which would require a lower $\alpha_{\rm CO}$. Indeed, \citet{renaud+19} argue that $\alpha_{\rm CO}$ varies depending on the merger phase and state of the molecular gas throughout the merger sequence, which our results support. Because both of these effects are likely present, it is difficult to disentangle one from the other. It would require extremely high resolution observations (i.e. those capable of resolving the scale height of the galaxy) that are able to trace the dust distribution, as well as map the velocity dispersion of the molecular gas to determine which is the source of this elevated ratio. Despite this, we are able to look at other properties of the molecular gas in these galaxies to provide us with additional information. We will revisit the gas-to-dust ratio in \S\ref{sec:starformation}.

%%%%%%%%%%%%%%% Table 3 %%%%%%%%%%%%%%%
\begin{table}[t]
\centering
\caption{CARMA GOALS Derived Properties} \vspace{-1mm}
\begin{tabular}{l c c c c}
\hline \hline
{\bf Name} & $M$(H$_2$) & $\Sigma_{\rm H_2}$\\
& 10$^9$\,$M_\odot$ & $M_\odot$~pc$^{-2}$\\
(1) & (2) & (3) \\
\hline \hline
MCG+12-02-001 & $8.63\pm0.01$ & $174.1\pm0.3$ \\
CGCG\,436-030 & $6.11\pm0.03$ & $461.9\pm2.1$ \\
III\,Zw\,35 & $7.96\pm0.05$ & $180.0\pm1.2$ \\
NGC\,695 & $16.18\pm0.04$ & $77.21\pm0.2$ \\
NGC\,958 & $15.34\pm0.02$ & $78.14\pm0.09$ \\
UGC\,2369N & $7.36\pm0.11$ & $140.4\pm2.0$ \\
UGC\,2369S & $14.00\pm0.11$ & $295.6\pm2.2$ \\
UGC\,02608 & $9.51\pm0.02$ & $118.4\pm0.3$ \\
IRAS\,03582+6012 & $6.01\pm0.02$ & $127.7\pm0.5$ \\
NGC\,1614 & $10.17\pm0.01$ & $357.1\pm0.4$ \\
CGCG\,468-002W & $1.15\pm0.01$ & $150.3\pm1.$ \\
CGCG\,468-002E & $2.04\pm0.01$ & $247.6\pm0.9$ \\
NGC\,2146 & $2.81\pm0.01$ & $995.3\pm0.6$ \\
NGC\,2623 & $6.78\pm0.01$ & $339.1\pm0.5$ \\
Arp\,55 & $37.03\pm0.08$ & $140.6\pm0.3$ \\
UGC\,05101 & $18.49\pm0.06$ & $612.2\pm2.1$ \\
Arp\,148 & $17.28\pm0.10$ & $295.1\pm1.8$ \\
Arp\,299W & $4.18\pm0.01$ & $463.5\pm0.7$ \\
Arp\,299E & $2.61\pm0.01$ & $232.9\pm0.6$ \\
NGC\,4418 & $0.468\pm0.01$ & $3011.6\pm7.7$ \\
NGC\,4922 & $4.12\pm0.03$ & $474.3\pm3.2$ \\
IC\,860 & $1.65\pm0.01$ & $668.9\pm3.3$ \\
UGC\,08335W & $3.96\pm0.04$ & $84.84\pm0.9$ \\
UGC\,08335E & $6.28\pm0.04$ & $151.3\pm1.$ \\
NGC\,5256 & $12.54\pm0.05$ & $234.1\pm0.8$ \\
CGCG\,142-034W & $5.31\pm0.01$ & $331.7\pm0.8$ \\
CGCG\,142-034E & $2.11\pm0.01$ & $88.94\pm0.5$ \\
NGC\,6670W & $54.50\pm0.07$ & $1587.7\pm2.0$ \\
NGC\,6670E & $14.35\pm0.07$ & $1314.2\pm6.7$ \\
NGC\,6786 & $10.44\pm0.03$ & $97.30\pm0.3$ \\
NGC\,6926 & $10.05\pm0.02$ & $69.36\pm0.1$ \\
II\,Zw\,96 & $20.24\pm0.10$ & $252.6\pm1.3$ \\
IC\,5298 & $4.90\pm0.01$ & $167.7\pm0.5$ \\
NGC\,7674 & $14.27\pm0.02$ & $86.19\pm0.1$ \\
 \\
\hline \hline
\end{tabular} \\
\label{tab:derived}
\raggedright{\footnotesize Column (1): Object name. 
Column (2): H$_2$ mass derived from $L_{\rm CO}$ in units of 10$^9$\,$M_\odot$ assuming $\alpha_{\rm CO}$\,=\,1.5 \citep{herrero-illana+19}.
Column (3): H$_2$ surface density, or $M({\rm H_2})$/area subtended by CO, in units of $M_\odot$~pc$^{-2}$.\\
}
\end{table}

\subsection{The molecular gas fraction of CARMA GOALS sources}
\label{sec:fmol}
We define the molecular gas fraction as the ratio between molecular gas mass and stellar mass:

\[ f_{\rm mol} = \frac{M_{\rm mol}}{M_{\rm mol}+M_\star} \]

There are many properties in these sources that naturally fall out of the assumptions we make when determining the proper conversion from $L_{\rm CO}$ and H$_2$ mass, including the molecular gas fractions that we measure in these systems. Indeed, if we were to use the Milky Way conversion, there are many cases where the molecular gas mass would exceed the stellar mass ($f_{\rm mol}$\,$\gtrsim$\,0.5), which while not impossible, is not common even in the most gas-rich mergers at low redshift \citep{downes+98}. Therefore, we have adopted $\alpha_{\rm CO}$\,=\,1.5, which is consistent with the value found from \citet{herrero-illana+19} for these sources. 

\begin{figure}[t!]
\includegraphics[width=0.47\textwidth]{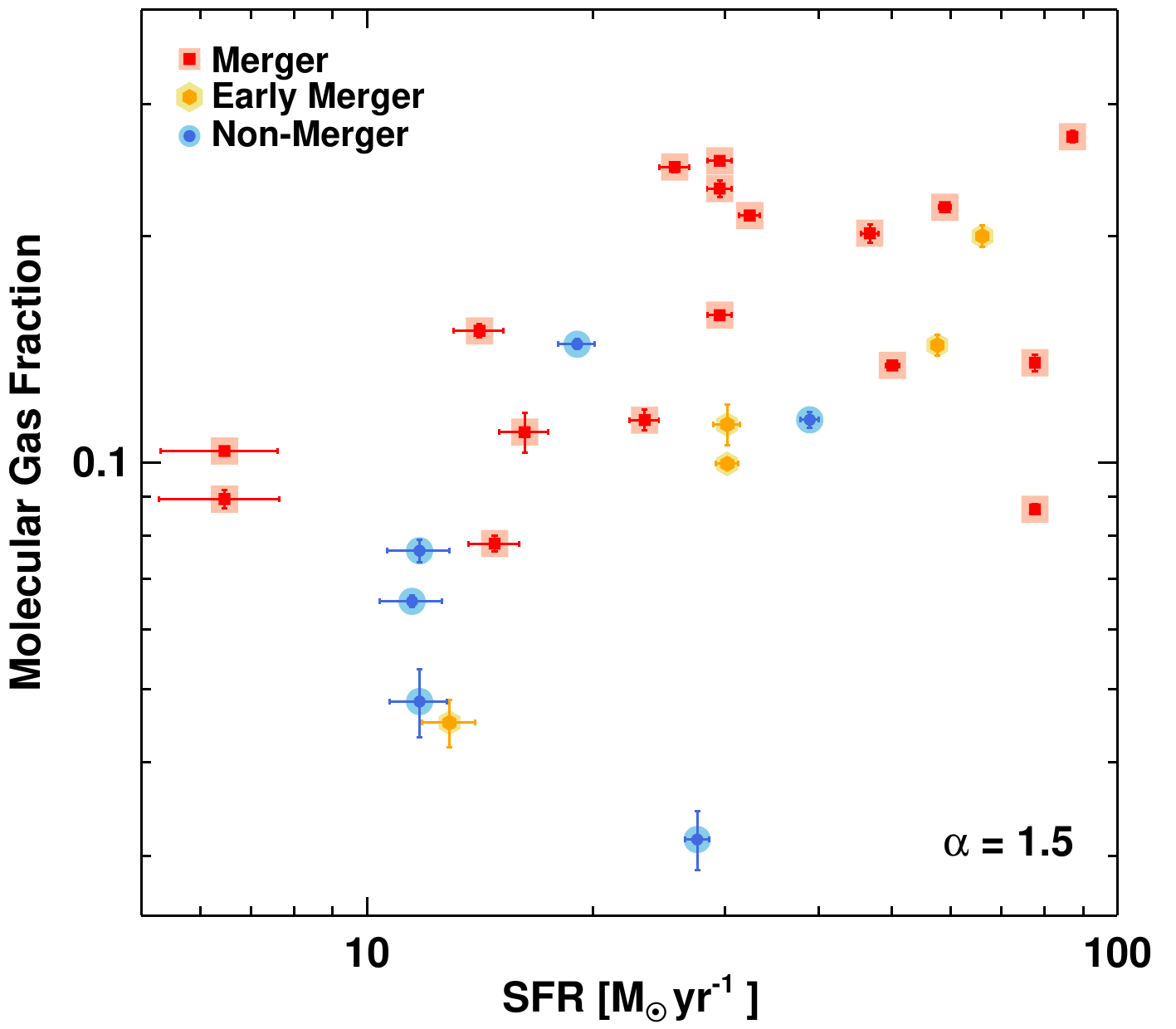} \vskip -1.75mm
\caption{The molecular gas fraction for all the CARMA GOALs sources, labeled based on their merger classification from \citet{petric+18}: red squares for "m"/mergers, orange hexagons for "em"/early-mergers and blue circles for "nm"/non-mergers. There is a noticeable trend that higher star formation rates also contain higher molecular gas fractions.}
\label{fig:fmol}
\end{figure}

We present the molecular gas fractions, derived from the stellar mass found in Table~\ref{tab:magphys} and the molecular gas mass found in Table~\ref{tab:derived}, in Figure~\ref{fig:fmol}. There are two trends visible in this plot: the first is that the molecular gas fraction correlates with star formation rate. Given that the stellar mass of these objects occupies a similar range, this result can be explained since more molecular gas naturally elicits more star formation. The second trend that we observe in this plot is that molecular gas fraction also correlates with merger stage, where those objects classified morphologically as being mergers are far more likely to have high molecular gas fractions, and those that are classified as non-mergers also have lower molecular gas fractions. Indeed, even when accounting for star formation rates, non-mergers tended to have lower molecular gas fractions than mergers \citep{iono+04}. It is well known from simulations that the molecular gas fraction increases as a merger progresses \citep{hopkins+08,he+23}, so our results are unsurprising.

%%%%%%%%%%  Figure 7  %%%%%%%%%%

%%%%%%%%%%  Figure 8  %%%%%%%%%%
\begin{figure*}[t!]
\centering
\includegraphics[width=\textwidth]{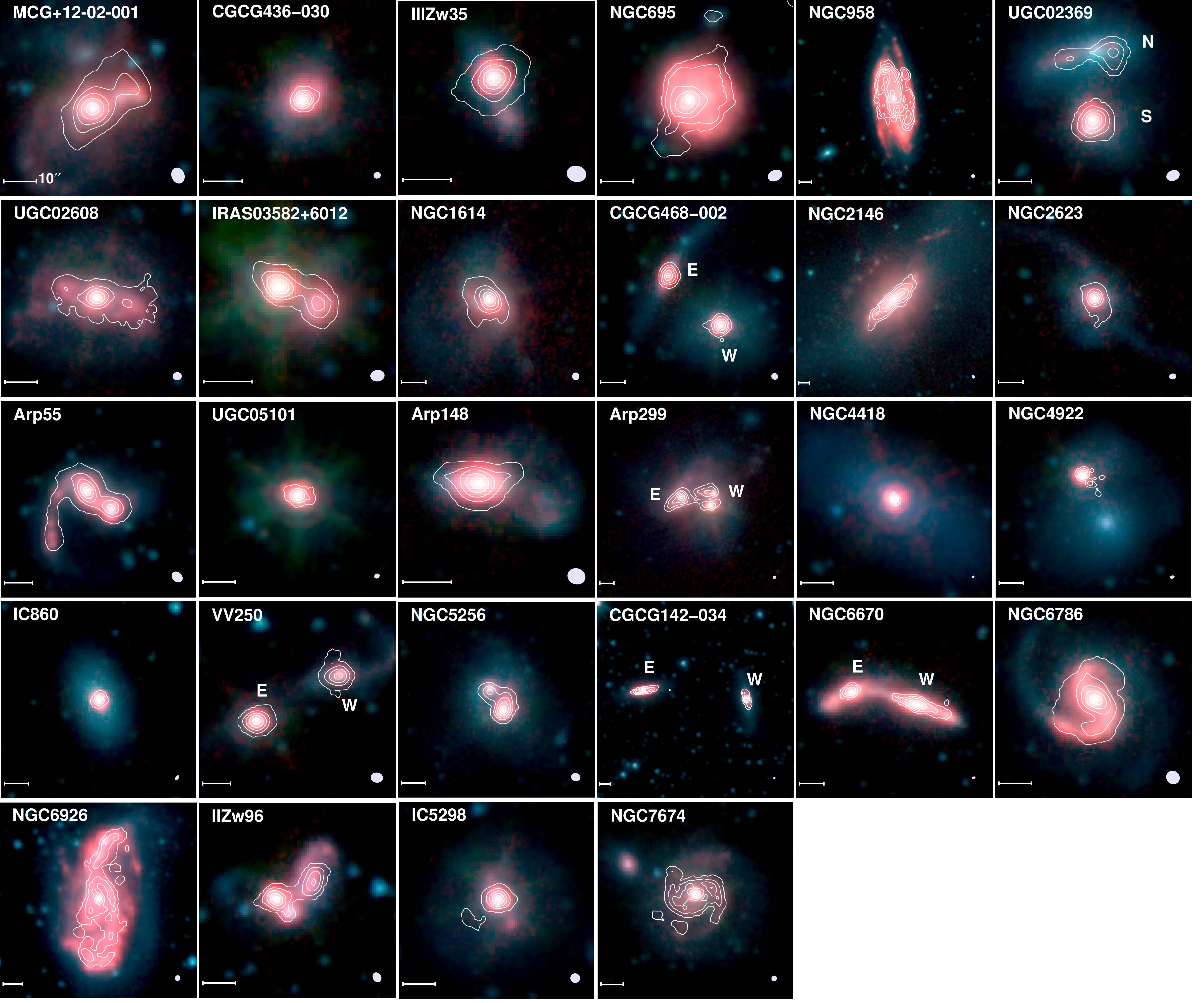} \vskip -1.75mm
\caption{The moment0 maps of CO(1--0) of the 28 CARMA GOALS galaxies overlaid atop the 3-color {\em Spitzer} 3.6-4.5-8.0\micron\ image. The white bar in the lower lefthand corner demarcates 10$''$ in each field. The beam is shown in the lower right corner in white. The red-hued portions of the {\em Spitzer} map show the locations of the PAH emission, which matches the positions of the molecular gas quite well.}
\label{fig:spitzer}
\end{figure*}

%%%%%%%%%%  Figure 11 %%%%%%%%%%
\begin{figure}[h]
\includegraphics[width=0.49\textwidth]{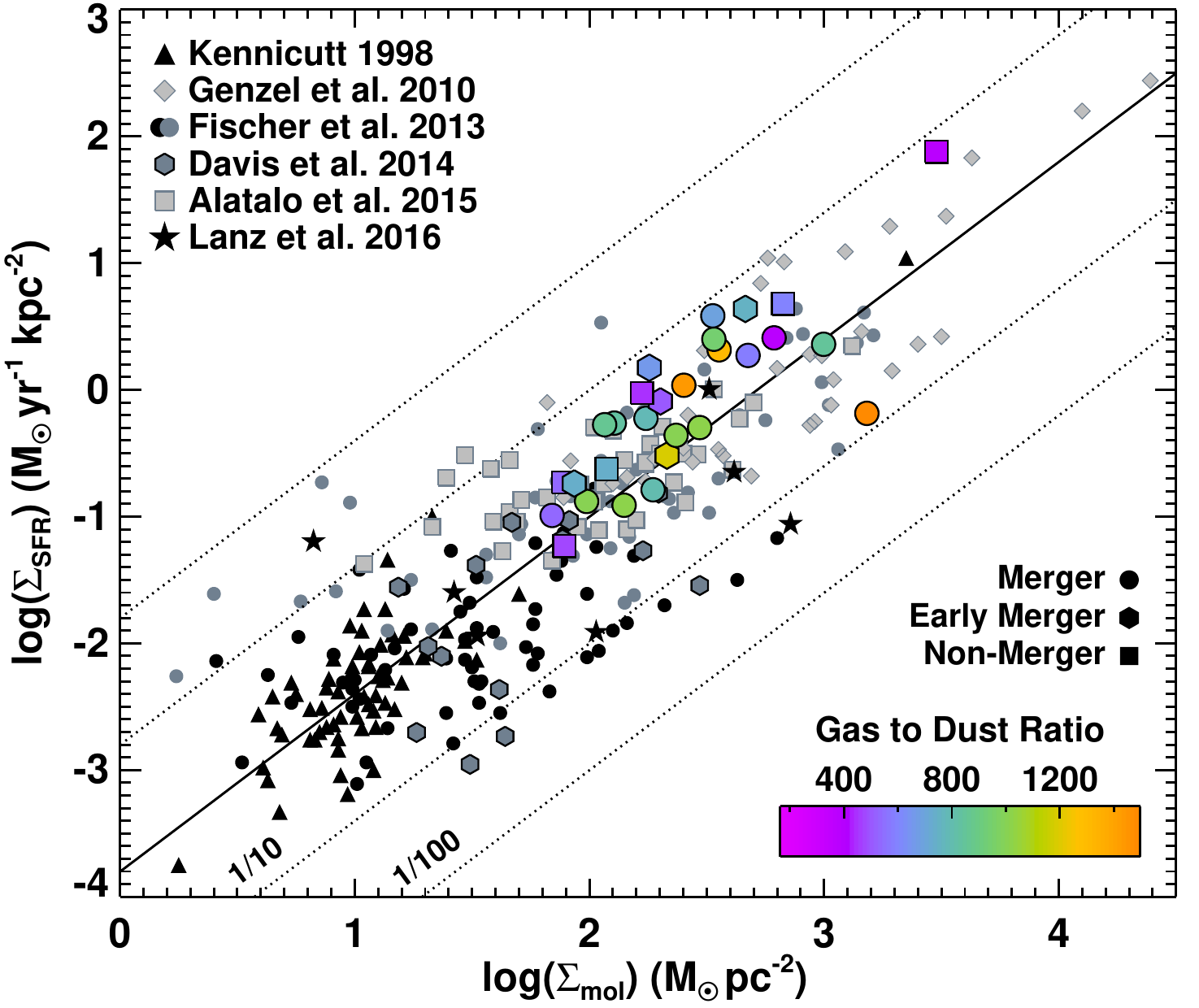}
\caption{The Schmidt-Kennicutt relation \citep{ken98} for the CARMA GOALS sample (color-coded by gas-to-dust ratio), with comparisons to normal star-forming galaxies (black triangles; \citealt{ken98} and black circles; \citealt{fisher+13}), bulges (gray circles; \citealt{fisher+13}), early-type galaxies (teal squares; \citealt{davis+14}), Hickson Compact Group galaxies (blue hexagons; \citealt{a15_hcgco}), high-redshift galaxies (gray diamonds; \citealt{genzel+10}), and radio galaxies (black squares; \citealt{ogle+10}). The diagonal lines represent the 10$\times$ and 100$\times$ deviations from the average relation. All star formation rates have been normalized to the Chabrier initial mass function \citep{chabrier03}.}
\label{fig:ks}
\end{figure}

\subsection{The star formation efficiency of CARMA GOALS sources}
\label{sec:starformation}
In order to compare the star formation -- molecular gas surface densities, we use the values derived for star formation via fitting of spectral energy distributions (SEDs) composed of ultraviolet (UV) to far-infrared (FIR) photometry, described in detail in Section~\ref{sec:sed}. We calculate the molecular gas mass using the CO luminosities presented in Table~\ref{tab:properties}, then assume a conversion factor informed by the gas-to-dust ratio results from \S\ref{sec:gastodust} of $\alpha_{\rm CO}$\,=\,1.5, which corresponds to M(H$_2$)/M$_\odot$ = 3.1$\times$10$^4$\,$L_{\rm CO}/L_\odot$. Finally, we use the footprint of the molecular gas present in the moment0 maps to determine the total area subtended by the molecular gas. Because the observations used for SED fitting includes wavebands in which the source is unresolved (e.g. far-infrared), we assume cospatiality of star formation and molecular gas. Figure~\ref{fig:spitzer} shows the CO intensity maps overlaid on a 3-color {\em Spitzer} image of each galaxy (the 8$\mu$m feature represented as red). The 7.7$\mu$m PAH feature, which is present in the 8$\mu$m broadband filter traces the gas remarkably well. If we assume that this feature is primarily tracing star formation in these systems \citep{calzetti+07,diaz-santos+11}, then cospatiality is an appropriate assumption to make.

Figure~\ref{fig:ks} shows the CARMA GOALS sample placed on ``Schmidt--Kennicutt'' relation (S-K; \citealt{schmidt59,ken98}), along with comparison samples, including normal star-forming galaxies (\citealt{ken98,fisher+13}), bulges (\citealt{fisher+13}), early-type galaxies (\citealt{davis+14}), Hickson Compact Group galaxies (\citealt{a15_hcgco}), high-redshift galaxies (\citealt{genzel+10}), and radio galaxies (\citealt{ogle+10}), all renormalized to a common \citet{chabrier03} initial mass function. We have also color-coded each CARMA GOALS source with its gas-to-dust ratio. On average, CARMA GOALS sources are found to be above the "standard" relation and in comparison to the selected samples but still well within the scatter. This suggests that the star formation in these sources is super-efficient (e.g. that more stars are forming in a given period of time than is seen in normal star-forming galaxies and the other comparisons we have made). There are many possible explanations for this. Firstly, this could be a result of the conversion factor that we have chosen. If we assumed a conversion factor more in line with what is observed in the Milky Way or normal star-forming galaxies \citep{sandstrom+13}, then these points would fall much closer to the average relation. While it is not improbable that some dust has been destroyed during the interaction, which could lead to an enhanced gas-to-dust ratio in these sources, it is much more likely that the diffuse component of the molecular gas compensates for this dust destruction, so it is less likely that our enhanced star formation efficiency in these sources can be explained simply as a misconversion.

It is worth considering that in these sources, the star formation efficiency is higher. \citet{krumholz+12} laid the theoretical groundwork to explain the S-K relation, in that approximately 1\% of dense gas is converted into stars per freefall time. \citet{salim+15} took this concept further, showing that the Mach number also has influence on the star formation efficiency by modifying the freefall time. In this way, more turbulent systems will appear to have enhanced star formation efficiencies compared to normal star-forming galaxies when in reality the efficiency per freefall time is consistent. Interactions are well known to be turbulent systems \citep{rich+11,lu+14,stierwalt+14,beaulieu+23}, and CARMA GOALS sources are no exception. Thus, it is possible that the turbulence that has made a determination of the conversion factor more difficult  has also impacted the freefall time of the star-forming gas, thus driving CARMA GOALS sources as a population "above" the S-K relation. In order to delve deeper into this question, an accurate determination of the intrinsic velocity dispersion of the molecular gas is necessary, which requires high resolution observations able to resolve the scale height of each galaxy, and is thus beyond the scope of this paper.

%The CARMA GOALS sources exhibit a linear relation with the same slope as derived by \citet{ken98}, with a slight offset toward sub-efficiency (with a mean suppression of $\approx$\,1.2). This offset from the S-K relation is sufficiently small that they can be accounted for based on the uncertainties in both the conversion from CO luminosity to $M$(H$_2$) \citep{downes+98,bolatto+13}, converting the observables to a star formation rate \citep{ken98,calzetti+07,calzetti+10}, and the uncertainties associated with the area subtended by the molecular gas.

\subsection{Molecular inflows and outflows in CARMA GOALS sources}
Multiple systems also show evidence of extended line wings (III\,Zw\,35, UGC\,02608, IRAS\,03582+6012, NGC\,1614, Arp\,148, Arp\,299, NGC\,4418, IC\,860, NGC\,5256, II\,Zw\,96). Some of these sources have confirmed AGN-driven molecular outflows already in the literature (NGC\,1614; \citealt{garcia-burillo+15}, III\,Zw\,35; \citealt{lutz+20}, IC\,860; \citealt{luo+22}). In the case of NGC\,4418, the wing may be associated with inflow of gas rather than outflow \citep{costagliola+13, gonzalez-alfonso+15}. In most cases, our CARMA-detected line wings are observed around $\approx$300\,km~s$^{-1}$, and often these wings are only seen on the redshifted or blueshifted side of the galaxy.

Pe\~naloza et al. (2023, in preparation) used ALMA Compact Array ($\theta_{\rm fwhm}$\,$\approx$\,4.7\arcsec) to study 18 LIRGs and ULIRGs from the GOALS sample and found that 12 of those 18 host molecular outflows involving more than 1\% of their molecular mass, though at velocities below the escape velocity. Our results also support their claim that molecular outflows are common in interacting galaxies (something seen in ionized gas as well; \citealt{harrison+14}), but that they are not a dominant mechanism in expelling the molecular gas. Indeed, it is possible that in these systems, an AGN jet punching through the interstellar medium has a more significant impact on the depletion of the gas reservoir via star formation, through the additional turbulence (and therefore shortened freefall time) than through the direct expulsion of the molecular gas, an effect that is seen in both Markarian~231 \citep{alatalo15} and NGC\,1266 \citep{a15_sfsupp}.

\section{Conclusions}
\label{sec:conclusions}
We have presented observations of 28 LIRGs and ULIRGs that are part of the GOALS sample with CARMA, resolving all sources in CO(1--0). 

\begin{itemize}

\item We have detected 16 of the 28 sources in 100\,GHz continuum, with the vast majority of those objects being unresolved. The star formation rates inferred from the 100\,GHz emission is factors of 2--10 too high (as compared with SED fitting) to be explained by star formation alone, suggesting that in most cases,  an AGN contributes non-negligibly to the emission.

\item Many of the CARMA GOALS systems show velocity gradients despite undergoing significant interactions. This implies that an ordered gradient observed in the molecular gas does not imply ordered rotation in the system.

\item The gas-to-dust ratio in the CARMA GOALS sources are much higher than expected if we use a L$_{\rm CO}$ to M(H$_2$) conversion from the Milky Way. We instead use the conversion derived by \citet{herrero-illana+19} for the CARMA GOALS sources. The gas-to-dust ratio is higher than what is seen in normal star-forming galaxies, but is consistent with the hypothesis that the turbulent medium in interacting galaxies may have destroyed some of the dust reservoir. This choice also influences the molecular gas fraction observed in these sources, which are in line with what is expected in interacting galaxies.

\item The CARMA GOALs sources as a group have an enhanced star formation efficiency as shown on the S-K relation. While this could be due to underestimating the $\alpha_{\rm CO}$ conversion factor, it is more likely that the turbulence present in these interacting systems has increased the Mach number above what is observed in star-forming galaxies, thus decreasing the freefall time, and therefore boosting the perceived efficiency of star formation.

\item Several of the CARMA GOALS sources show evidence of extended line wings, suggesting the presence of molecular outflows in many, supporting a growing number of studies that show that these outflows are ubiquitous in systems with molecular gas proximate to the AGN. In spite of their ubiquity, it does not appear that they play a substantial role in the depletion of molecular gas from these systems.

\end{itemize}

It is clear that the study of molecular gas in merging galaxies is able to shed light on many other physical processes that are taking place, including the interaction between turbulence and star formation, between the AGN and the host galaxy, and even the fundamental question of how a galaxy ultimately quenches its star formation, depletes its star forming fuel, and evolves passively as a quiescent galaxy. Higher resolution molecular gas measurements capable of accurately determining the total molecular mass, paired with resolved maps able to trace star formation uncontaminated will be able to further our understanding of the intricacies of this important phase in quenching galaxies.

\section{Acknowledgements}
We thank the anonymous referee for excellent suggestions that have improved this manuscript. KA thanks the participants of the {\sl Behind the Curtain of Dust II} in Sexten for enlightening conversations that have helped develop and improve this manuscript. Support for KA is provided by NASA through Hubble Fellowship grant \hbox{\#HST-HF2-51352.001} awarded by the Space Telescope Science Institute, which is operated by the Association of Universities for Research in Astronomy, Inc., for NASA, under contract NAS5-26555. ET acknowledges support from: ANID through Millennium Science Initiative Program NCN19\_058, CATA-BASAL ACE210002 and FB210003, FONDECYT Regular 1190818 and 1200495. YL and AP acknowledge support from SOFIA grant \#08-0226 (PI: Petric). YL acknowledges support from the Space Telescope Science Institute Director's Discretionary Research Fund grant D0101.90281. JAO acknowledges support from the Space Telescope Science Institute Director's Discretionary Research Fund grants D0101.90296 and D0101.90311. MS acknowledges support from the William H. Miller III Graduate Fellowship. VU acknowledges funding support from NSF Astronomy and Astrophysics Research Grant \# AST-2408820, NASA Astrophysics Data Analysis Program (ADAP) grant \# 80NSSC23K0750, and STScI grant \# HST-AR-17063.005-A, HST-GO-17285.001-A, and JWST-GO-01717.001-A. AMM acknowledges support from NASA ADAP grant \#80NSSC23K0750 and from NSF AAG grant \#2009416 and NSF CAREER grant \#2239807.

Support for CARMA construction was derived from the Gordon and Betty Moore Foundation, the Kenneth T. and Eileen L. Norris Foundation, the James S. McDonnell Foundation, the Associates of the California Institute of Technology, the University of Chicago, the states of California, Illinois, and Maryland, and the National Science Foundation. Ongoing CARMA development and operations are supported by the National Science Foundation under a cooperative agreement, and by the CARMA partner universities. This research has made use of the NASA/IPAC Extragalactic Database (NED) which is operated by the Jet Propulsion Laboratory, California Institute of Technology, under contract with the National Aeronautics and Space Administration. {\em Herschel} is an ESA space observatory with science instruments provided by European-led Principal Investigator consortia and with important participation from NASA. This publication makes use of data from the Galaxy Evolution Explorer, retrieved from the Mikulski Archive for Space Telescopes (MAST), part of the Space Telescope Science Institute, which is operated by the Association of Universities for Research in Astronomy, Inc., under NASA contract NAS5-26555. The Pan-STARRS1 Surveys (PS1) and the PS1 public science archive have been made possible through contributions by the Institute for Astronomy, the University of Hawaii, the Pan-STARRS Project Office, the Max-Planck Society and its participating institutes, the Max Planck Institute for Astronomy, Heidelberg and the Max Planck Institute for Extraterrestrial Physics, Garching, The Johns Hopkins University, Durham University, the University of Edinburgh, the Queen's University Belfast, the Harvard-Smithsonian Center for Astrophysics, the Las Cumbres Observatory Global Telescope Network Incorporated, the National Central University of Taiwan, the Space Telescope Science Institute, the National Aeronautics and Space Administration under Grant No. NNX08AR22G issued through the Planetary Science Division of the NASA Science Mission Directorate, the National Science Foundation Grant No. AST-1238877, the University of Maryland, Eotvos Lorand University (ELTE), the Los Alamos National Laboratory, and the Gordon and Betty Moore Foundation. This publication makes use of data products from the TwoMicron All Sky Survey, which is a joint project of the University of Massachusetts and the Infrared Processing and Analysis Center (IPAC)/Caltech, funded by NASA and the NSF. This publication used observations made with the Spitzer Space Telescope, which is operated by the Jet Propulsion Laboratory (JPL)/California Institute of Technology (Caltech) under a contract with NASA. 

\facilities{CARMA}

\bibliographystyle{mnras}
\bibliography{master}

\clearpage
\appendix
\section{Details of Additional Photometry Measurements}
\label{app:photometry}
We present the photometry and errors of the sources not originally in \citet{U+12} or \citet{chu+17} for SED fitting in \S\ref{sec:sed}. All UV and optical band photometry have been extinction corrected using the values of Table~\ref{tab:Ecorr} using: \[S_{\rm corr}  = S_{\rm uncorr}*10^{A_\lambda/2.5} \]

\begin{deluxetable*}{lcccccccc}
\tabletypesize{\scriptsize}
\tablecaption{Additional Photometry\label{tab:photData}}
\tablewidth{0pt}
 \centering
\tablehead{
\colhead{Name} & \colhead{FUV} & \colhead{NUV} & \colhead{g}  &
\colhead{r}  & \colhead{i} & \colhead{J} & 
\colhead{H}  & \colhead{K} }
\startdata
MCG+12-02-001 & ... & 1.92-03 & 8.74-03 & 1.57-02 & 3.09-02 & 4.38-02 & 6.75-02 & 6.45-02 \\
       & (...) & (4.49-04) & (8.74-04) & (1.57-03) & (-9.9e+0) & (5.11-03) & (4.39-03) & (4.00-03) \\
UGC02608 & ... & ... & 1.13-02 & 1.89-02 & 2.26-02 & 5.05-02 & 6.51-02 & 5.99-02 \\
       & (...) & (...) & (1.13-03) & (1.89-03) & (-9.9e+0) & (2.49-03) & (3.21-03) & (3.39-03) \\
IRAS03582+6012 & ... & 1.81-03 & 2.98-03 & 4.24-03 & 4.89-03 & 8.20-03 & 1.31-02 & 1.39-02 \\
       & (...) & (5.67-04) & (5.96-04) & (8.49-04) & (-9.9e+0) & (7.38-04) & (1.17-03) & (1.11-03) \\
CGCG468-002 & ... & 9.53-04 & 1.29-02 & 2.17-02 & 2.79-02 & 6.25-02 & 9.13-02 & 8.28-02 \\
       & (...) & (2.12-04) & (1.29-03) & (2.17-03) & (-9.9e+0) & (4.04-03) & (6.35-03) & (5.91-03) \\
NGC2146 & 3.67-03 & 9.97-03 & 9.64-02 & 1.93-01 & 2.66-01 & 9.23-01 & 1.19+00 & 1.18+00 \\
       & (3.69-04) & (9.97-04) & (9.64-03) & (1.93-02) & (-9.9e+0) & (7.14-02) & (8.83-02) & (7.48-02) \\
NGC4418 & ... & ... & 1.04-02 & 1.83-02 & 2.50-02 & 6.10-02 & 6.54-02 & 6.13-02 \\
       & (...) & (...) & (1.04-03) & (1.83-03) & (-9.9e+0) & (3.31-03) & (4.10-03) & (5.52-03) \\
CGCG142-034 & ... & ... & 4.61-03 & 9.70-03 & 1.51-02 & 4.32-02 & 5.46-02 & 5.59-02 \\
       & (...) & (...) & (4.61-04) & (9.70-04) & (-9.9e+0) & (2.34-03) & (3.14-03) & (3.53-03) \\
NGC6670 & 2.68-04 & 6.03-04 & 5.56-03 & 1.05-02 & 1.48-02 & 4.32-02 & 5.46-02 & 5.59-02 \\
       & (2.72-05) & (6.04-05) & (5.56-04) & (1.05-03) & (-9.9e+0) & (2.34-03) & (3.14-03) & (3.53-03) \\
NGC6786 & 1.51-03 & 2.82-03 & 1.19-02 & 1.95-02 & 2.35-02 & 4.76-02 & 6.23-02 & 5.38-02 \\
       & (1.59-04) & (2.85-04) & (1.19-03) & (1.95-03) & (-9.9e+0) & (2.55-03) & (3.79-03) & (3.57-03) \\
NGC6926 & 2.90-03 & 5.33-03 & 2.97-02 & 4.82-02 & 6.05-02 & 1.34-01 & 1.56-01 & 1.43-01 \\
       & (2.99-04) & (5.37-04) & (2.97-03) & (4.82-03) & (-9.9e+0) & (5.67-03) & (7.56-03) & (8.18-03) \\
IIZw96 & 1.91-03 & 2.91-03 & 7.73-03 & 1.04-02 & 1.13-02 & 2.02-02 & 2.19-02 & 2.12-02 \\
       & (1.99-04) & (2.94-04) & (7.73-04) & (1.04-03) & (-9.9e+0) & (1.44-03) & (1.95-03) & (1.89-03) 
\enddata
\tablecomments{All fluxes are given in Jy, with uncertainties given on the row below.}
\end{deluxetable*}

\begin{deluxetable*}{lccccc}
\tabletypesize{\scriptsize}
\tablecaption{Additional Photometry: {\em Spitzer} \label{tab:mirData}}
\tablewidth{0pt}
 \centering
\tablehead{
\colhead{Name} & \colhead{IRAC1}  &
\colhead{IRAC2}  & \colhead{IRAC3} & \colhead{IRAC4} & \colhead{MIPS1} }
\startdata
MCG+12-02-001 & 5.65-02 & 4.85-02 & 1.53-01 & 5.39-01 & 4.44+00 \\
       & (1.88-03) & (1.75-03) & (6.15-03) & (2.97-02) & (3.65-01) \\
UGC02608 & 4.74-02 & 5.76-02 & 1.03-01 & 2.88-01 & 1.21+00 \\
       & (1.65-03) & (2.22-03) & (5.01-03) & (1.18-02) & (9.75-02) \\
IRAS03582+6012 & 3.81-02 & 6.57-02 & 1.84-01 & 2.46-01 & 8.19-01 \\
       & (1.19-03) & (2.08-03) & (6.07-03) & (8.41-03) & (6.55-02) \\
CGCG468-002 & 5.69-02 & 5.24-02 & 7.06-02 & 1.65-01 & 9.96-01 \\
       & (2.27-03) & (2.57-03) & (6.43-03) & (1.19-02) & (8.32-02) \\
NGC2146 & 8.62-01 & 6.59-01 & 2.45+00 & 7.06+00 & 1.70+01 \\
       & (4.14-02) & (4.59-02) & (1.92-01) & (1.44+00) & (1.38+00) \\
NGC4418 & 2.94-02 & 3.34-02 & 1.31-01 & 2.92-01 & 6.10+00 \\
       & (1.38-03) & (1.81-03) & (7.68-03) & (2.97-02) & (4.88-01) \\
CGCG142-034 & 4.04-02 & 3.19-02 & 8.28-02 & 3.13-01 & 8.59-01 \\
       & (1.37-03) & (1.23-03) & (3.33-03) & (1.04-02) & (6.92-02) \\
NGC6670 & 4.04-02 & 3.19-02 & 8.28-02 & 3.13-01 & 8.59-01 \\
       & (1.37-03) & (1.23-03) & (3.33-03) & (1.04-02) & (6.92-02) \\
NGC6786 & 3.13-02 & 2.22-02 & 4.60-02 & 1.63-01 & 4.64-01 \\
       & (1.16-03) & (1.08-03) & (2.72-03) & (6.34-03) & (3.81-02) \\
NGC6926 & 8.86-02 & 6.37-02 & 1.40-01 & 4.52-01 & 6.74-01 \\
       & (3.34-03) & (3.12-03) & (8.46-03) & (1.72-02) & (6.01-02) \\
IIZw96 & 1.76-02 & 1.75-02 & 3.70-02 & 1.42-01 & 1.90+00 \\
       & (6.17-04) & (6.58-04) & (1.58-03) & (4.78-03) & (1.52-01) 
\enddata
\tablecomments{All fluxes are given in Jy, with uncertainties given on the row below.}
\end{deluxetable*}

\begin{deluxetable*}{lcccccccccc}
%\tabletypesize{\scriptsize}
\tablecaption{Extinction Corrections\label{tab:Ecorr}}
\tablewidth{0pt}
 \centering
\tablehead{
\colhead{Name} & \colhead{A$_{\rm FUV}$} & \colhead{A$_{\rm NUV}$} & \colhead{A$_{\rm U}$}  & \colhead{A$_{\rm B}$}  &
\colhead{A$_{\rm g}$}  & \colhead{A$_{\rm V}$} & \colhead{A$_{\rm r}$} &
\colhead{A$_{\rm R}$}  & \colhead{A$_{\rm i}$} & \colhead{A$_{\rm I}$}  }
\startdata
MCG+12-02-001 & 4.564 & 4.762 & 2.669 & 2.233 & 2.034 & 1.689 & 1.407 & 1.336 & 1.046 & 0.927 \\ 
UGC02608 & 1.192 & 1.243 & 0.697 & 0.583 & 0.531 & 0.441 & 0.367 & 0.349 & 0.273 & 0.242 \\ 
IRAS03582+6012 & 5.307 & 5.538 & 3.104 & 2.597 & 2.366 & 1.964 & 1.637 & 1.554 & 1.216 & 1.078 \\ 
CGCG468-002 & 2.610 & 2.724 & 1.527 & 1.277 & 1.164 & 0.966 & 0.805 & 0.764 & 0.598 & 0.530 \\ 
NGC2146 & 0.711 & 0.711 & 0.415 & 0.348 & 0.317 & 0.263 & 0.219 & 0.208 & 0.163 & 0.144 \\ 
NGC4418 & 0.173 & 0.173 & 0.101 & 0.085 & 0.077 & 0.064 & 0.053 & 0.051 & 0.040 & 0.035 \\ 
CGCG142-034 & 1.013 & 1.057 & 0.593 & 0.497 & 0.452 & 0.375 & 0.313 & 0.297 & 0.233 & 0.206 \\ 
NGC6670 & 0.354 & 0.354 & 0.208 & 0.174 & 0.158 & 0.131 & 0.109 & 0.104 & 0.081 & 0.072 \\ 
NGC6786 & 1.048 & 1.094 & 0.614 & 0.513 & 0.468 & 0.388 & 0.324 & 0.307 & 0.240 & 0.213 \\ 
NGC6926 & 1.205 & 1.258 & 0.705 & 0.590 & 0.538 & 0.446 & 0.372 & 0.353 & 0.276 & 0.245 \\ 
IIZw96 & 0.619 & 0.619 & 0.362 & 0.303 & 0.276 & 0.229 & 0.191 & 0.181 & 0.142 & 0.126 \\ 
CGCG436-030 & 0.270 & 0.270 & 0.158 & 0.132 & 0.120 & 0.100 & 0.083 & 0.079 & 0.062 & 0.055 \\ 
IIIZw35 & 0.465 & 0.465 & 0.272 & 0.228 & 0.207 & 0.172 & 0.143 & 0.136 & 0.107 & 0.094 \\ 
NGC695 & 0.667 & 0.667 & 0.390 & 0.326 & 0.297 & 0.247 & 0.206 & 0.195 & 0.153 & 0.135 \\ 
NGC958 & 0.224 & 0.224 & 0.131 & 0.109 & 0.100 & 0.083 & 0.069 & 0.065 & 0.051 & 0.045 \\ 
UGC02369 & 0.754 & 0.754 & 0.441 & 0.369 & 0.336 & 0.279 & 0.232 & 0.221 & 0.173 & 0.153 \\ 
NGC1614 & 1.140 & 1.190 & 0.667 & 0.558 & 0.508 & 0.422 & 0.352 & 0.334 & 0.261 & 0.232 \\ 
NGC2623 & 0.305 & 0.305 & 0.178 & 0.149 & 0.136 & 0.113 & 0.094 & 0.089 & 0.070 & 0.062 \\ 
Arp55 & 0.127 & 0.127 & 0.075 & 0.062 & 0.057 & 0.047 & 0.039 & 0.037 & 0.029 & 0.026 \\ 
UGC05101 & 0.246 & 0.246 & 0.144 & 0.120 & 0.110 & 0.091 & 0.076 & 0.072 & 0.056 & 0.050 \\ 
Arp148 & 0.065 & 0.068 & 0.038 & 0.032 & 0.029 & 0.024 & 0.020 & 0.019 & 0.015 & 0.013 \\ 
Arp299 & 0.124 & 0.124 & 0.072 & 0.060 & 0.055 & 0.046 & 0.038 & 0.036 & 0.028 & 0.025 \\ 
NGC4922 & 0.084 & 0.087 & 0.048 & 0.041 & 0.037 & 0.031 & 0.026 & 0.024 & 0.019 & 0.017 \\ 
IC860 & 0.100 & 0.100 & 0.058 & 0.048 & 0.044 & 0.037 & 0.030 & 0.029 & 0.023 & 0.020 \\ 
VV250 & 0.168 & 0.168 & 0.097 & 0.081 & 0.074 & 0.062 & 0.051 & 0.049 & 0.038 & 0.034 \\ 
NGC5256 & 0.097 & 0.10 & 0.057 & 0.047 & 0.043 & 0.036 & 0.030 & 0.028 & 0.022 & 0.020 \\ 
IC5298 & 0.635 & 0.635 & 0.372 & 0.311 & 0.284 & 0.235 & 0.196 & 0.186 & 0.146 & 0.129 \\ 
NGC7674 & 0.440 & 0.440 & 0.257 & 0.215 & 0.196 & 0.163 & 0.135 & 0.129 & 0.101 & 0.089  
\enddata
\tablecomments{Extinction corrections for the UV and optical bands used in the SED fitting. 
The optical extinctions are from \citep{schlafly+11}, as given in NED. The UV extinctions are
calculated from A$_{\rm V}$ based on the relations of \citep{wyder+05}.}
\end{deluxetable*}

\section{Position-velocity diagrams of individual objects}
\label{app:pv}
The position-velocity (PV) diagrams of each of the CARMA GOALs objects. Position-velocity diagrams were constructed by estimating the angle across the galaxy with the largest gradient, and slicing the cube at that point, to create a plane. Flux perpendicular to this plane was then summed to create a position-velocity diagram that subtended the majority of the source emission, creating an integrated intensity map across the plane slice. The PV diagrams of each CARMA GOALS source is shown below. Double sources (i.e. sources with discrete emission) were broken out individually, including UGC\,02369, CGCG\,468-002, VV\,250, CGCG\,142-034, and NGC\,6670, shown in Figure\,\ref{fig:pv_doubles}. The left-hand figure for each shows the moment0 map overlaid with the moment1 contours, as well as the chosen rotation angle for the PV slice. The right-hand figure shows the corresponding PV slice. 
\begin{figure*}[t!]
\raggedright
\subfigure{\includegraphics[width=0.49\textwidth,clip,trim=0cm 0cm 2cm 0cm]{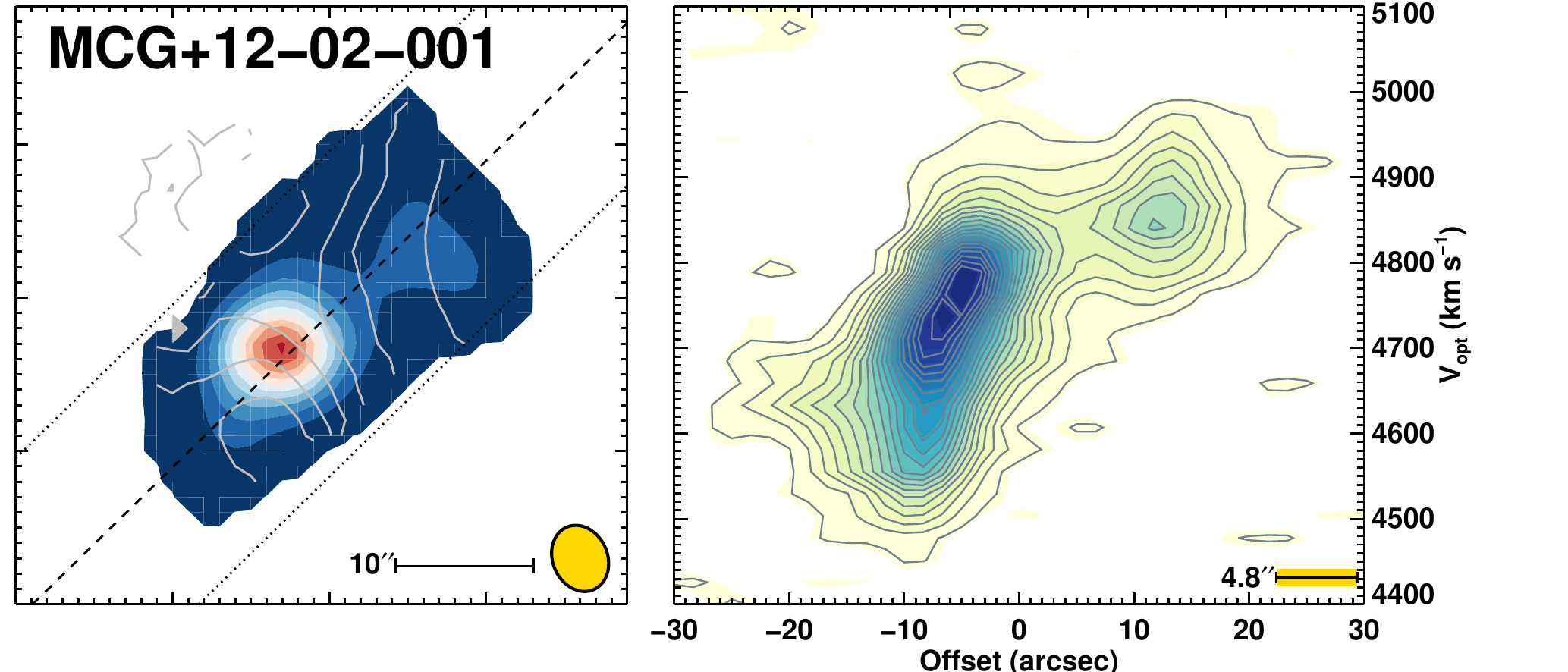}}
\subfigure{\includegraphics[width=0.49\textwidth,clip,trim=0cm 0cm 2cm 0cm]{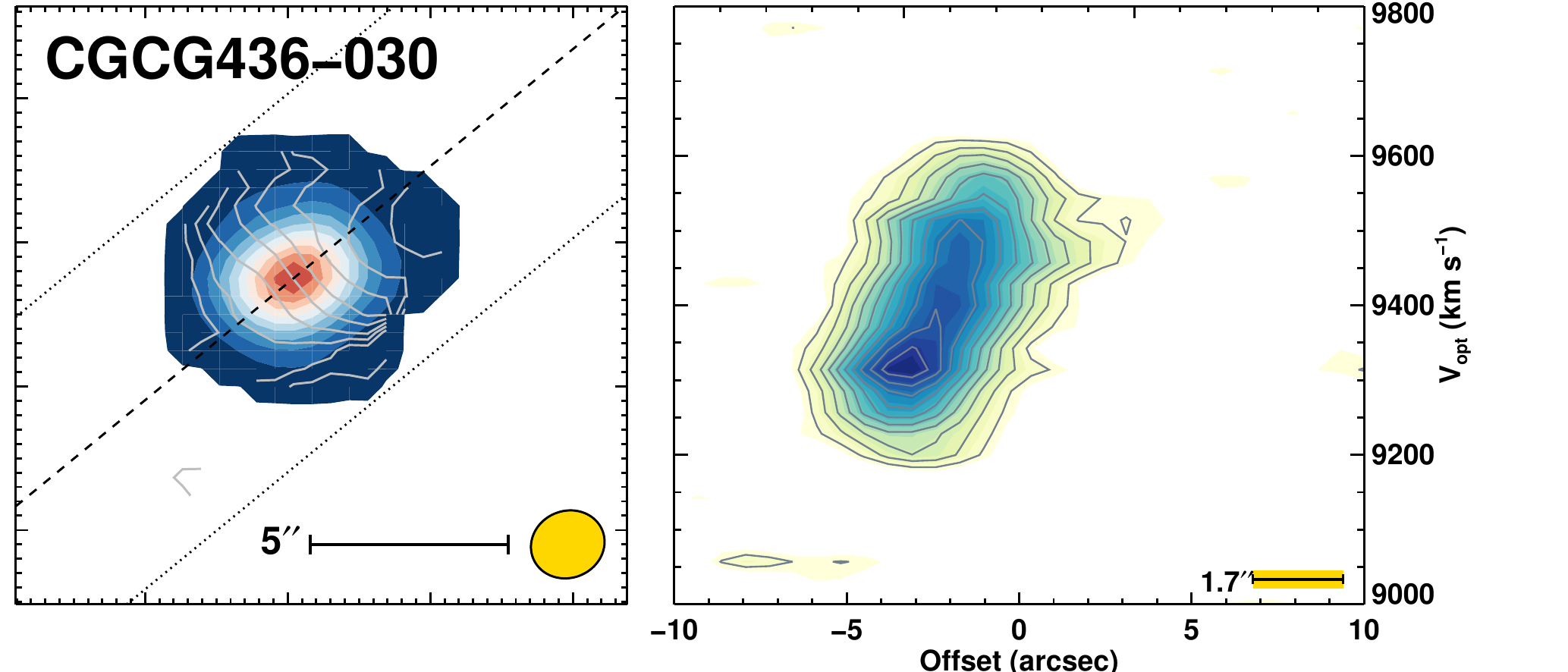}} \vskip -1mm
MCG+12-02-001 (left) with 5$\sigma$ contour levels and CGCG\,436-030 (right) with 2$\sigma$ contour levels. \vskip 1mm

\subfigure{\includegraphics[width=0.49\textwidth,clip,trim=0cm 0cm 2cm 0cm]{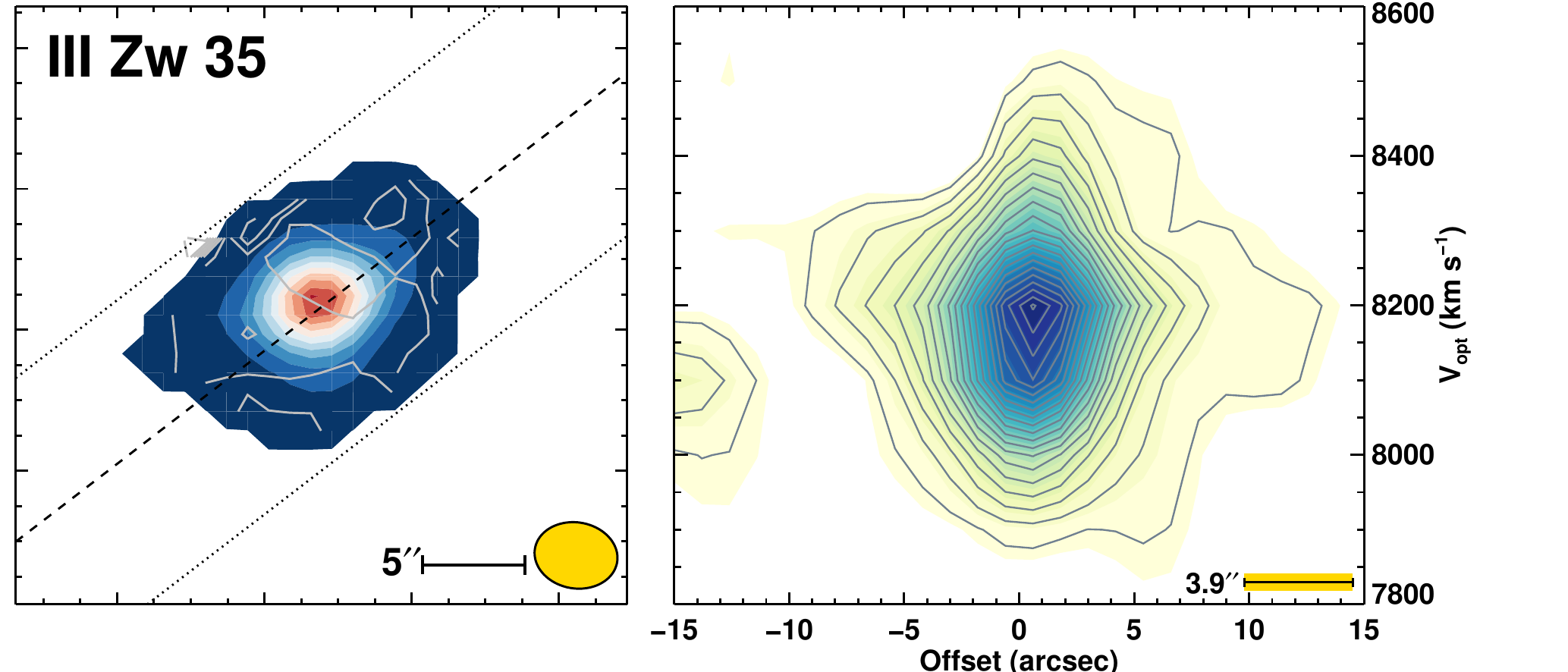}} 
\subfigure{\includegraphics[width=0.49\textwidth,clip,trim=0cm 0cm 2cm 0cm]{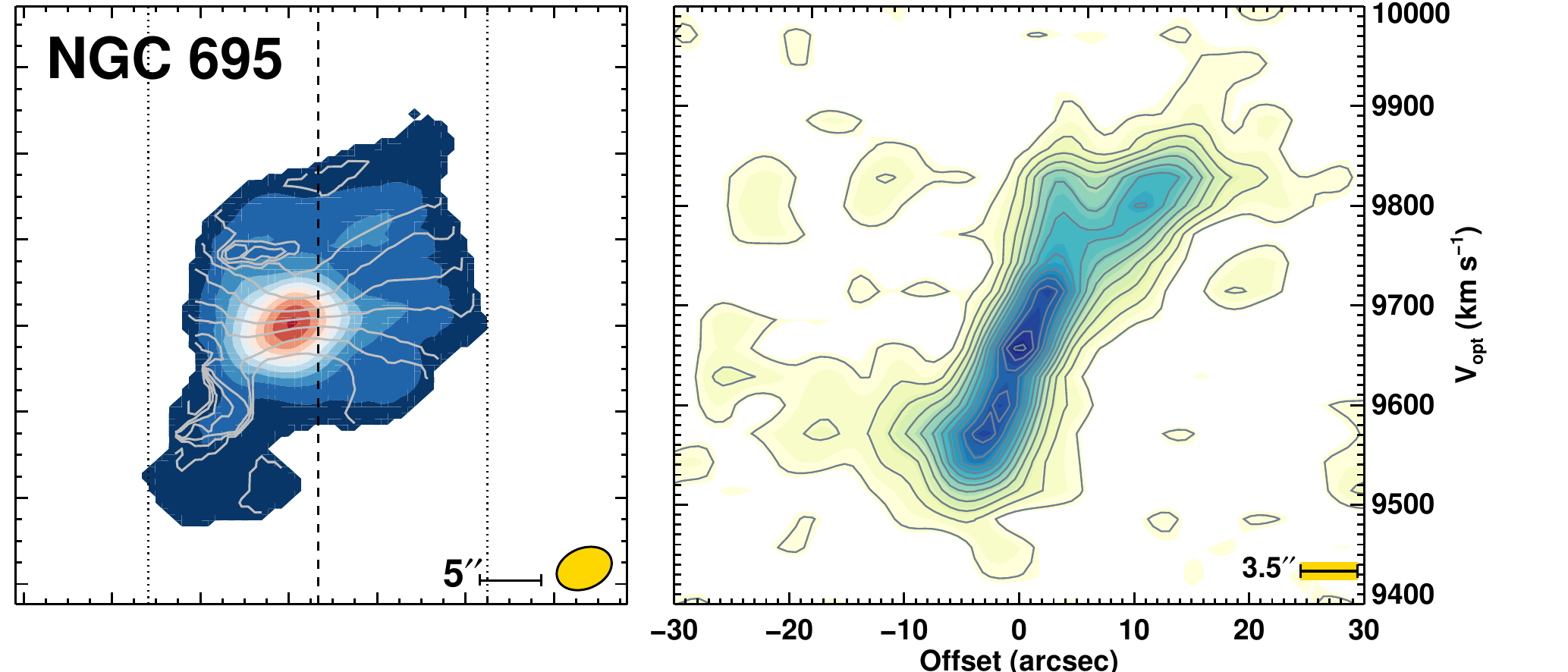}} \vskip -1mm
III\,Zw\,35 (left) with 3$\sigma$ contour levels and NGC\,695 (right) with 6$\sigma$ contour levels. \vskip 1mm

\subfigure{\includegraphics[width=0.49\textwidth,clip,trim=0cm 0cm 2cm 0cm]{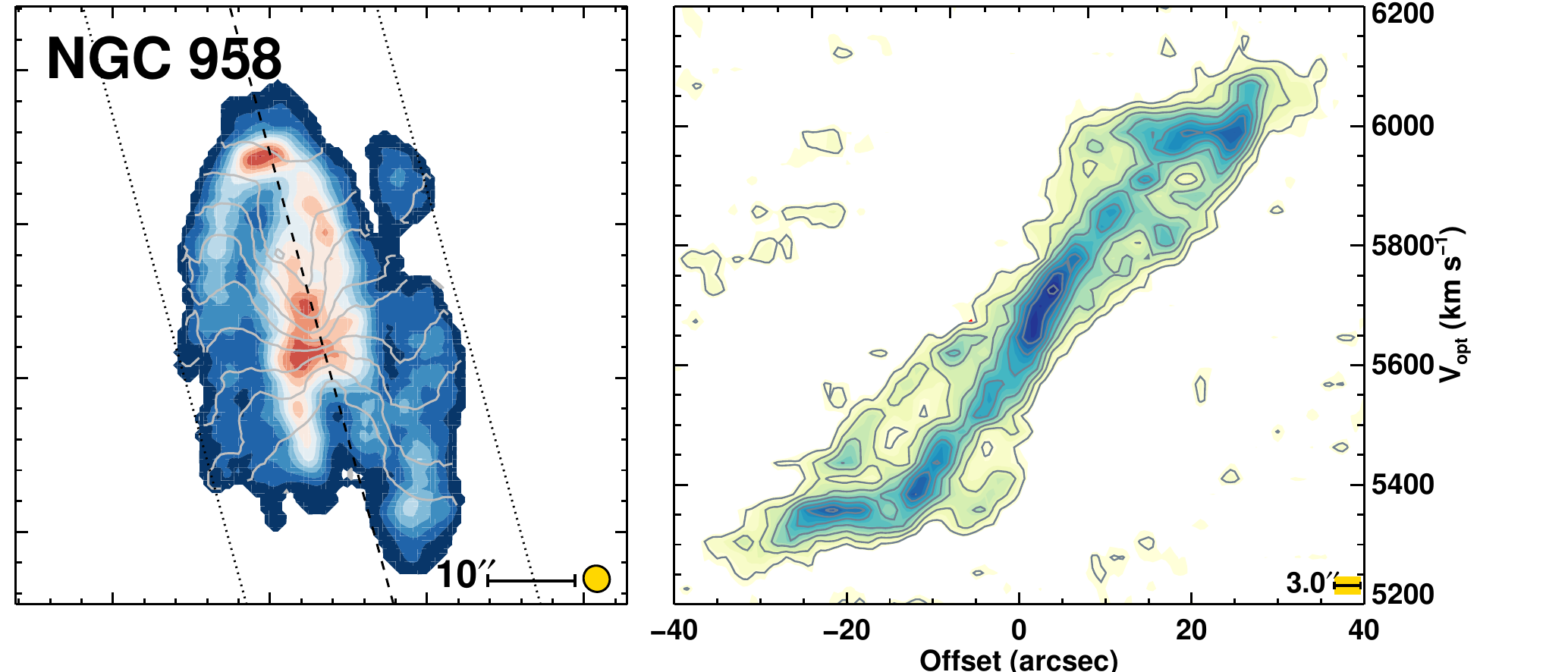}}
\subfigure{\includegraphics[width=0.49\textwidth,clip,trim=0cm 0cm 2cm 0cm]{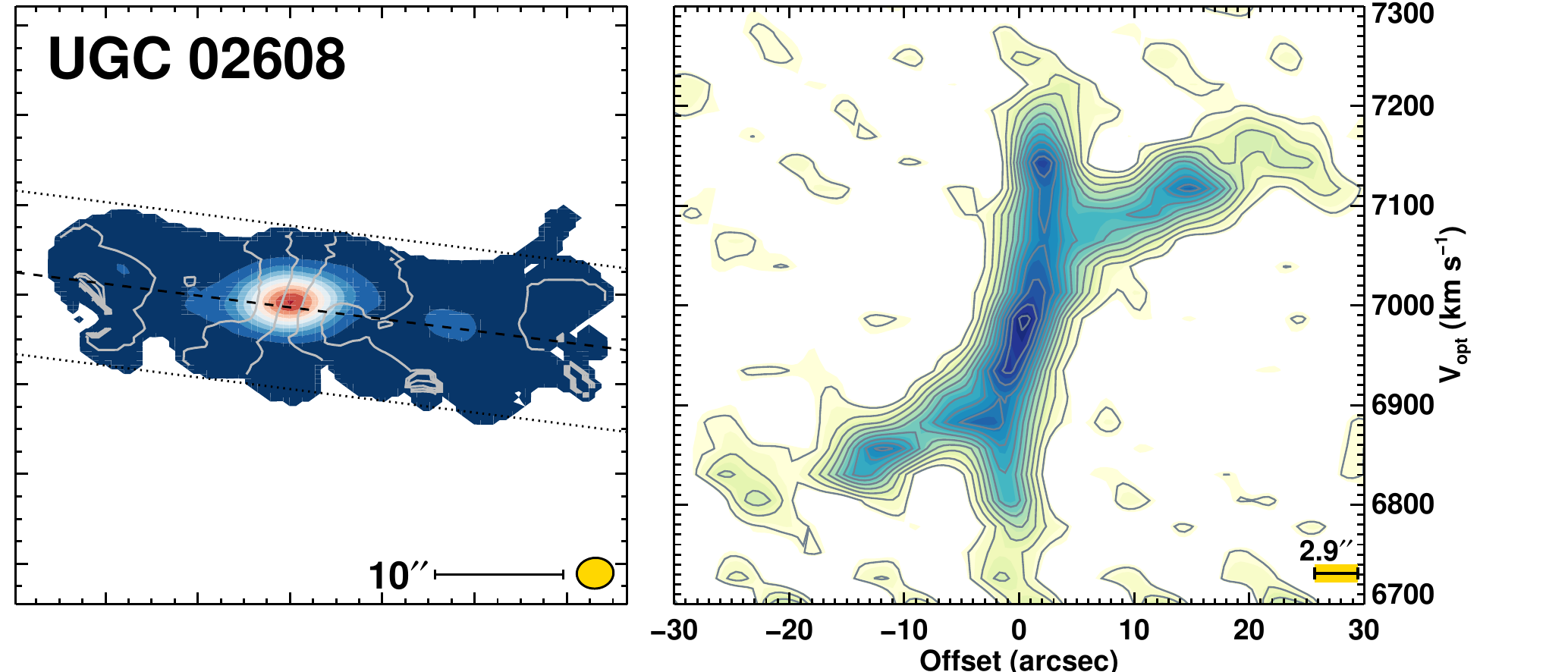}} \vskip -1mm
NGC\,958 (left) with 4$\sigma$ contour levels and UGC\,02608 (right) with 4$\sigma$ contour levels. \vskip 1mm

\subfigure{\includegraphics[width=0.49\textwidth,clip,trim=0cm 0cm 2cm 0cm]{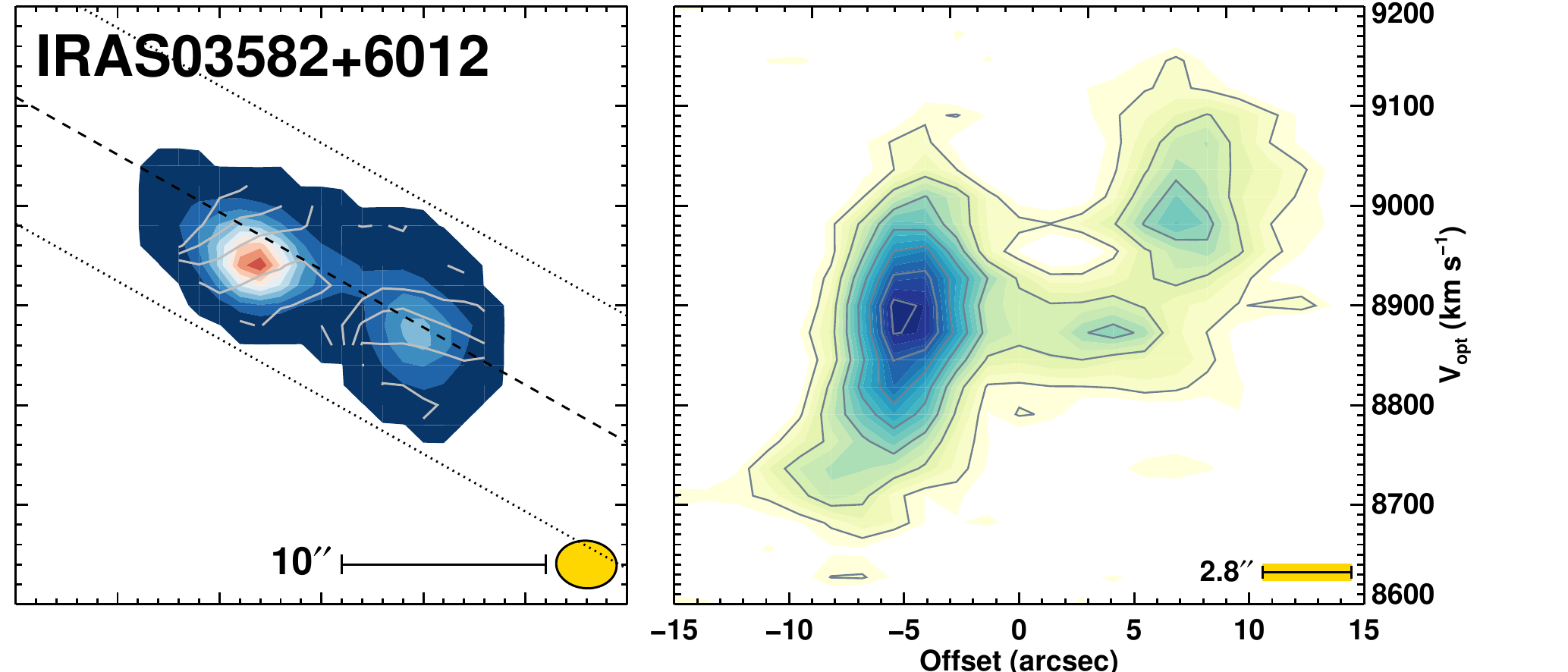}}
\subfigure{\includegraphics[width=0.49\textwidth,clip,trim=0cm 0cm 2cm 0cm]{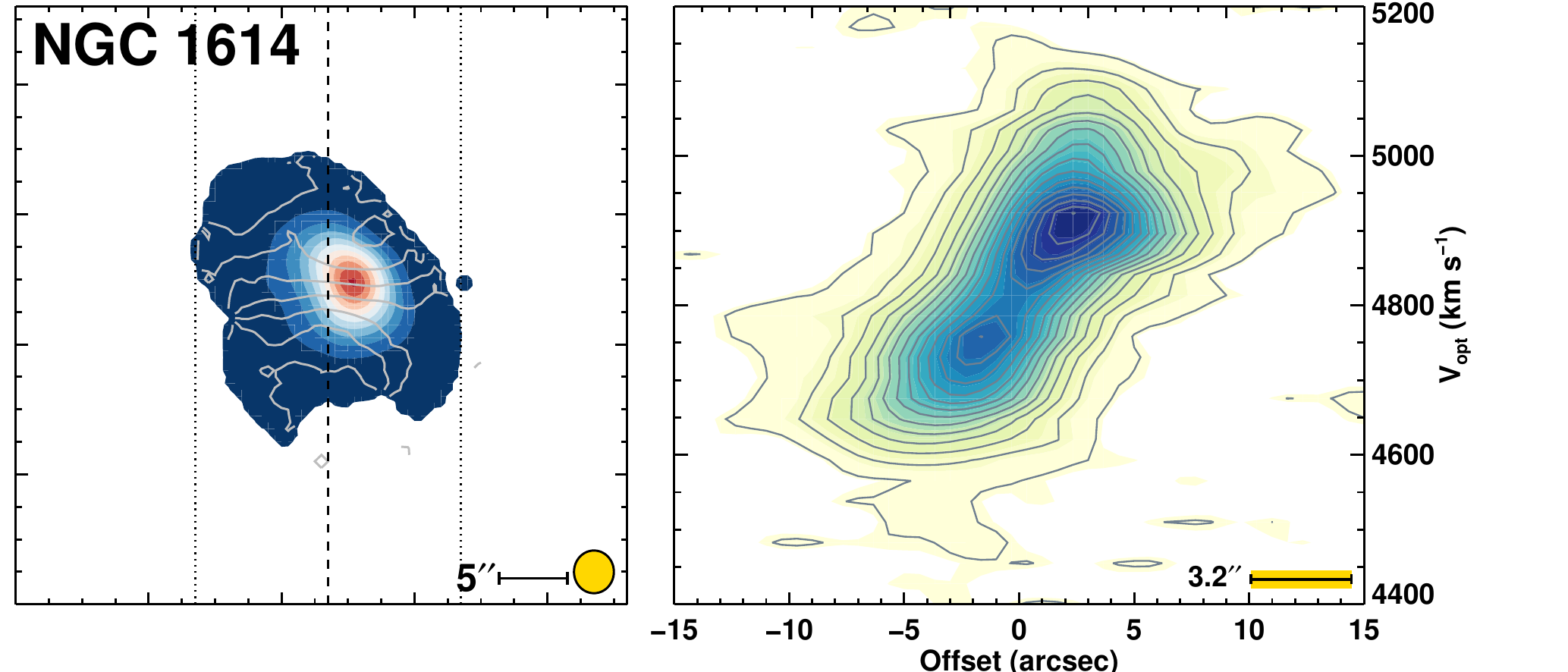}}
\vskip -1mm
IRAS\,03582+6012 (left) with 3$\sigma$ contour levels and NGC\,1614(right) with 8$\sigma$ contour levels. \vskip 1mm
\caption{Position-velocity (PV) diagrams for the CARMA GOALS sample}
\end{figure*}

\renewcommand{\thefigure}{A\arabic{figure} (Cont)}
\addtocounter{figure}{-1}

\begin{figure*}[t!]
\raggedright
\subfigure{\includegraphics[width=0.49\textwidth,clip,trim=0cm 0cm 2cm 0cm]{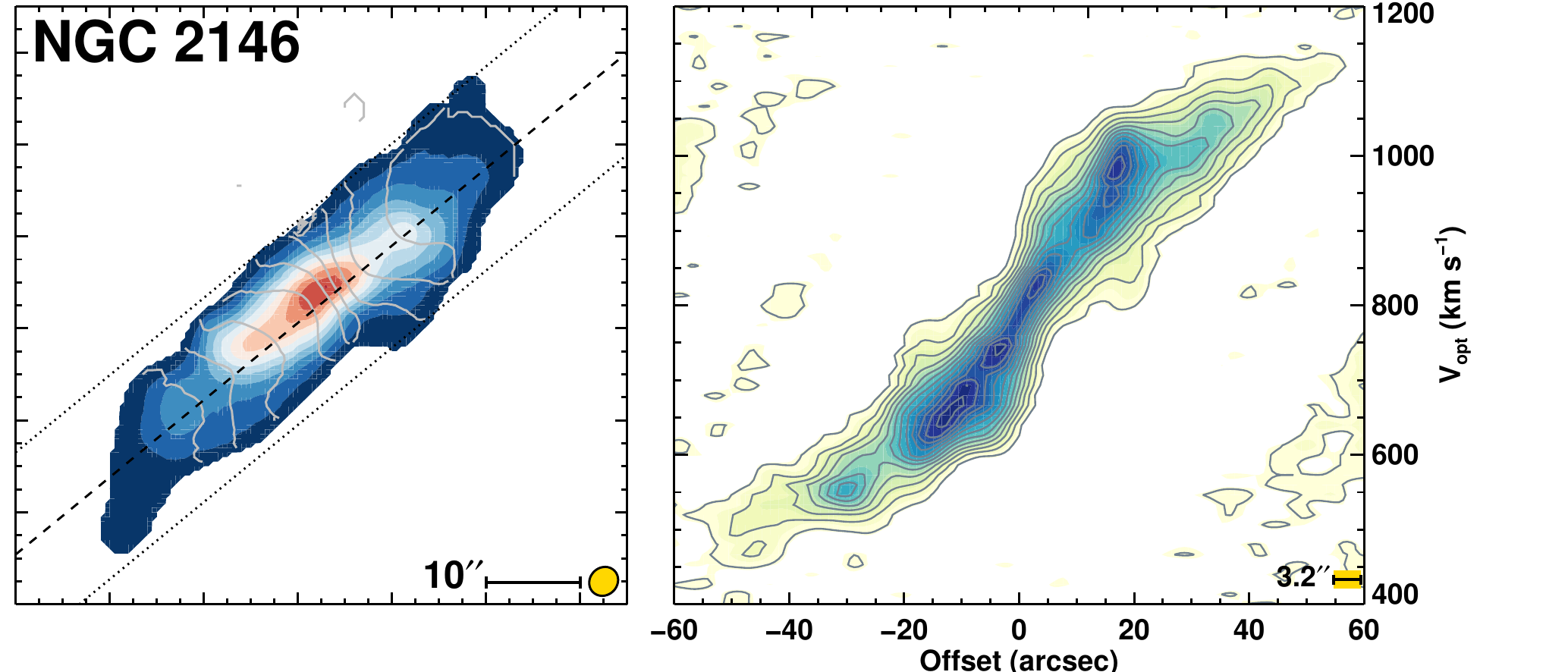}} 
\subfigure{\includegraphics[width=0.49\textwidth,clip,trim=0cm 0cm 2cm 0cm]{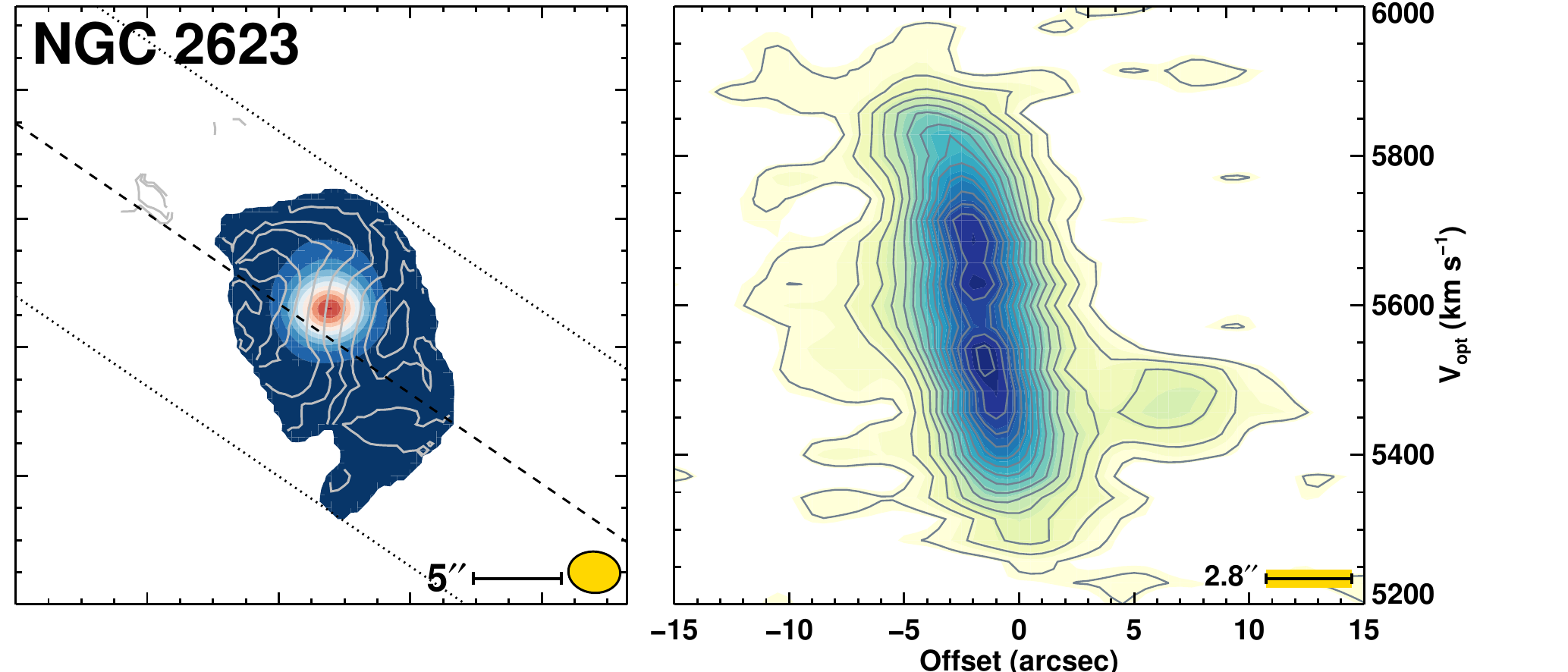}} \vskip -1mm
NGC\,2146 (left) with 8$\sigma$ contour levels and NGC\,2623 (right) with 6$\sigma$ contour levels. \vskip 1mm

\subfigure{\includegraphics[width=0.49\textwidth,clip,trim=0cm 0cm 2cm 0cm]{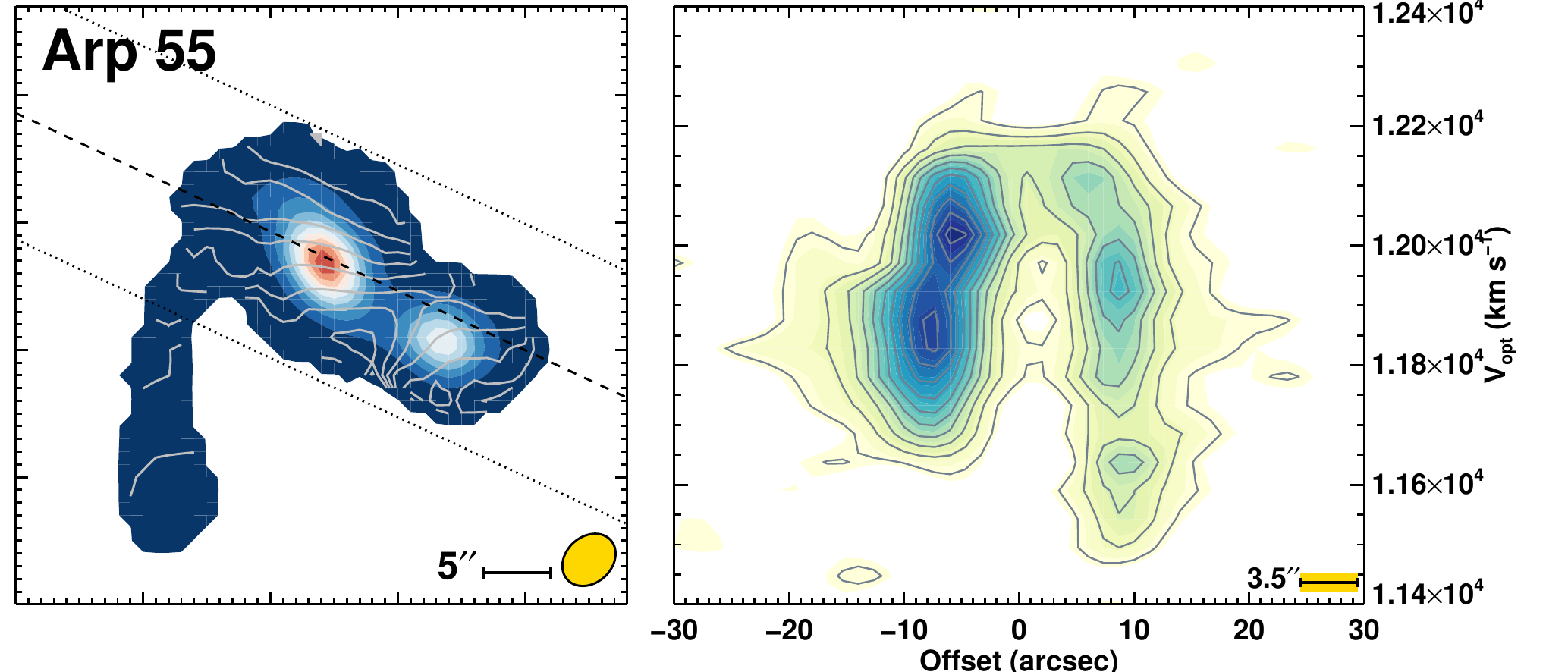}} 
\subfigure{\includegraphics[width=0.49\textwidth,clip,trim=0cm 0cm 2cm 0cm]{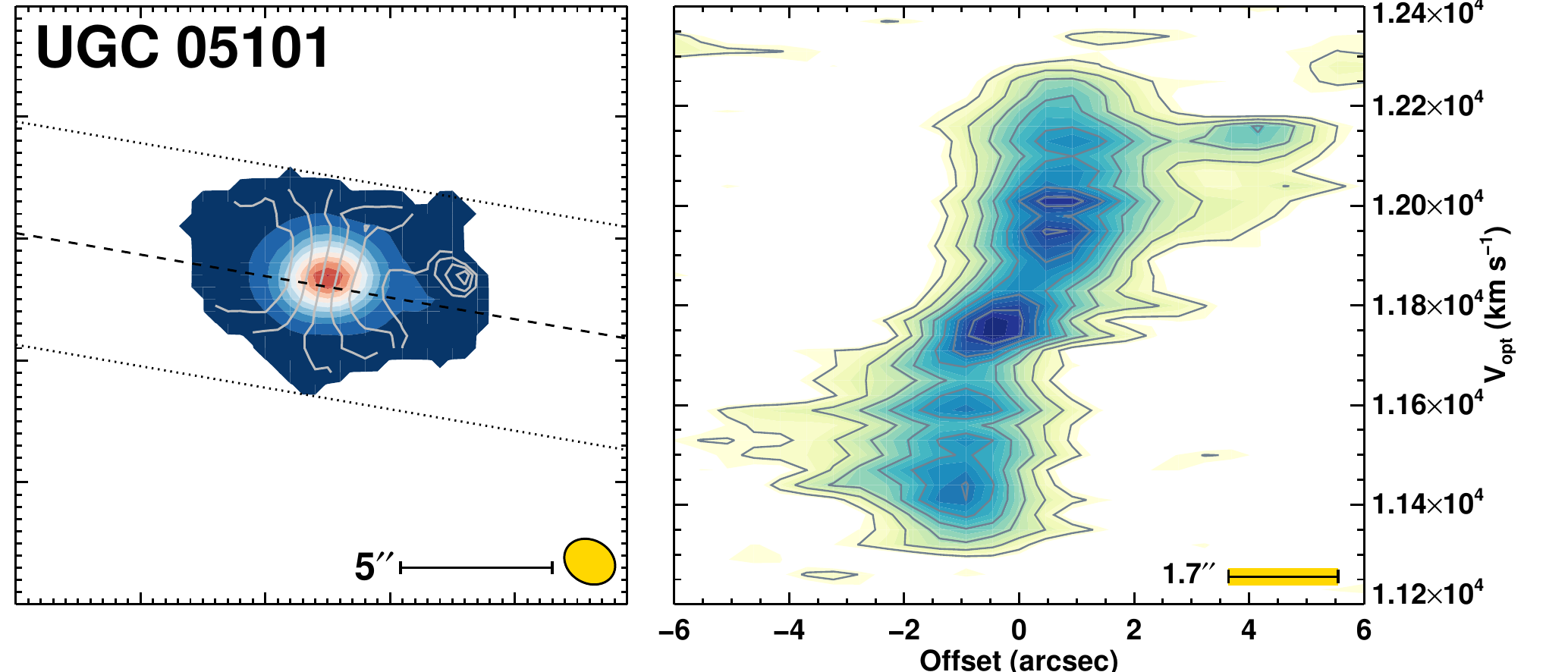}}  \vskip -1mm
Arp\,55 (left) with 4$\sigma$ contour levels and UGC\,05101 (right) with 3$\sigma$ contour levels. \vskip 1mm

\subfigure{\includegraphics[width=0.49\textwidth,clip,trim=0cm 0cm 2cm 0cm]{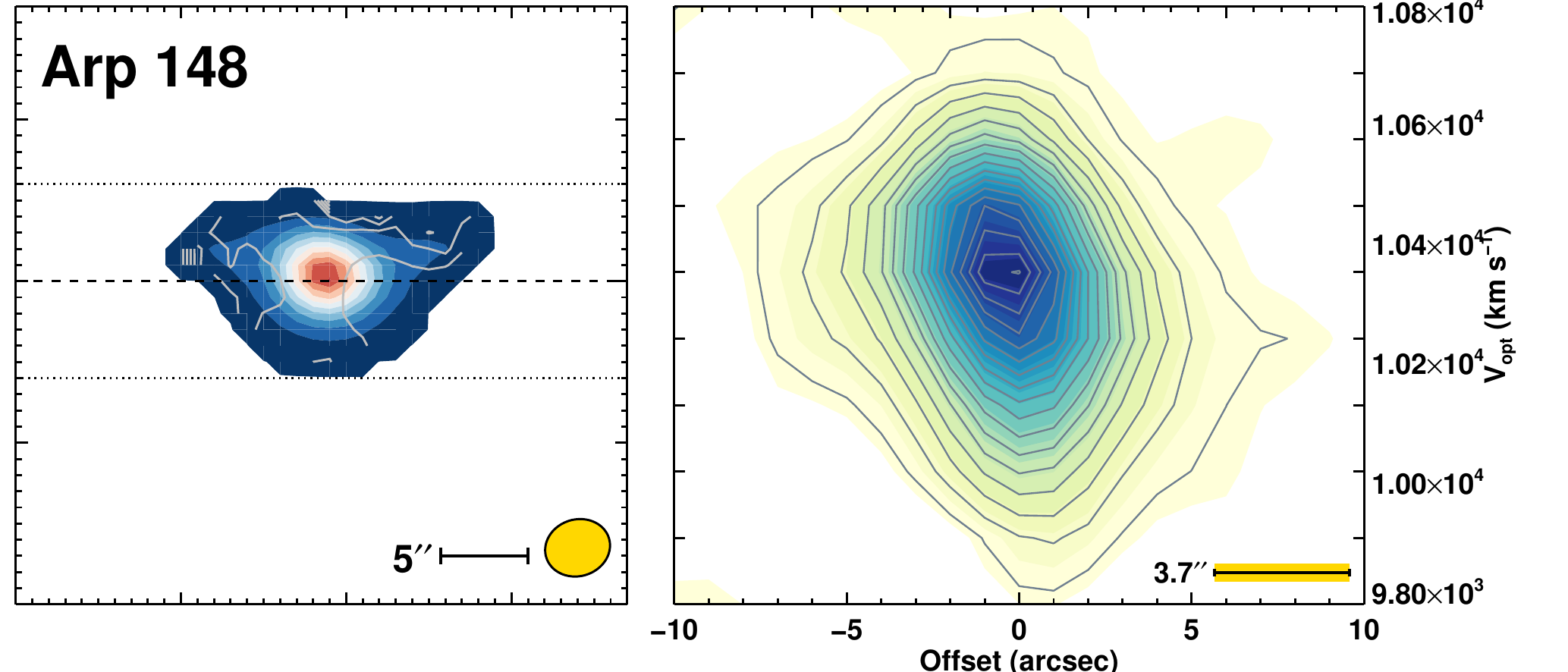}} 
\subfigure{\includegraphics[width=0.49\textwidth,clip,trim=0cm 0cm 2cm 0cm]{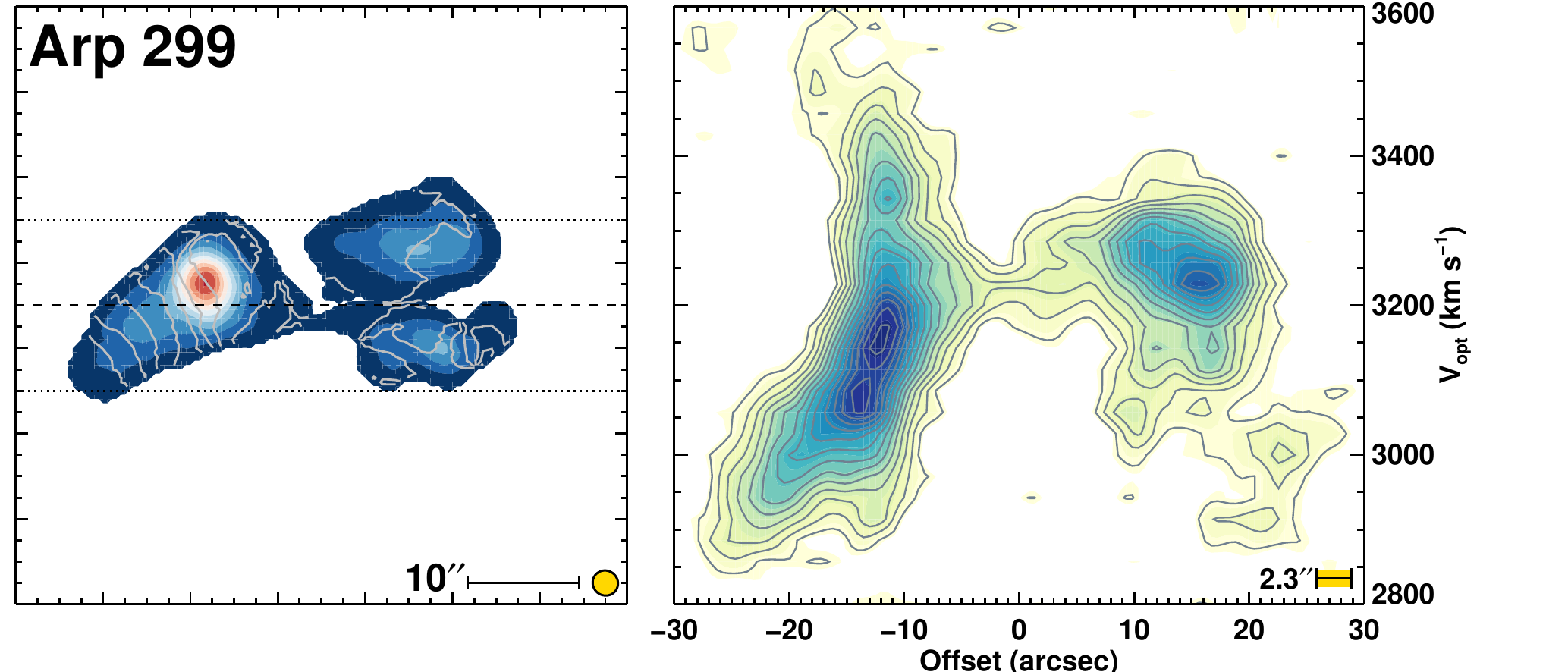}} \vskip -1mm
Arp\,148 (left) with 3$\sigma$ contour levels and  4$\sigma$ contour levels. \vskip 1mm

\subfigure{\includegraphics[width=0.49\textwidth,clip,trim=0cm 0cm 2cm 0cm]{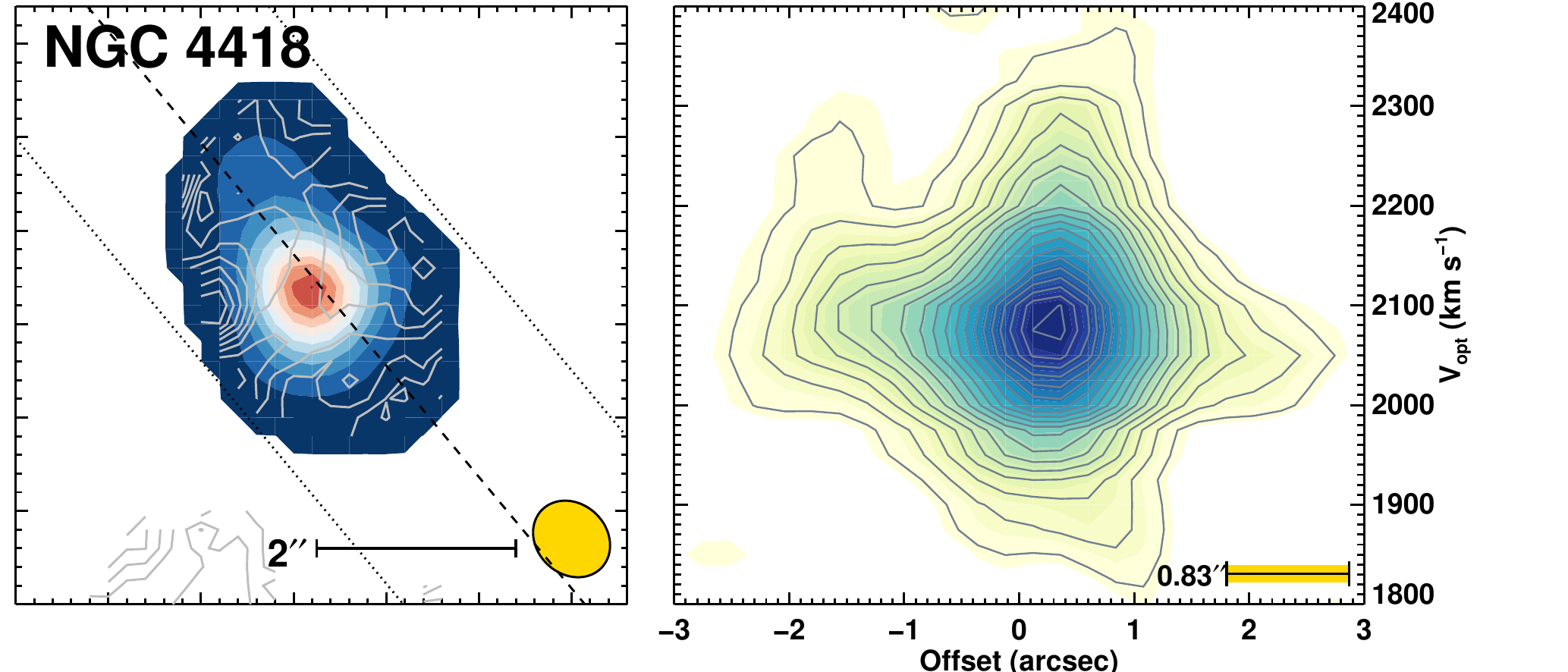}}
\subfigure{\includegraphics[width=0.49\textwidth,clip,trim=0cm 0cm 2cm 0cm]{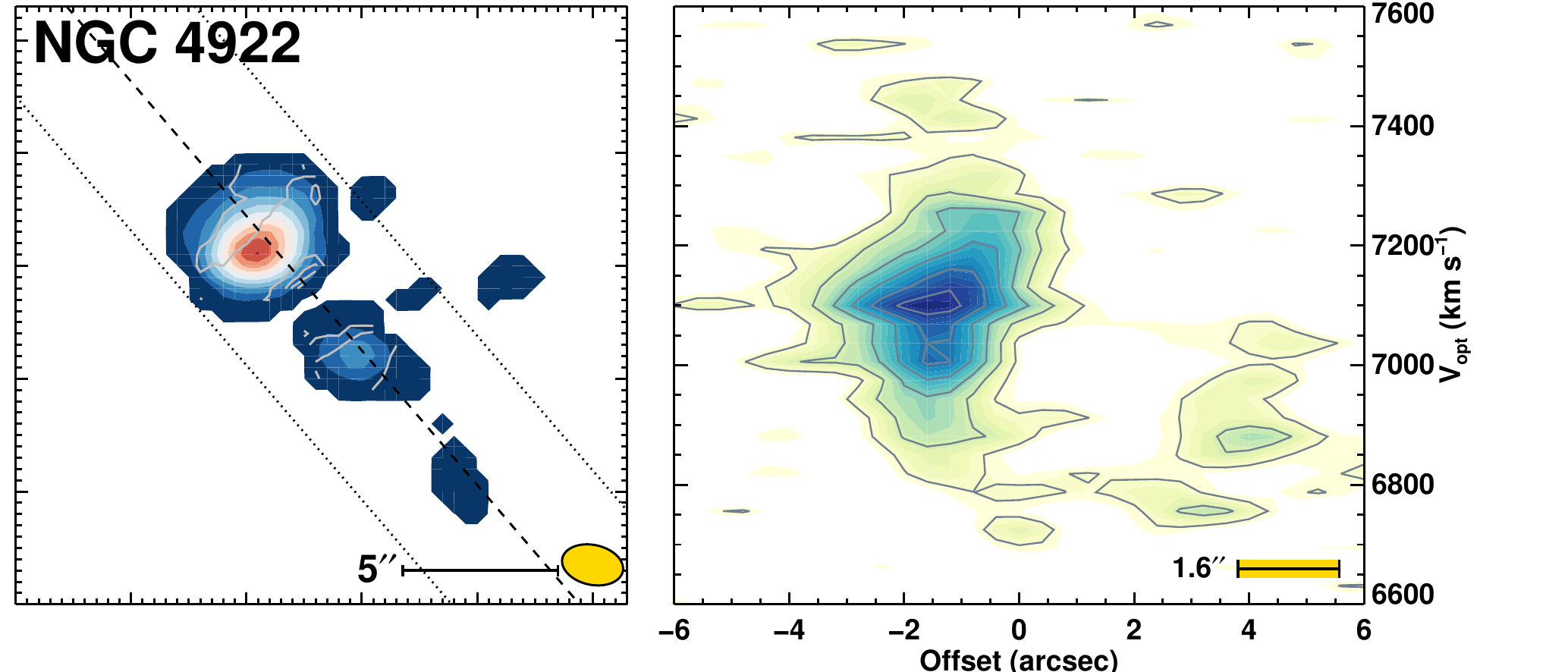}}
%\subfigure{\includegraphics[width=0.49\textwidth,clip,trim=0cm 0cm 2cm 0cm]{figures/mrk231_pv.pdf}}  \vskip -1mm
NGC\,4418 (left) and NGC\,4922 (right) with 3$\sigma$ contour levels. \vskip 1mm

\caption{ }
\end{figure*}
\addtocounter{figure}{-1}

\begin{figure*}[t!]
\subfigure{\includegraphics[width=0.49\textwidth,clip,trim=0cm 0cm 2cm 0cm]{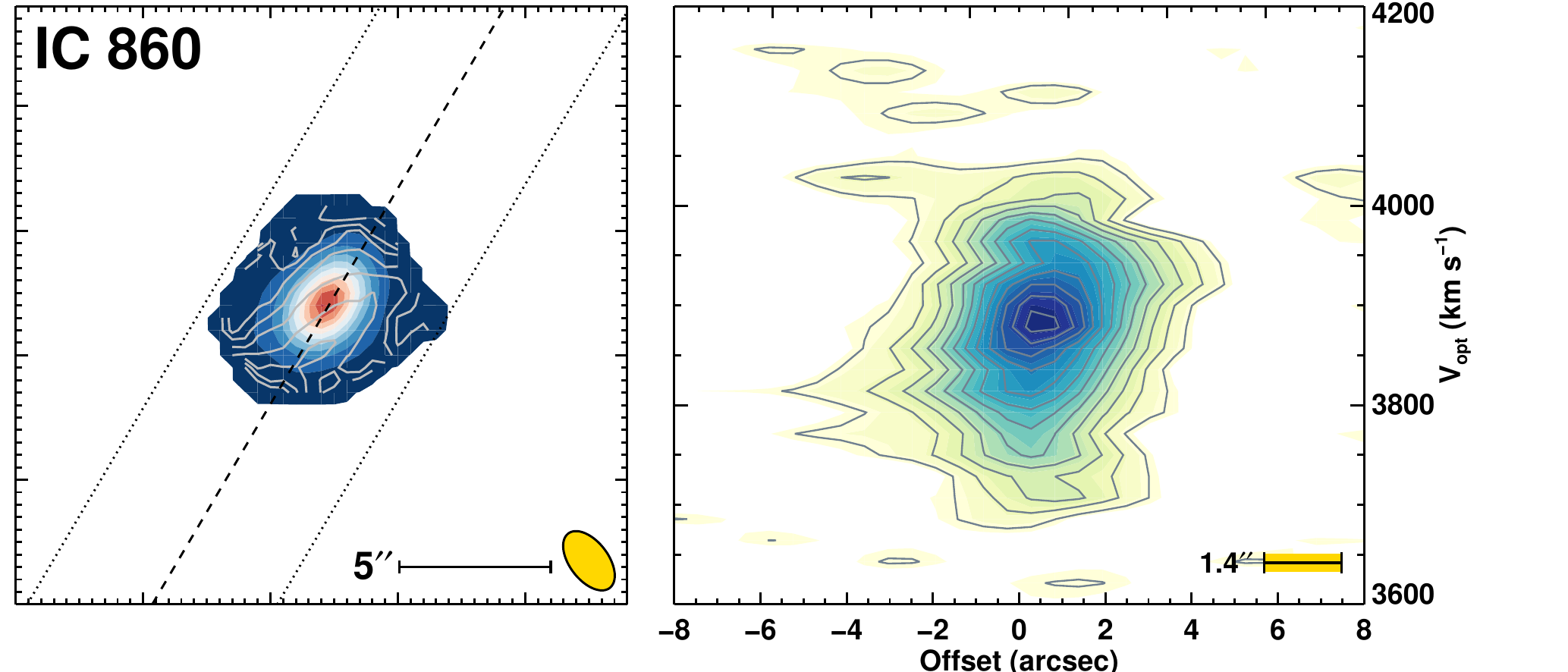}} 
\subfigure{\includegraphics[width=0.49\textwidth,clip,trim=0cm 0cm 2cm 0cm]{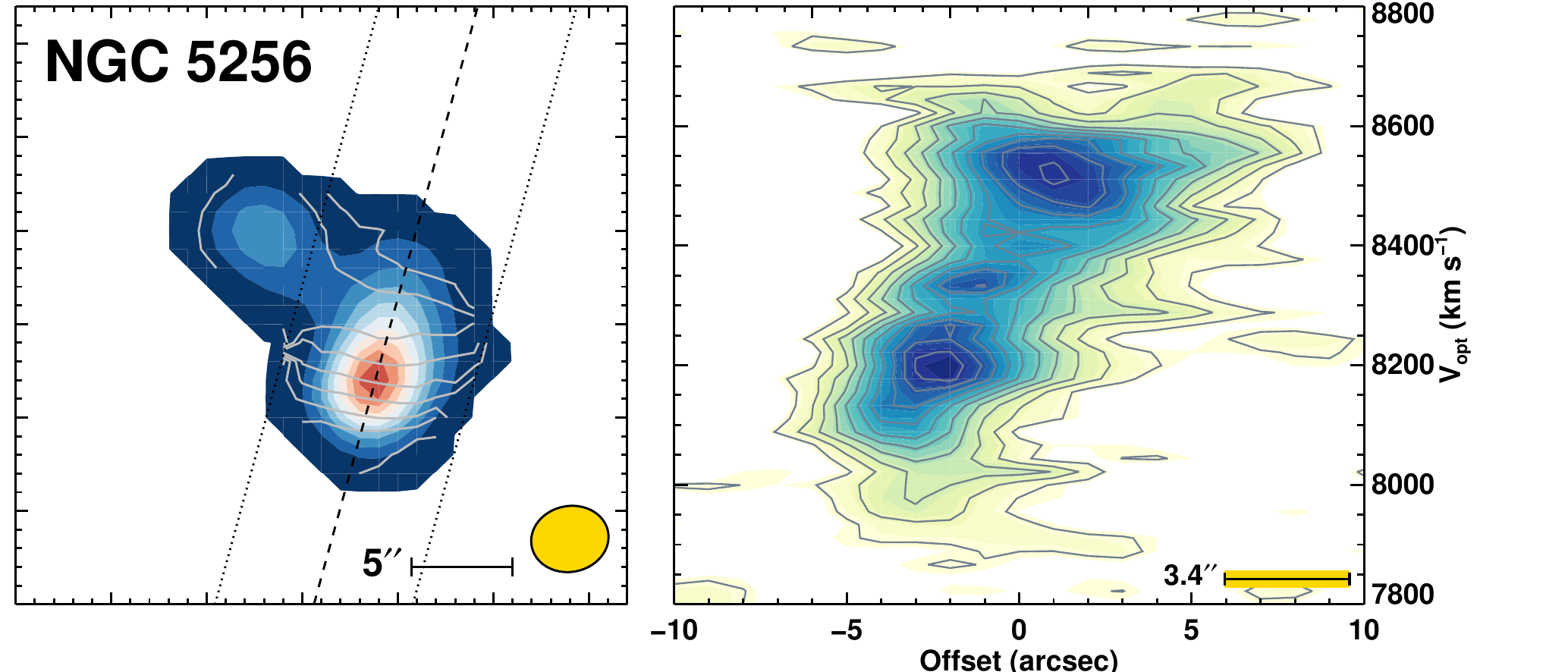}} \vskip -1mm
IC\,860 (left) with 3$\sigma$ contour levels and NGC\,5256 (right) with 4$\sigma$ contour levels. \vskip 1mm

\subfigure{\includegraphics[width=0.49\textwidth,clip,trim=0cm 0cm 2cm 0cm]{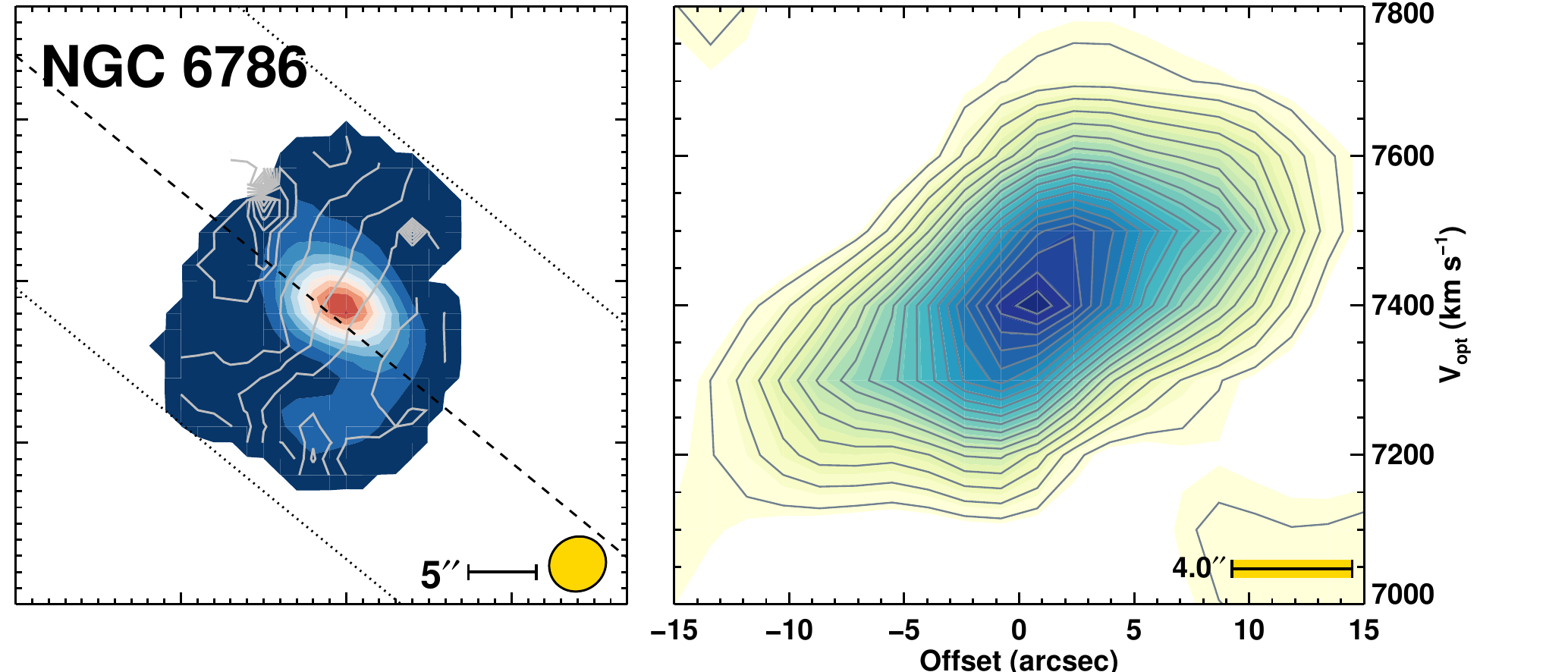}}
\subfigure{\includegraphics[width=0.49\textwidth,clip,trim=0cm 0cm 2cm 0cm]{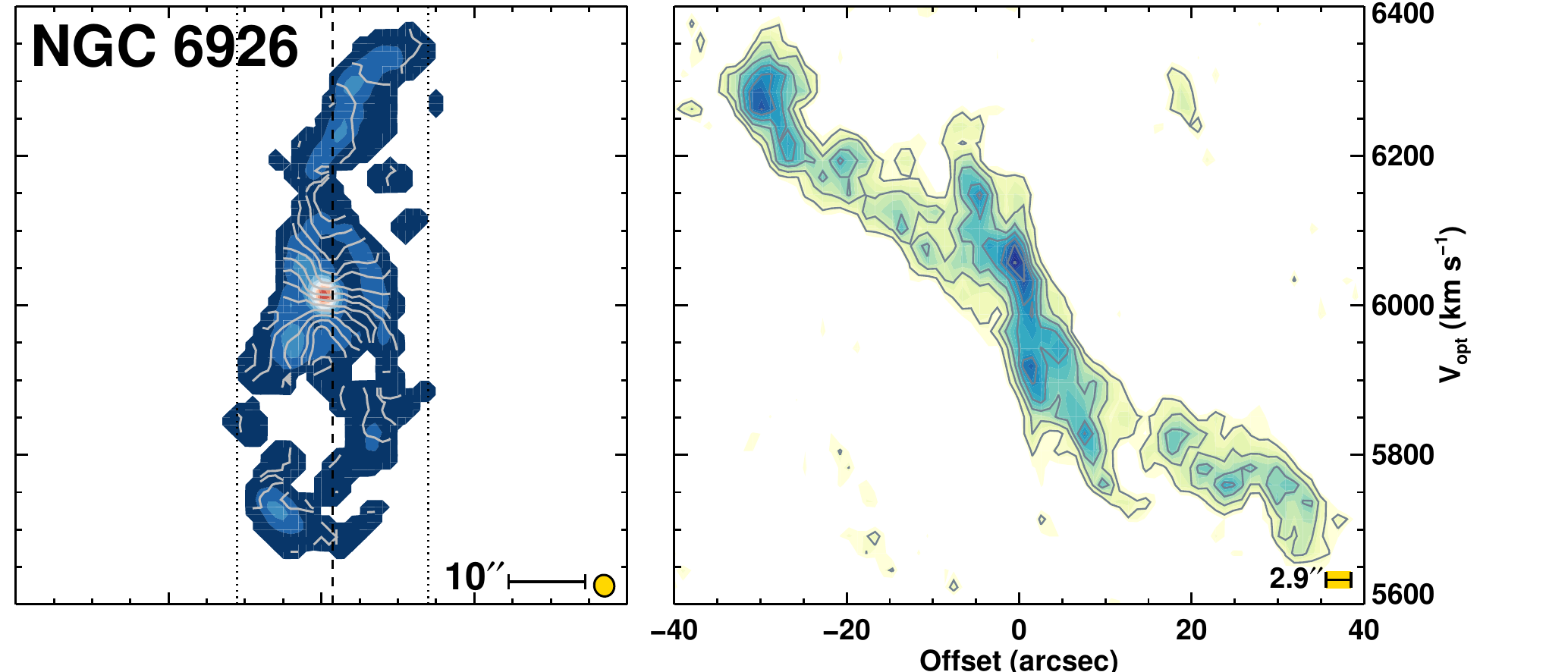}} \vskip -1mm
NGC\,6786 (left) with 3$\sigma$ contour levels and NGC\,6926 (right) with 2$\sigma$ contour levels. \vskip 1mm

\subfigure{\includegraphics[width=0.49\textwidth,clip,trim=0cm 0cm 2cm 0cm]{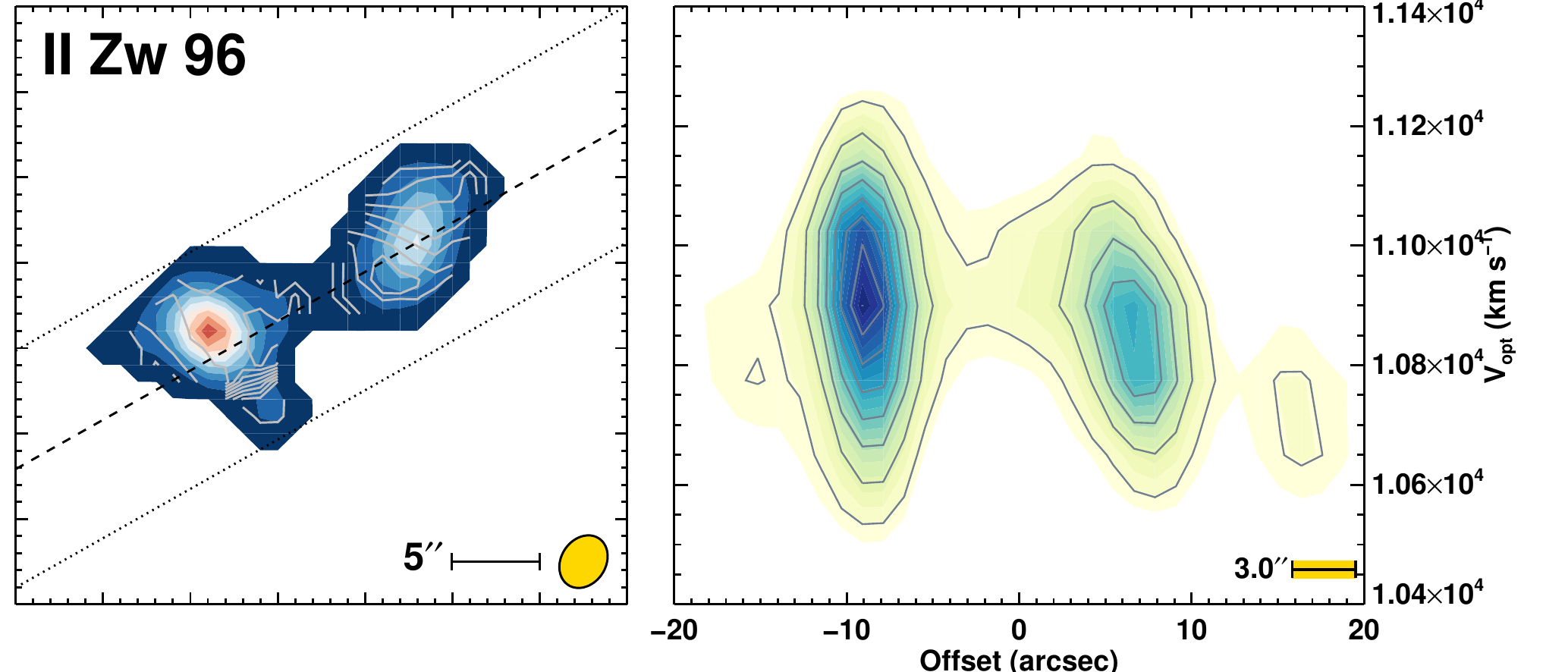}}
\subfigure{\includegraphics[width=0.49\textwidth,clip,trim=0cm 0cm 2cm 0cm]{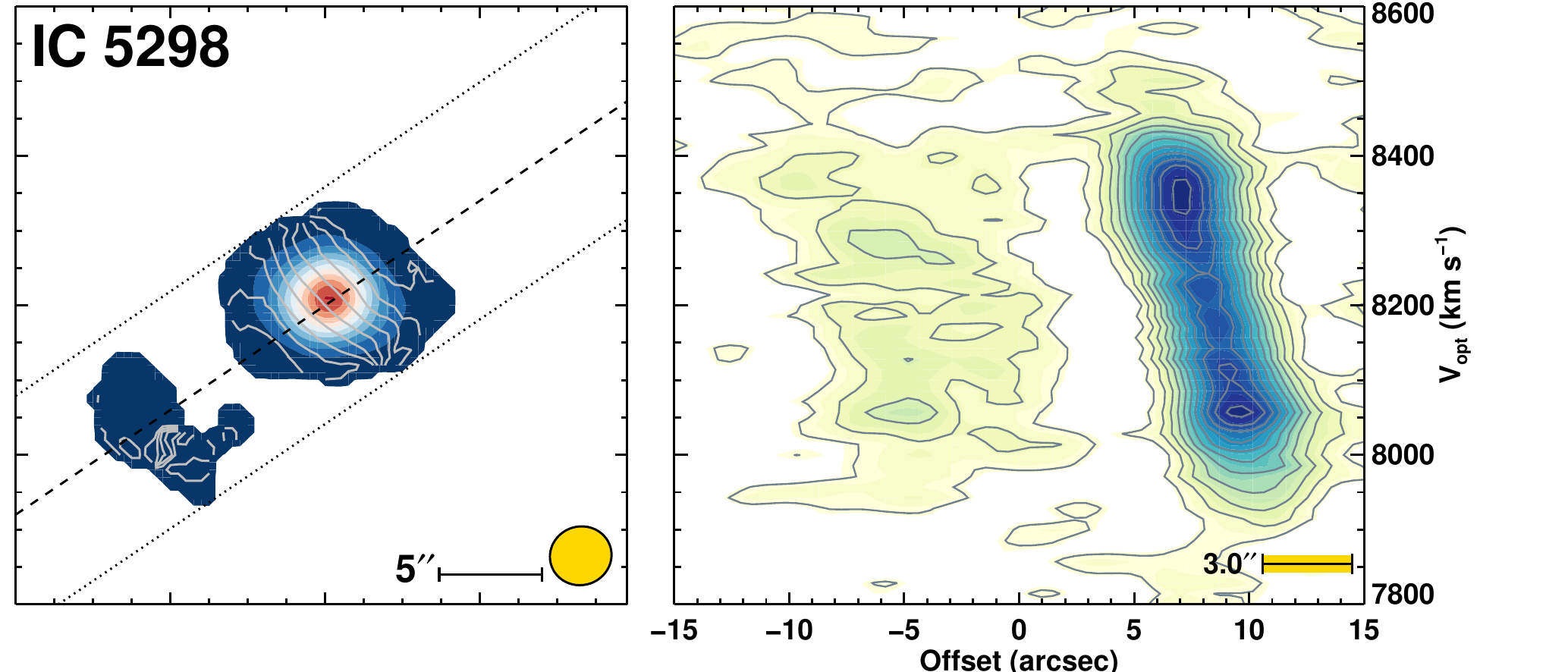}} \vskip -1mm
II\,Zw\,96 (left) with 3$\sigma$ contour levels and IC\,5298 (right) with 6$\sigma$ contour levels. \vskip 1mm

\subfigure{\includegraphics[width=0.49\textwidth,clip,trim=0cm 0cm 2cm 0cm]{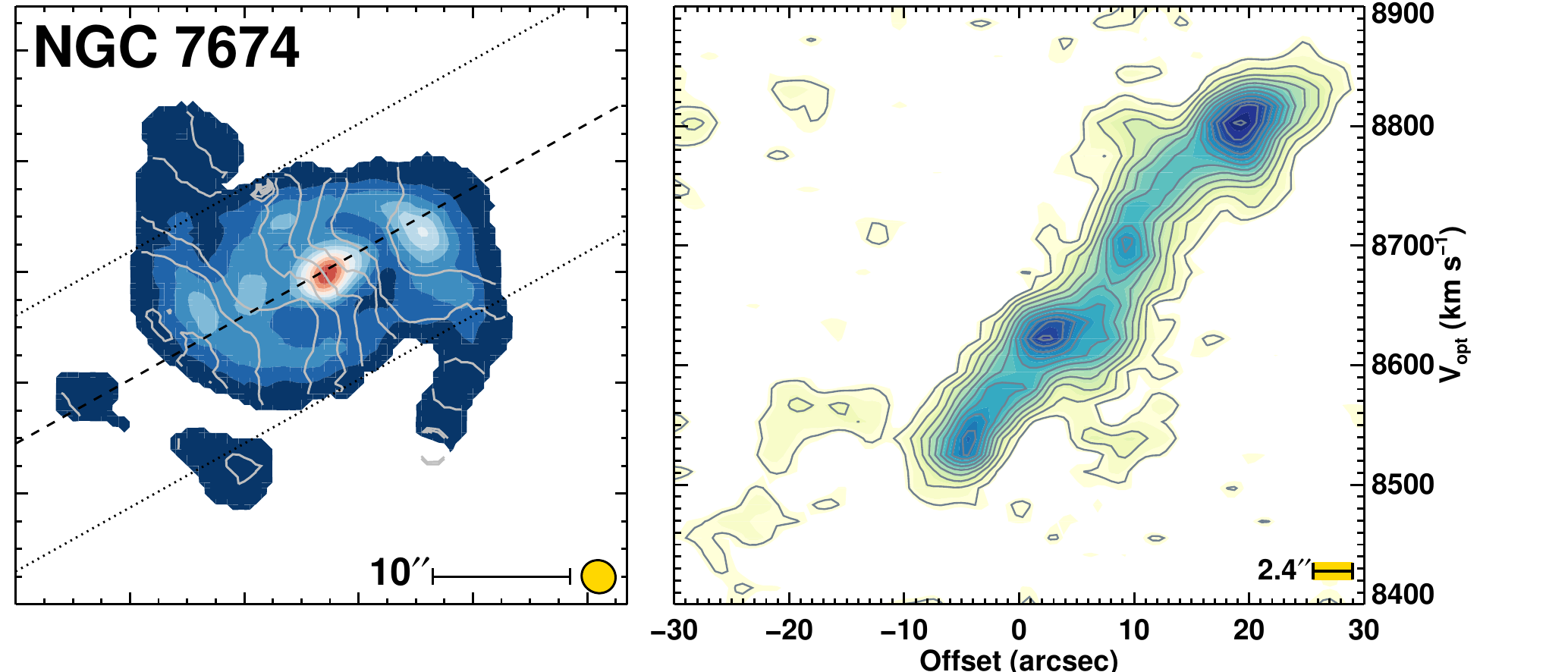}}\vskip -1mm
NGC\,7674 with 4$\sigma$ contour levels. \vskip 1mm
\caption{ }
\end{figure*}

\renewcommand{\thefigure}{A\arabic{figure}}
\begin{figure*}[t!]
\subfigure{\includegraphics[width=0.99\textwidth,clip,trim=0cm 0cm 4.2cm 0cm]{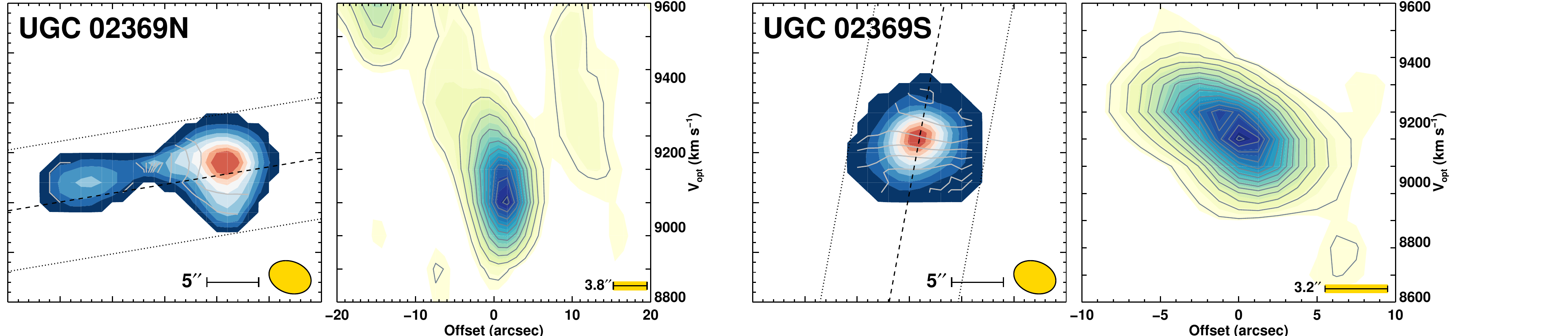}}\vskip -1mm
%UGC\,2369N (left) with 3$\sigma$ contour levels and UGC\,2369S (right) with 5$\sigma$ contour levels. \vskip 1mm

\subfigure{\includegraphics[width=0.99\textwidth,clip,trim=0cm 0cm 4.2cm 0cm]{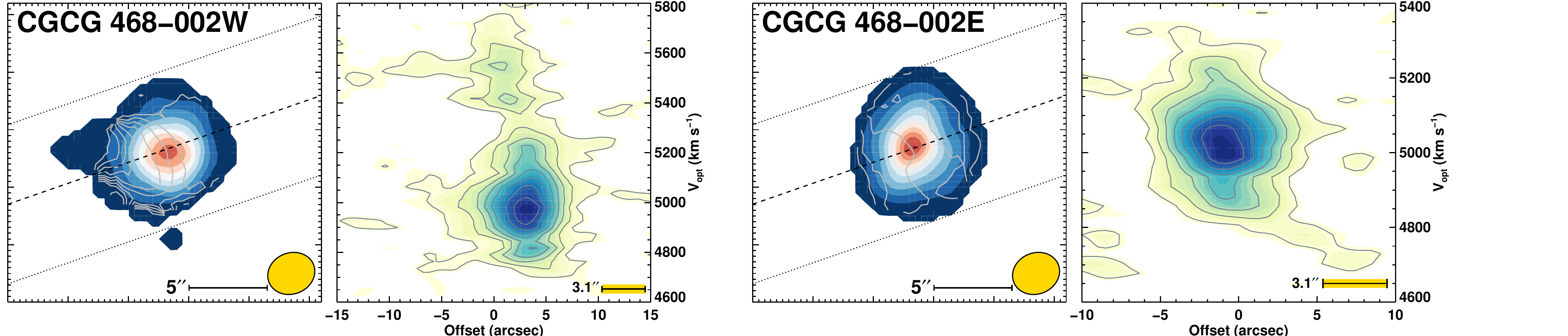}}\vskip -1mm
%CGCG\,468-002W (left) with 5$\sigma$ contour levels and CGCG\,468-002E (right) with 5$\sigma$ contour levels. \vskip 1mm

\subfigure{\includegraphics[width=0.99\textwidth,clip,trim=0cm 0cm 4.2cm 0cm]{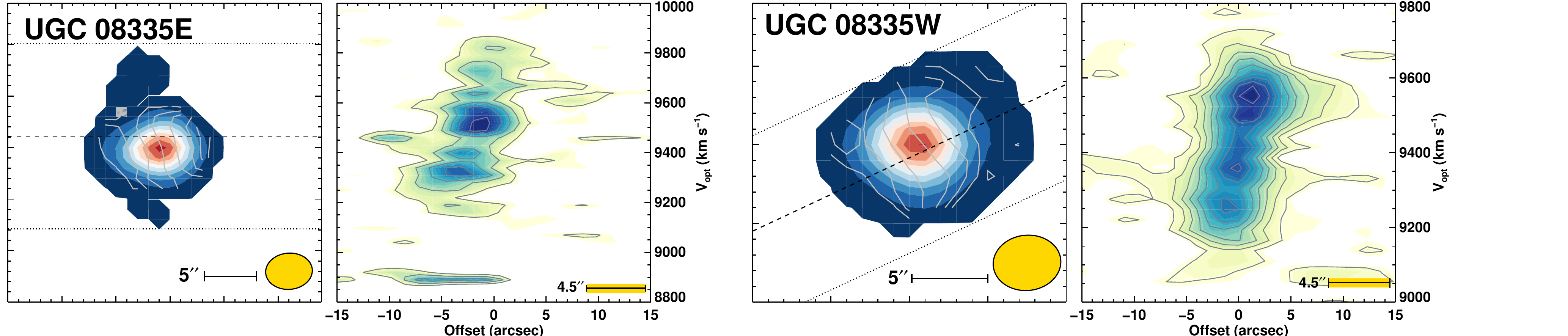}}\vskip -1mm
%UGC\,08335E (left) with 3$\sigma$ contour levels and UGC\,08335W (right) with 3$\sigma$ contour levels, part of VV\,250. \vskip 1mm

\subfigure{\includegraphics[width=0.99\textwidth,clip,trim=0cm 0cm 4.2cm 0cm]{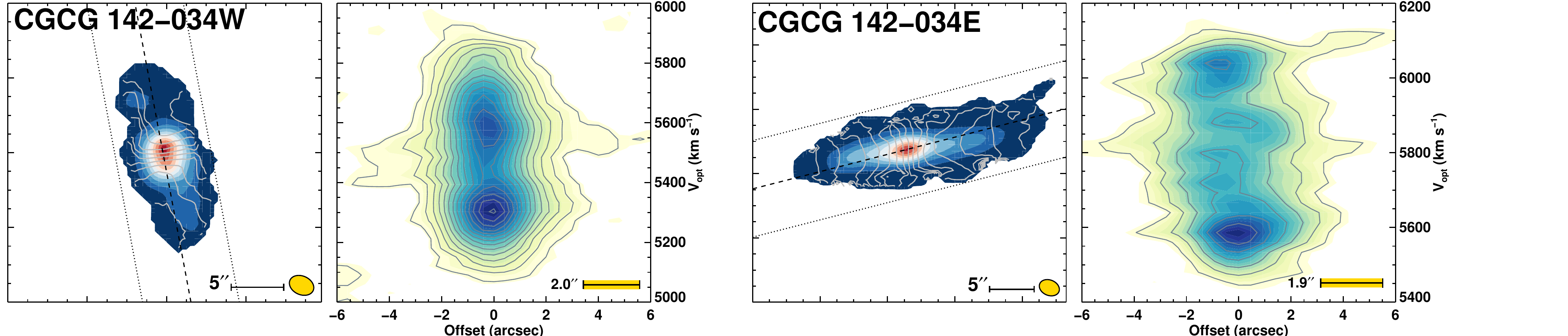}}\vskip -1mm
%CGCG\,142-034W (left) with 5$\sigma$ contour levels and CGCG\,142-034E (right) with 3$\sigma$ contour levels. \vskip 1mm

\subfigure{\includegraphics[width=0.99\textwidth,clip,trim=0cm 0cm 4.2cm 0cm]{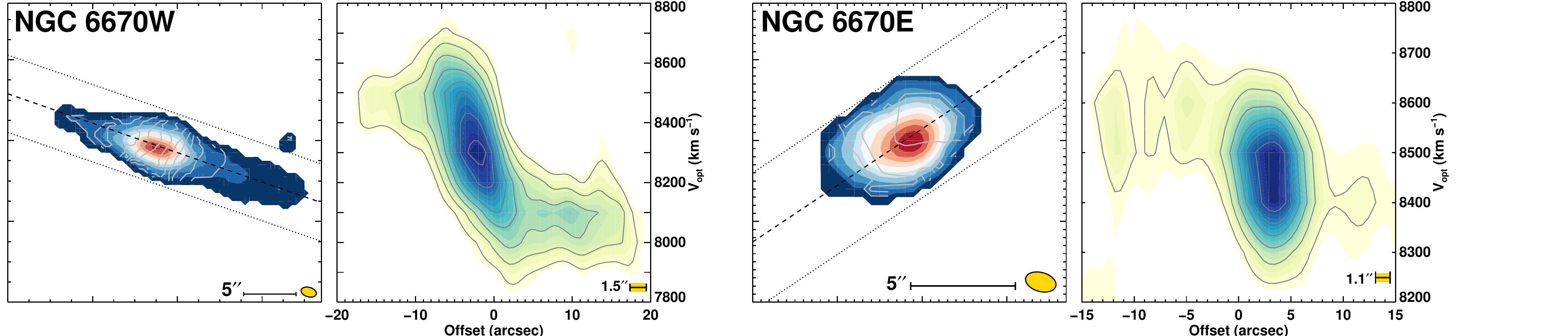}}\vskip -1mm
%NGC\,6670W (left) with 3$\sigma$ contour levels and NGC\,6670E (right) with 3$\sigma$ contour levels. \vskip 1mm

\caption{Position velocity diagrams of double sources, including UGC\,02369 (top row), CGCG\,468-002 (2$^{\rm nd}$ row), VV\,250 (3$^{\rm rd}$ row), CGCG\,142-034 (4$^{\rm th}$ row), and NGC\,6670 (bottom row). The correlator edge is seen in UGC\,02369N, and it is likely that some of the flux is missing from that source.}
\label{fig:pv_doubles}
\end{figure*}

\end{document}